%% file: main.tex
\documentclass[fleqn,10pt]{wlscirep}
\usepackage[utf8]{inputenc}
\usepackage[T1]{fontenc}
\usepackage{caption}
\usepackage{subcaption}
\usepackage{siunitx}
\usepackage{multirow}
\usepackage{amsmath}
\usepackage{array}
\usepackage{booktabs}
\usepackage[ruled]{algorithm2e}
\usepackage{algpseudocode}
\usepackage{svg}
\usepackage{cleveref}
\usepackage{comment}

\usepackage{pdflscape}
\usepackage{rotating} 

\usepackage{multirow}

\usepackage{tokcycle}[2021-03-10]
\newcounter{wordcount}
\newcounter{lettercount}
\newcounter{wordlimit}
\newif\ifinword
\newif\ifrunningcount
\newif\ifsummarycount
\def\limitcolor{red}
\makeatletter
\newcommand\addtomacro[2]{\tc@defx#1{#1#2}}
\newcommand\changecolor[1]{\tctestifx{.#1}{}{\addcytoks{\color{#1}{}}%
  \tc@defx\currentcolor{#1}}}
\makeatother
\newcommand\dumpword{%
  \addcytoks[1]{\accumword}%
  \ifinword\stepcounter{wordcount}
    \ifrunningcount\addcytoks[x]{$^{\thewordcount,\thelettercount}$}\fi
    \ifnum\thewordcount=\value{wordlimit}\relax\changecolor{\limitcolor}\fi
  \fi%
  \inwordfalse
  \def\accumword{}}
\newcommand\addletter[1]{%
  \tctestifcatnx A#1{\stepcounter{lettercount}\inwordtrue}{\dumpword}%
  \addtomacro\accumword{#1}}
\xtokcycleenvironment\countem
  {\addletter{##1}}
  {\dumpword\groupedcytoks{\processtoks{##1}\dumpword\expandafter}\expandafter
    \changecolor\expandafter{\currentcolor}}
  {\dumpword\addcytoks{##1}}
  {\dumpword\addcytoks{##1}}
  {\stripgroupingtrue\def\accumword{}\def\currentcolor{.}
    \setcounter{wordcount}{0}\setcounter{lettercount}{0}}
  {\dumpword\ifsummarycount\tcafterenv{%
    \par(Wordcount=\thewordcount, Lettercount=\thelettercount)}\fi}

\definecolor{YB}{RGB}{0,150,255}
\definecolor{CL}{RGB}{181, 101, 29}

\title{Efficient Nonlinear Function Approximation in Analog Resistive Crossbars for Recurrent Neural Networks}

\input{sections/authors}

\input{sections/00_abstract}

\begin{document}
\captionsetup[figure]{name={Fig.},labelsep=period}
\captionsetup[table]{name={Tab.},labelsep=period}
\setlength{\abovedisplayskip}{10pt}
\setlength{\belowdisplayskip}{10pt}
\flushbottom
\maketitle

\thispagestyle{empty}


\input{sections/01_introduction}
\input{sections/02_results}

\input{sections/03_discussion}
\input{sections/04_methods}

\section*{Acknowledgments}
This work was supported in part by CityU SGP grant 9380132 and ITF MSRP grant ITS/018/22MS; in part by RGC (27210321, C1009-22GF, T45-701/22-R), NSFC (62122005) and Croucher Foundation\\
Any opinions, findings, conclusions or recommendations expressed in this material do not reflect the views of the
Government of the Hong Kong Special Administrative Region, the Innovation and Technology Commission or the Innovation and Technology Fund Research Projects Assessment Panel.
\section*{Author contributions statement}
J.Y and A.B conceived the idea. J.Y performed software experiments with help from Y.C and P.S.V.S with software baselines for KWS and NLP tasks.  M.R performed hardware experiments with help from X.S, G. P and J. I on device fabrication, IC design and system setup respectively. S. D helped with system simulations using Neurosim. J.Y, R.M, C.L and A.B wrote the manuscript with inputs from all authors.
\section*{Data availability}
The data supporting plots within this paper and other findings of this study are available with reasonable requests made to the corresponding author.
\section*{Code availability}
The code used to train the model and perform the simulation on crossbar arrays is publicly available in an online repository\cite{CODELINK}.

\bibliography{sample} 



\newpage
\input{Supplementary}
\end{document}

%% file: sections/authors.tex
\author[1,*]{Junyi Yang}
\author[2,*]{Ruibin Mao}
\author[2]{Mingrui Jiang}
\author[1]{Yichuan Cheng}
\author[1]{Pao-Sheng Vincent Sun}
\author[1]{Shuai Dong}
\author[3]{Giacomo Pedretti}
\author[3]{Xia Sheng}
\author[3]{Jim Ignowski}
\author[1]{Haoliang Li}
\author[2,+]{Can Li}
\author[1,+]{Arindam Basu}

\affil[1]{City University of Hong Kong}
\affil[2]{The University of Hong Kong}
\affil[3]{Hewlett Packard Labs, Hewlett Packard Enterprise, Milpitas, CA, USA.}
\affil[*]{Equal Contribution}
\affil[+]{Correspondence to arinbasu@cityu.edu.hk and canl@hku.hk}

%% file: sections/00_abstract.tex
\begin{abstract}
Analog In-memory Computing (IMC) has demonstrated energy-efficient and low latency implementation of convolution and fully-connected layers in deep neural networks (DNN)  by using physics for computing in parallel resistive memory arrays. However, recurrent neural networks (RNN) that are widely used for speech-recognition and natural language processing have tasted limited success with this approach. This can be attributed to the significant time and energy penalties incurred in implementing nonlinear activation functions that are abundant in such models. In this work, we experimentally demonstrate the implementation of a non-linear activation function integrated with a ramp analog-to-digital conversion (ADC) at the periphery of the memory to improve in-memory implementation of RNNs. Our approach uses an extra column of memristors to produce an appropriately pre-distorted ramp voltage such that the comparator output directly approximates the desired nonlinear function. We experimentally demonstrate programming different nonlinear functions using a memristive array and simulate its incorporation in RNNs to solve keyword spotting and language modelling tasks. Compared to other approaches, we demonstrate manifold increase in area-efficiency, energy-efficiency and throughput due to the in-memory, programmable ramp generator that removes digital processing overhead.

\end{abstract}

%% file: sections/01_introduction.tex
\section*{Introduction} 
Artificial Intelligence (AI) algorithms, spurred by the growth of deep neural networks (DNN), have produced the state-of-the-art solutions in several domains ranging from computer vision\cite{vit}, speech recognition\cite{graves2013speech}, game playing\cite{silver2016mastering} to scientific discovery\cite{senior2020improved}, natural language processing\cite{vaswani2017attention} and more. The general trend in all these applications has been increasing the model size by increasing the number of layers and the number of weights in each layer. This trend has, however, caused growing concern in terms of energy efficiency for both edge applications and servers for training; power is scarce due to battery limits in edge devices while the total energy required for training large models in the cloud raises environmental concerns. Edge devices have a further challenge posed by strong latency requirements in applications such as keyword spotting to turn on mobile devices, augmented reality and virtual reality platforms, anti-collision systems in driverless vehicles etc.

The bottleneck for implementing DNNs on current hardware arises due to the frequent memory access necessitated by the von Neumann architecture and the high memory access energy for storing the parameters of a large model\cite{horowitz20141}. As a solution to this problem, a new architecture of In-memory Computing (IMC) has become increasingly popular. Instead of reading and writing data from memory in every cycle, IMC allows neuronal weights to remain stationary in memory with inputs being applied to it in parallel and the final output prior to neuronal activation being directly read from memory. Among the IMC techniques explored, analog/mixed-signal IMC using non-volatile memory devices such as memristive\cite{memristor_intro} ones have shown great promise in improving latency, energy and area efficiencies of DNN training\cite{strukov_training1,strukov_training2,can_training} and inference\cite{sebastian2020memory,memristor_conv,memristor_mlp}, combinatorial optimization\cite{yang2020transiently,jiang2023efficient}, hardware security\cite{john2021halide,sebastian2020memory}, content addressable memory\cite{mao2022experimentally}, signal processing\cite{sheridan2017sparse,zidan2018general,memristor_lineareq} etc. It should be noted that analog IMC does not refer to the input and output signals being analog; rather, it refers to the storage of multi-bit or analog weights in each memory cell (as opposed to using memristor for 1-bit storage\cite{digital_imc_1,digital_imc_2}) and using analog computing techniques (such as Ohm's law, Kirchoff's law etc.) for processing inputs. Analog weight storage\cite{natIBM64core} enables higher density of weights as well as higher parallelism (by enabling multiple rows simultaneously) compared to digital counterparts.

\begin{figure}[t]
    \centering 
    \includegraphics[width=0.95\textwidth]{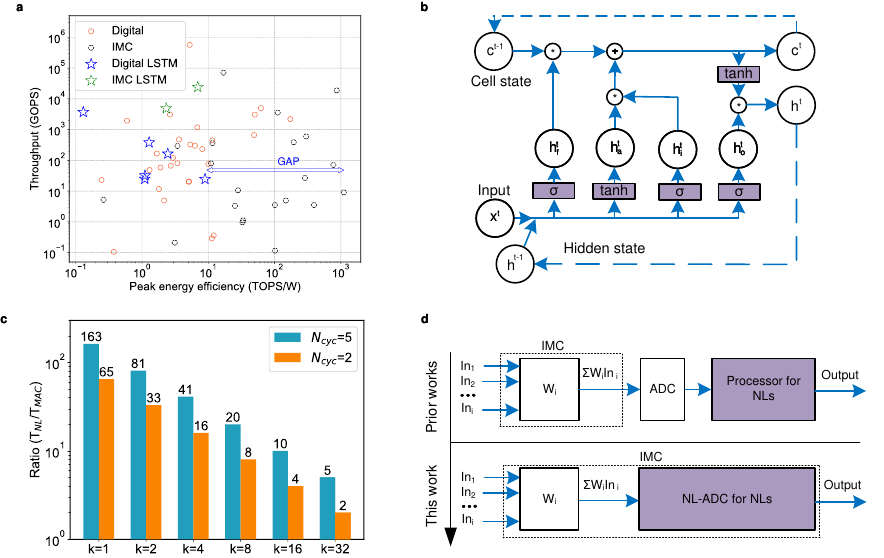}
    \caption{
    \textbf{Limitation of current In-memory computing (IMC) for Recurrent Neural Networks and our proposed solution. a} A survey of DNN accelerators show the improvement in energy efficiency offered by IMC over digital architectures. However, the improvement does not extend to recurrent neural networks (RNN) such as LSTM and there exists a gap in energy efficiency between RNNs and feedforward architectures. Details of the surveyed papers available here\cite{survey}.
    \textbf{b} Architecture of a LSTM cell showing a large number of nonlinear (NL) activations such as sigmoid and hyperbolic tangent which are absent in feedforward architectures that mostly use simple nonlinearities like rectified linear unit (ReLU).
    \textbf{c} Digital implementation of the NL operations causes a bottleneck in latency and energy efficiency since the linear operations are highly efficient in time and energy usage due to inherent parallelism of IMC. For a LSTM layer with $512$ hidden unit and with $k=32$ parallel digital processors for the NL operations, the NL operations still take $2-5$X longer time for execution due to the need of multiple clock cycles ($N_{cyc}$) per NL activation.
    \textbf{d} Our proposed solution creates an In-memory analog to digital converter (ADC) that combines NL activation with digitization of the dot product between input and weight vectors. 
    }
    \label{fig:nladc_motivation}
\end{figure}

Comparing the energy efficiency and throughput of recently reported DNN accelerators (Fig. \ref{fig:nladc_motivation}\textbf{a}) shows the improvements provided by IMC approaches over digital architectures. However, taking a closer look based on DNN architecture exposes an interesting phenomenon--while analog IMC has improved energy efficiency of convolutional and fully connected layers in DNNs, the same cannot be said for recurrent neural network (RNN) implementations such as long short term memory (LSTM)\cite{giraldo2018laika,kadetotad20208,shin201714,conti2018chipmunk,yin20171}. Resistive memories store the layer weights in their resistance values, inputs are typically provided as pulse widths or voltage levels, multiplications between input and weight happen in place by Ohm's law, and summation of the resulting current occurs naturally by Kirchoff's current law. This enables an efficient implementation of linear operations in vector spaces such as dot products between inputs and weight vectors. Early implementations of LSTM using memristors have focussed on achieving acceptable accuracy in network ouptut in the presence of programming errors. A $128\times 64$ 1T1R array\cite{li2019long} was shown to be able to solve real-life regression and classification problems while 2.5M phase-change memory devices\cite{tsai2019inference} have been programmed to implement LSTM for language modeling. While these were impressive demonstrations, energy efficiency improvements were limited, since  RNNs such as LSTMs have a large fraction of nonlinear (NL) operations such as sigmoid and hyperbolic tangent being applied as neuronal activations (Fig. \ref{fig:nladc_motivation}\textbf{b}). With the dot products being very efficiently implemented in the analog IMC, the conventional digital implementation of the NL operations now serves as a critical bottleneck. 

As an example, an RNN transducer was implemented on a 34-tile IMC system with 35 million phase-change memory (PCM) devices and efficient inter-tile communication\cite{Nature2023analogAIchipIBM}. While the system integration and scale of this effort\cite{Nature2023analogAIchipIBM} is very impressive, the NL operations are performed off-chip using low energy efficiency digital processing reducing the overall system energy efficiency. Another pioneering research\cite{natIBM64core} integrated 64 cores of PCM arrays for IMC operations with on-chip digital processing units for NL operations. However, the serial nature of the digital processor, which is shared across the neurons in 8 cores, reduced both the energy efficiency and throughput of the overall system. This work used look-up tables (LUT), similar to other works\cite{xie2020twofold,arvind2020hardware,kwon20221ynm}; alternate techniques using cordic\cite{cordic} or piece wise linear approximations\cite{ime_pwl,giraldo2018laika,pasupuleti2019low,feng2021high} or quadratic polynomial approximation\cite{li2020low} have also been proposed to reduce overhead and latency ($N_{cyc}$) of computing one function. However, it is the big difference in parallelism of crossbars versus serial digital processors which causes this inherent bottleneck. Even in a hypothetical situation with an increased number of parallel digital computing engines for the NL activations (Fig. \ref{fig:nladc_motivation}\textbf{c}, Supplementary Section \ref{supsec:latency_comparison_NL_Mac}), albeit at a large area penalty, the latency of the NL operations still dominates the overall latency due to the extremely fast implementation of vector-matrix-multiplication (VMM) in memristive crossbars.  

In this work, we introduce a novel in-memory analog to digital conversion (ADC) technique for analog IMC systems that can combine nonlinear activation function computations in the data conversion process (Fig. \ref{fig:nladc_motivation}\textbf{d}), and experimentally demonstrate its benefits in an analog memristor array. Utilizing the sense amplifiers (SA) as a comparator in ramp ADC, and creating a ramp voltage by integrating the current from an independent column of memristors which are activated row by row in separate clock cycles, an area-efficient in-memory ADC for memristive crossbars is demonstrated. However, instead of generating a linear ramp voltage as in conventional ADCs, we generated a nonlinear ramp voltage by appropriately choosing different values of memristive conductances such that the shape of   the ramp waveform matches that of the inverse of the desired NL activation function. Using this method, we demonstrate energy-efficient 5-bit implementations of commonly used NL functions such as sigmoid, hyperbolic tangent, softsign, softplus, elu, selu etc. A one-point calibration scheme is shown to reduce the integral nonlinearity (INL) from 0.948 to 0.886 LSB for various NL functions. Usage of the same IMC cells for ADC and dot-product also gives added robustness to read voltage variations, reducing INL to $\approx 0.04$ LSB compared to $\approx 5.0$ LSB for conventional methods. Using this approach combined with hardware aware training\cite{hw_aware}, we experimentally demonstrate a keyword spotting (KWS) task on the Google speech commands dataset\cite{GSD2018speech}. With a $32$ hidden neuron LSTM layer (having $128$ nonlinear gating functions) that uses $9216$ memristors from the $3 \times 64 \times 64$ memristor array on our chip, we achieve $88.5\%$ accuracy using a 5-bit NL-ADC with a $\approx 9.9X$ and $\approx 4.5X$ improvement of area and energy efficiencies respectively at the system level for the LSTM layer over previous reports. Moreover, compared to a conventional approach using the exact same configuration (input and output bit-widths) as ours, the estimated area and energy efficiency advantages are still retained at $\approx 6.2X$ and $\approx 1.46X$ respectively for system level of evaluation. Finally, we demonstrate the scalability of our system by performing a character prediction task on the Penn Treebank dataset\cite{PTB1993} using a LSTM model $\approx 100$X bigger than the one for KWS using experimentally validated nonideality models and achieving software equivalent accuracy. The improvements in area efficiency are estimated to be $6.6X$ over a conventional approach baseline and $125X$ over earlier work\cite{Nature2023analogAIchipIBM} at the system level, with the drastic increase in performance due to the much higher number of nonlinear functions in the larger model.


%% file: sections/02_results.tex
\section*{Results} 
\label{sec:results}
\subsection*{Nonlinear function Approximation by Ramp ADC}
\label{subsec:theory}
A conventional ramp ADC operates on an input voltage $V_{in}\in\{V_{min},V_{max}\}$ and produces a binary voltage $V_{out}$ whose time of transition from low to high, $t_{in}\in\{0,T_S\}$, encodes the value of the input. As shown in Fig. \ref{fig:nladc_overview}\textbf{a}, $V_{out}$ is produced by a comparator whose positive input is connected to a time-varying ramp signal, $V_{ramp}(t)=f(t)$ and the negative input is connected to $V_{in}$. For the conventional case of linearly increasing ramp voltage, and denoting the comparator's operation by a Heaviside function $\Theta$, we can mathematically express $V_{out}$ as:
\begin{equation}
\label{eq:ramp}
    V_{out}=\Theta (V_{ramp}(t) - V_{in})= \Theta (f(t) - V_{in})=\Theta (Kt - V_{in})
\end{equation}
The threshold crossing time $t_{in}$ can be obtained as $t_{in}=f^{-1}(V_{in})=\frac{1}{K}V_{in}$ by setting \Cref{eq:ramp} equal to zero and solving for `t'. The pulse width information in $t_{in}$ may be directly passed to the next layers as pulse-width modulated input \cite{Nature2023analogAIchipIBM} or can be converted to a digital code using a time-to-digital converter (TDC)\cite{AllenHolberg}. Now, suppose we want to encode the value of a nonlinear function of $V_{in}$, denoted by $g(V_{in})$ in $t_{in}$. Comparing with the earlier equation of $t_{in}$, we can conclude this is possible if:
\begin{equation}
\label{eq:ramp_inverse}
    t_{in}=f^{-1}(V_{in})=g(V_{in})\implies f()=g^{-1}()
\end{equation}
where we assume that the desired nonlinear function g() is bijective and an inverse exists in the defined domain of the function. \ref{supsec:nlfunc_compare} shows the required ramp function for six different nonlinear activations. For the case of non-monotonic functions, this method can still be applied by splitting the function into sections where the function is monotonic. Examples of such cases are shown in \ref{supsec:Non-monotonic nonlinear function approximation by ramp ADC} for two common non-monotonic activations--Gelu and Swish.

In practical situations, the ramp function is a discrete approximation to the continuous function $f(t)$ mentioned earlier. For a $b$-bit ADC, the domain of the function $V_{ramp}=f(t)$ is split into $P=2^b$ segments using $P+1$ points $t_k$ such that $t_0=0$, $t_k=\frac{kT_s}{P}$ and $t_P=T_s$. The initial voltage of the ramp, $f(0)=V_0=V_{init}$, defines the starting point while the other voltages $V_k$ ($k=1$ to $P$) can be obtained recursively as follows:
\begin{equation}
\label{eq:ramp_discrete}
    V_k=V_{k-1}+\Delta V_k = V_{k-1}+(g^{-1}(t_k)-g^{-1}(t_{k-1}))
\end{equation}
Here, $V_{init}$ may be selected appropriately to maximize the dynamic range of the function represented by the limited $b$-bits. Supplementary Tab. \ref{suptab:nlfunc_compare2} demonstrates the choice of $(t_k,\Delta V_k)$ tuples for six different nonlinear functions commonly used in neural networks.

\subsection*{In-memory Implementation of Nonlinear ADC and Vector Matrix Multiply in a Crossbar Array}
\begin{figure}[p]
    \centering 
    \includegraphics[width=0.95\textwidth]{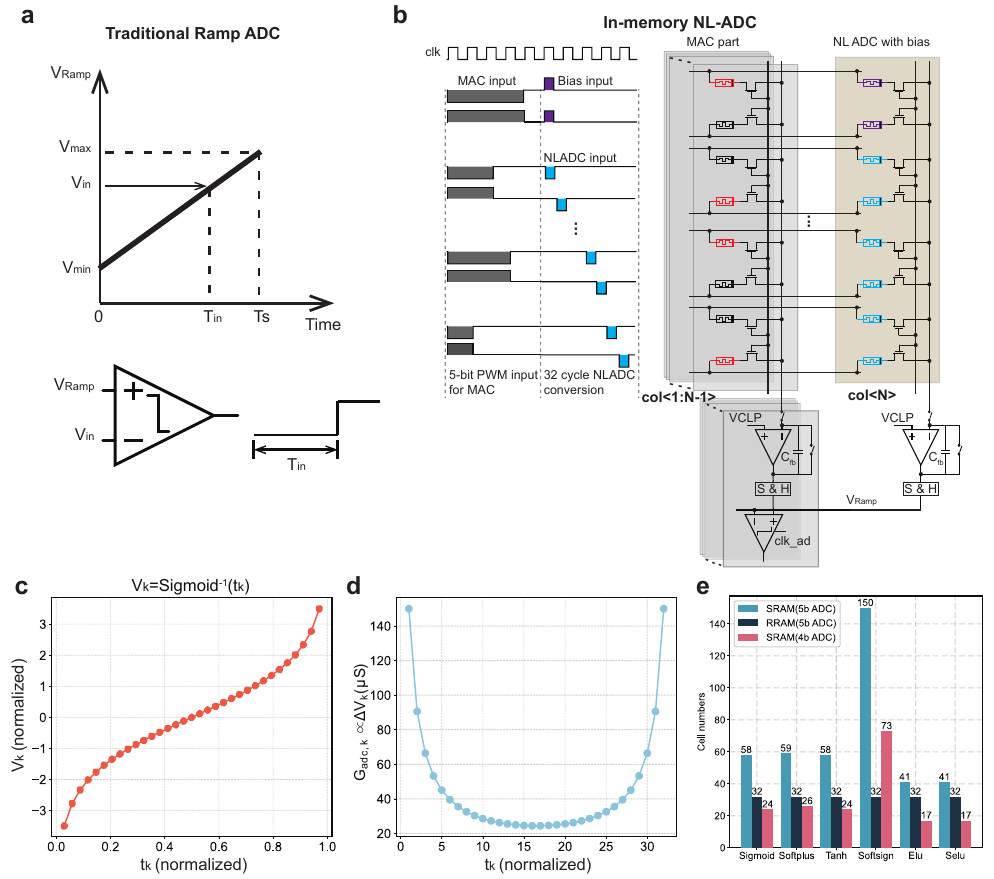}
    \caption{
    \textbf{Overview of in-memory nonlinear ADC.}
    \textbf{a} The concept of traditional ramp-based ADC.
    \textbf{b} The schematic and timing of in-memory computing circuits with embedded nonlinear activation function generation.
    \textbf{c} The Inverse of the sigmoid function illustrates the shape of the required ramp voltage.
    \textbf{d} The value of each step of the ramp voltage $V_{ramp}$ denoted by $\Delta V_k$  is proportional to memristor conductances $G_{adc,k}$ used to program the nonlinear ramp voltage. The desired conductances for a 5-bit implementation of a sigmoid nonlinear activation is shown.
    \textbf{e} Comparison of used cell numbers between 5-bit and 4-bit in-SRAM with 5-bit in-RRAM nonlinear function. The RRAM-based nonlinear function has an approximation error between the two SRAM-based ones due to write noise while using a smaller area due to its compact size.}
    \label{fig:nladc_overview}
\end{figure}

Different from traditional in-memory computing systems where the ADCs and nonlinear functions calculation are separated from the memory core, our proposed hardware implements the ADC with nonlinear calculation ability inside the memory along with the computation part. 
The nonlinear function approximating ADC described earlier is implemented using memristors with the following unique features:
\begin{enumerate}
    \item We utilize the memristors to generate the ramp voltages directly within the memory array which incurs very low area overhead with high flexibility.
    \item By leveraging the multi-level state of memristors, we can generate the nonlinear ramp voltage according to the x-y relationship of the nonlinear function as described in \Cref{eq:ramp_discrete}.
\end{enumerate}
Take the sigmoid function, ($y=g(x)=1/(1+e^{-x})$) as an example. 
To extract the step of the generated ramp voltages, we first take the inverse of the sigmoid function ($x=g^{-1}(y)=\ln{\frac{y}{1-y}}$) as shown in Fig. \ref{fig:nladc_overview}\textbf{c}. This function is exactly the ramp function that needs to be generated during the conversion process, as described in detail in the earlier section. The figure shows the choice of $32$ (x,y) tuples for $5$-bit nonlinear conversion.
The voltage difference between successive points ($\Delta v_k$ in \Cref{eq:ramp_discrete}) is shown in Fig. \ref{fig:nladc_overview}\textbf{d} highlighting the unequal step sizes. Memristors proportional to these step values need to be programmed in order to generate the ramp voltage, as described next.
We show our proposed in-memory nonlinear ADC circuits in Fig. \ref{fig:nladc_overview}\textbf{b}. 
In each memory core, only a single column of memristors will be utilized to generate $V_{ramp}(t)=f(t)$ with very low hardware cost as shown in Fig. \ref{fig:nladc_overview}\textbf{e}.
The memory is separated into $N$ columns for the multiplication-and-accumulation (MAC) part and one column for the nonlinear ADC (NL-ADC) part. For the MAC part, the inputs are quantized to $b$-bit ($b=3$ to $5$ in experiments) if necessary and transferred into pulse width modulation (PWM) signals sent to the SL of the array. 
To adapt to the positive and negative weights or inputs in the neural networks, we encode one weight or input into differential 1T1R pairs and inputs shown in Supplementary Fig. \ref{supfig:weight_mapping}.
We propose a charge-based approach for sensing the MAC results where the feedback capacitor of an integrator circuit is used to store the charge accumulated on the BL. Denoting the feedback capacitor by $C_{fb}$, the voltage on the sample and hold (S\&H) for the k-th column is:$$V_{mac,k}=V_{CLP}+\frac{1}{C_{fb}}\sum_i{V_{read}G_{ik}T_{in,i}}$$
where $V_{CLP}$ is the clamping voltage on the bitline enforced by the integrator, $T_{in,i}$ denotes the pulse width encoding the i-th input and $G_{ik}$ is the memristive conductance on the i-th row and k-th column.

In the column of the NL-ADC part, we have two sets of memristors. One is the memristors for generating the initial bias voltage ($V_{init}$ in the earlier section), and another set of memristors is for generating the nonlinear ramp voltages.
The addition of bias memristors is used to set the initial ramp value as well as calibrate the result after programming the NL-ADC memristors due to programming not being accurate. The `one point' calibration will move the ramp function generated by the NL-ADC column to intersect the desired, theoretical ramp function at the zero-point which leads to minimal error. This is described in detail in the next sub-section.
As shown in the timing diagram in Fig. \ref{fig:nladc_overview}\textbf{b}, after the MAC results are latched on the S\&H, the positive input with one cycle pulse width is first sent into NL-ADC column to generate a bias voltage (corresponding to the most negative voltage at the starting point $V_{init}$) for the ramp function on $V_{ramp}$. Then for each clock cycle, negative input is sent to the SL of NL-ADC memristors to generate the ramp voltages on $V_{ramp}$. Since the direction of each step voltage is known, only one memristor is used corresponding to the magnitude of the step while the polarity of the input sets the direction of the ramp voltage. Using less memristors for every step provides the flexibility to use more devices for calibration or error correction if stuck devices are found. The ramp voltage generated at the q-th clock cycle, $V_{ramp}^q$, is given by the following equation: $$V_{ramp}^q=V_{CLP}+V_{init}+\frac{1}{C_{fb}}\sum_{i=1}^q{V_{read}G_{adc,i}T_{adc}}$$ where $T_{adc}$ are the pulse width of ADC read pulses and $\frac{V_{read}G_{adc,k}T_{adc}}{C_{fb}}$ equals $\Delta V_k$ from \Cref{eq:ramp_discrete}. An example of the temporal evolution of the ramp voltage is shown in the \ref{supsec:one-point calibration scheme} The comparison between the $V_{ramp}$ and MAC result is done by enabling the comparator using the \textit{clk\_ad} signal after each step of the generation of ramp voltages. 
The conversion continues for $P=2^b$ cycles producing a thermometer code or pulse width proportional to the nonlinear activation applied to the MAC result. The pulse-width can be directly transferred to other layers as the input as done in other works\cite{Nature2023analogAIchipIBM}; else, the thermometer code can be converted to binary code using a ripple counter\cite{natIBM64core}.

To prove the effectiveness of this approach, we simulate it at the circuit level and the results are shown in Fig. \ref{figs:SPICE sim result}. The fabricated chip does not include the integrators and the comparator which are implemented in software after obtaining the crossbar output. In-memory ramp ADC using SRAM has been demonstrated earlier\cite{yu202265} in a different architecture with a much larger ($100\%$) overhead compared to the memory used for MAC operations; however, nonlinear functions have not been integrated with the ADC. Implementing the proposed scheme using SRAM would require many cells for each step due to the different step sizes. Since an SRAM cell can only generate two step sizes (+1 or -1) intrinsically, other step sizes need to be quantized and represented in terms of this unit step size (or LSB). Denoting this unit step by $min(\Delta V_{k})$, $round(\frac{\Delta V_{k}}{min(\Delta V_{k})})$ is the number of SRAM cells needed for a single step while the total number of steps needed is given by $\sum_{k=1}^{2^n}round(\frac{\Delta V_{k}}{min(\Delta V_{k})})$. Fig. \ref{fig:nladc_overview}\textbf{e} illustrates this for several common nonlinear functions.
However, thanks to the analog tunability of the conductance of memristors, we can encode each step of the ramp function $\Delta V_k$ into only one memristor $G_{adc,k}$ which leads to the usage of much lower number of bitcells (1.28X - 4.68X) compared with nonlinear  SRAM-based ramp function generation (Fig. \ref{fig:nladc_overview}\textbf{e}) for the 5-bit case. However, the write noise in memristors can result in higher approximation error than the SRAM version. Using Monte-Carlo simulations for the SRAM case, we estimate the mean squared error (MSE) for a memristive 5-bit version is generally in between that of the 5-bit and 4-bit SRAM versions (e.g. MSE for sigmoid nonlinearity is $\approx 0.0008$ and $\approx 0.0017$ respectively for the 5-bit and 4-bit SRAM versions, while that of the memristive 5-bit version is $\approx 0.0014$). Hence, we also plot in Fig. \ref{fig:nladc_overview}\textbf{e} the number of SRAM bitcells needed in 4-bit versions as well. However, combined with the inherently smaller ($>2X$) bitcell size of memristors compared to SRAM, our proposed approach still leads to more compact implementations with very little overhead due to the ramp generator. The approximation accuracy for the memristive implementation is expected to improve in future with better devices\cite{rao_nature} and programming methods\cite{song_science}. Details of the number of SRAM cells needed for six different nonlinear ramp functions are shown in Supplementary Tab. \ref{suptab:nlfunc_compare2}. The topology shown here is restricted to handle inputs limited to the dimension of the crossbar. We show in \ref{supsec: Capacitor-based accumulation method for large model} how it can be extended to handle input vectors with dimensions larger than the number of rows in the crossbar by using the integrator to store partial sums and splitting the weight across multiple columns.

\subsection*{Calibration Procedures for Accurate Programming of nonlinear functions in crossbars}
Mapping the NL-ADC into memristors has many challenges including device-to-device variations, programming errors, etc such that the programmed conductance $G_{adc,k}'$ will deviate from the desired $G_{adc,k}$. To tackle these problems, we introduce the adaptive mapping method to calibrate the programming errors. For each nonlinear function, we first extract the steps $\Delta V_k$ of the function as illustrated in Fig. \ref{fig:nladc_overview}\textbf{d}. Then we normalize them and map them to the conductances with a maximum conductance of \SI{150}{\micro\siemens}. Detailed characterization of our memristive devices were presented earlier\cite{rram_our_characterize}. We program the conductances using the iterative-write-and-verify method explained in methods. However, the programming will introduce errors, and hence, we reuse the $N_{cali}$ (=5 in experiments) bias memristors to calibrate the programming error according to the mapped result. These memristors are also used to create the starting point of the ramp $V_{init}$, which is negative in general (assuming domain of g() spans negative and positive values). Hence, positive voltage pulses are applied to these bias/calibration conductances while negative ones are applied to the $G_{adc,k}$ for the ramp voltage as shown in Fig. \ref{fig:nladc_overview}(a). Based on this, the one-point calibration strategy is to match the zero crossing point of the implemented ramp voltage with the desired theoretical one by changing the value of $V_{init}$ to $V_{init,new}$. The total calibration conductance is first found as: 
\begin{equation*}
    G_{\text{cali,tot}} = \sum_{i=1}^m G_{adc,i}'  \quad \text{where $m\in \mathbb{N}$ s.t.} \quad V_m = 0
\end{equation*}
Here, $m$ is the index where the function $g()=f^{-1}()$ crosses the zero point of the x-axis. We then represent this calibration term using only a few memristors with the largest conductance on the chip ($G_{\text{max}}=\SI{150}{\micro\siemens}$). The number of memristors used for calibration is $N_\text{cali}=[G_\text{cali,tot} / \SI{150}{\micro\siemens}] + 1$ in which $N_\text{cali} - 1$ memristors are \SI{150}{\micro\siemens} and the one left is $G_\text{cali} \: mod \: \SI{150}{\micro\siemens}$. More details about the calibration process including timing diagrams are provided in the \ref{supsec:one-point calibration scheme}

\begin{figure}[p]
    \centering 
    \includegraphics[width=0.8\textwidth]{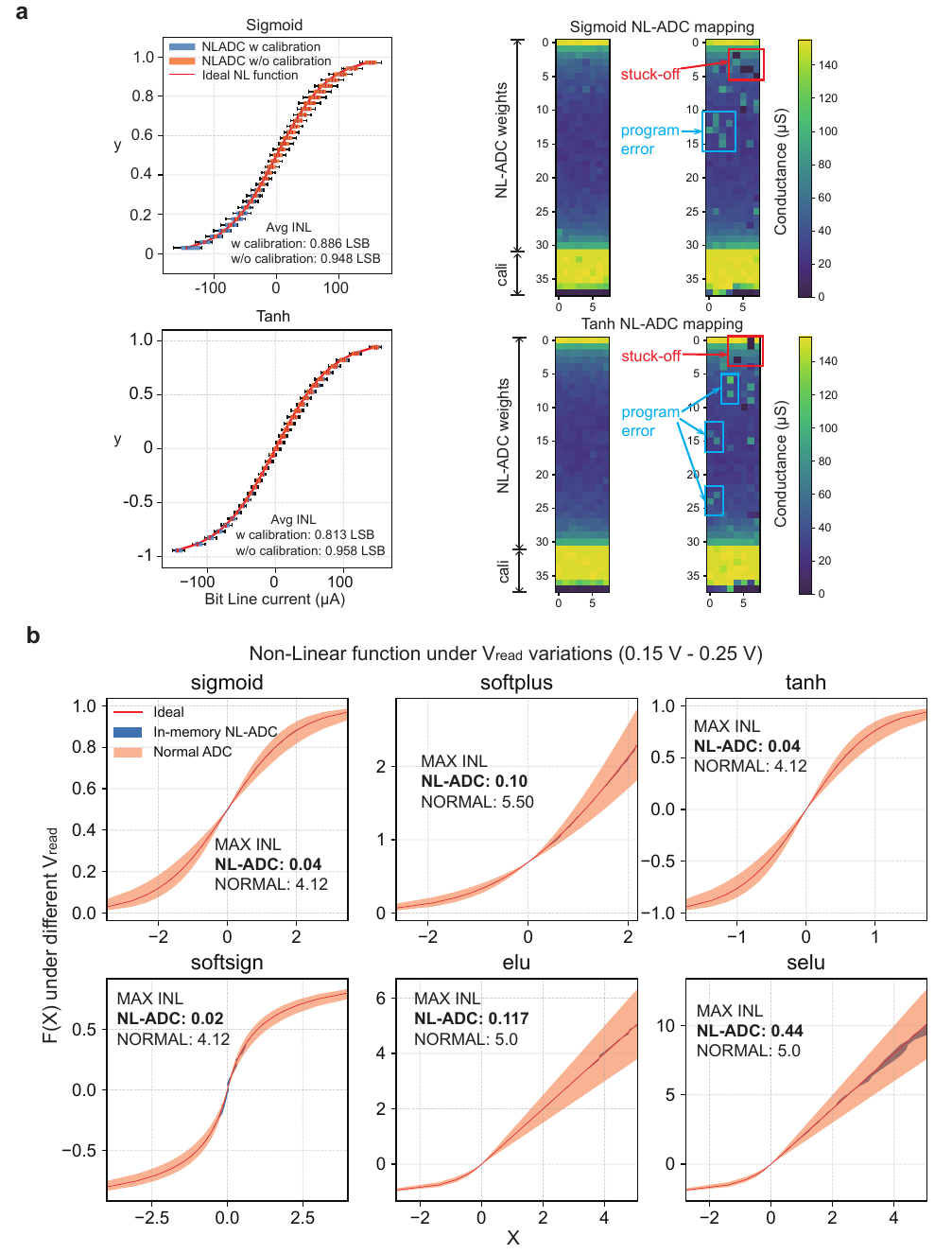}
    \caption{
    \textbf{Experimentally demonstrated NL-ADC on crossbar arrays}
    \textbf{a} Calibration process for accurate NL-ADC programming. The left panel shows the ramp function of the ideal case, programming without bias calibration and with bias calibration. The case with bias calibration shows better INL performance. The right panel shows the actual conductance mapping on the crossbar arrays on two blocks of $8$ arbitrary selected columns. The lower $5$ conductances are for bias calibration while the top $32$ are for the ramp generation.
    We show the cases when mapping of NL-ADC weights doesn't have stuck-at-OFF devices and low programming error (left block), and the cases which have stuck-at-OFF devices and high programming error (right block). The results show that both cases can be calibrated by the additional 5 memristors.
    \textbf{b} Robustness of our proposed in-memory NL-ADC under $V_{\text{read}}$ variations. We sweep the $V_{\text{read}}$ from \SI{0.15}{\volt} to \SI{0.25}{\volt} to simulate noise induced variations in read voltage.
    Normal ADC has large variations while our in-memory NL-ADC can track the $V_{\text{read}}$.
    }
    \label{fig:nladc_performance}
\end{figure}

To demonstrate the effectiveness of our approach, we experimentally programmed two frequently used nonlinear functions in neural network models: sigmoid function and tanh function, with the calibration methods mentioned above, as shown Fig. \ref{fig:nladc_performance}\textbf{a}. We program $64$ columns containing the same NL-ADC weights and calibration terms in one block of our memristor chip and set the read voltage to \SI{0.2}{\volt}. In the left panel, we compare the transfer function with 3 different cases: (1) Ideal nonlinear function (2) Transfer function without on-chip calibration (3) Transfer function with on-chip calibration. We can find that the curve with calibration matches well with the ideal curve while the one without calibration has some deviation. We also show the programmed conductance map in the right panel for a block of $8$ arbitrarily selected columns to showcase different types of programming errors that can be corrected. The first $32$ rows are the NL-ADC weights representing 5-bit resolution. The remaining 5 rows are the calibration memristors after reading out the NL-ADC weights to mitigate the program error. In the conductance map on the right, we can also find some devices which are either stuck at OFF or have higher programming error. Fortunately,these errors happening during the NL-ADC weights programming can be compensated by the calibration part as evidenced in the reduced values of average INL. In addition to these two functions, we also show other nonlinear functions with in-memory NL-ADC in Supplementary Fig. \ref{supfig:nladc_prog} which covers nearly all activation functions used in neural networks. Further reductions in the INL may be achieved through redundancy. Briefly, the entire column used to generate the ramp has many unused memristors which may be used to program redundant copies of the activation function. The best out of these may be chosen to further reduce the effect of programming errors. Details of this method along with measured results from two examples of nonlinear activations are shown in \ref{supsec:redundancy}.

In a real in-memory computing system, the voltage variations on-chip could harm the performance. This can be a challenge when using traditional ADC due to the lack of ability to track the voltage variation. For example, if the voltage we set for VMM is \SI{0.2}{\volt} but the actual voltage sent to SL is \SI{0.25}{\volt} (due to supply noise, voltage drop etc.), the final result read out from the ADC will deviate a lot from the real VMM result. Fortunately, our in-memory NL-ADC has the natural ability to track the voltage variations on-chip since the LSB of the ADC is generated using circuits matching those generating the MAC result. Hence, the read voltage is canceled out during the conversion and only the relative values of MAC and ADC conductances will affect the final result. To demonstrate that our proposed in-memory NL-ADC is robust under different read voltages, we run the experiment by setting different read voltages from \SI{0.15}{\volt} to \SI{0.25}{\volt} and measure the transfer function. From the result shown in Fig. \ref{fig:nladc_performance}\textbf{b}, we can see that the conventional ADC has large variations (maximum INL of $4.12-5.5$ LSB) due to $V_\text{read}$ variations while our in-memory NL-ADC only experiences a little effect (maximum INL of $0.02-0.44$ LSB). 

\subsection*{Long-Short-Term Memory experiment for Keyword Spotting implemented in Memristor Hardware}
After verifying the performance of the nonlinear activations on chip, we proceed to assess the inference accuracy obtained when executing neural networks using NL-ADC model on chip. The Google Speech Commands Dataset (GSCD)\cite{GSD2018speech}, a common benchmark for low-power RNNs\cite{ISSCCKWS8603,KWS202223}, is used to train and test for 12-class KWS task. The KWS  training network consists of  Mel-frequency
cepstral coefficient (MFCC), standardization, LSTM layer ($9,216$ weight parameters in a $72\times 128$ crossbar) and fully connected (FC) layer; detailed parameters pertaining to the task are provided in Methods. Further training with weight noise-aware and NL-ADC noise-aware (Methods) is implemented.

\begin{figure}[p]  
    \centering 
    \includegraphics[width=0.85\textwidth]{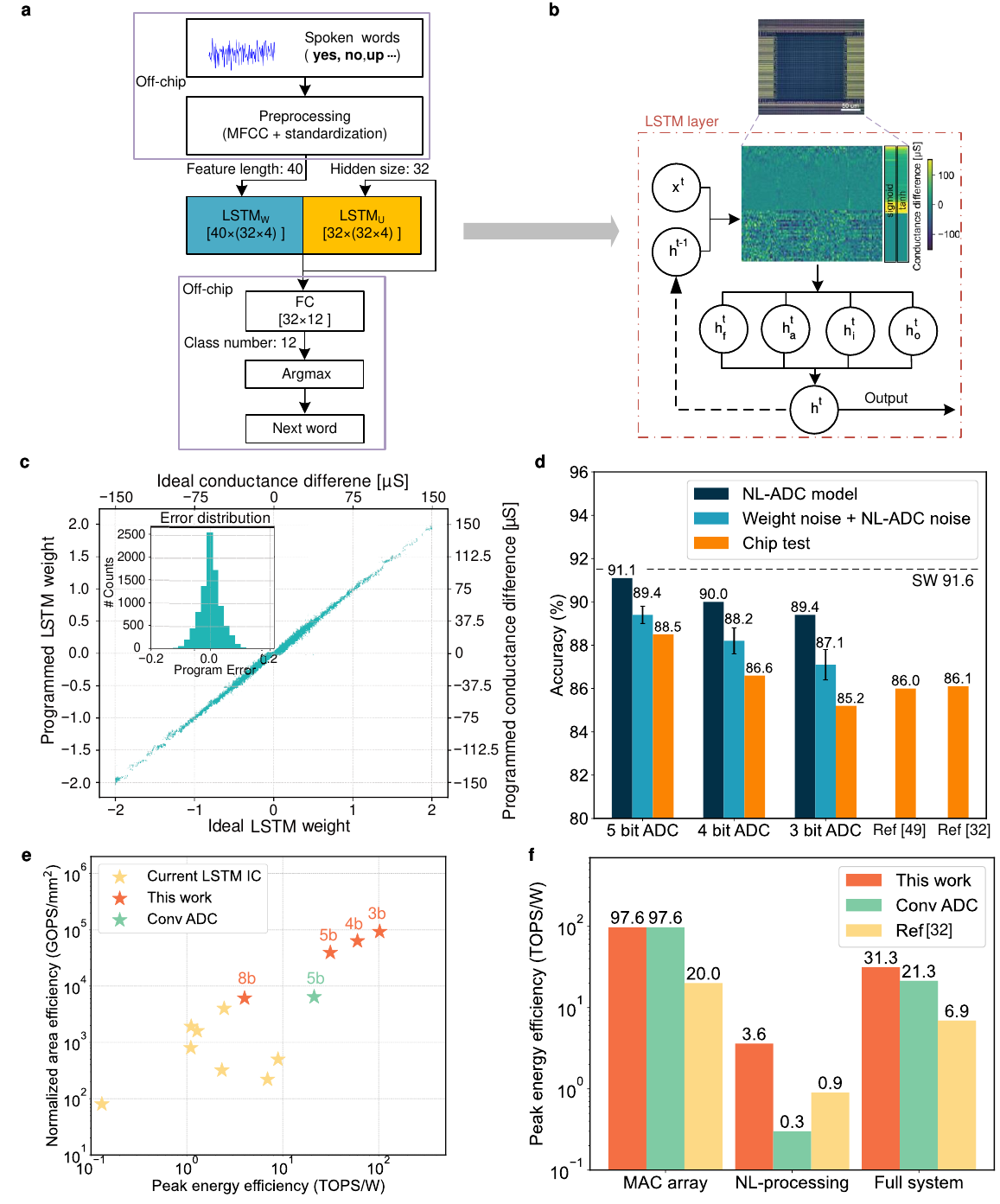}
    \caption{
    \textbf{| LSTM for KWS task.}
    \textbf{a} Architecture of LSTM network on-chip inference.
    \textbf{b} Mapping of LSTM network onto the chip. Weights and nonlinearities (Sigmoid and Tanh) of LSTM layer are programmed crossbar arrays as conductance. Input and output (I/O) data of LSTM layer are sent from/to the integrated chip through off-chip circuits.
    \textbf{c} Weight conductance distribution curve and error.
    \textbf{d} The measured inference accuracy results obtained on the chip are compared with the software baseline using the ideal model, as well as simulation results under different bit NL-ADC models and hardware-measured weight noise.
    \textbf{e} Energy efficiency and area efficiency comparison: our LSTM IC, conventional ADC model and recently published LSTM ICs from research papers\cite{yue20197,kadetotad20208,shin201714,yin20171,conti2018chipmunk,natIBM64core,Nature2023analogAIchipIBM,jouppi2017datacenter}. Energy efficiency and throughput under 8 bit, }5 bit, 4 bit and 3 bit NL-ADC are calculated based on 16 nm CMOS technology and clock frequency of 1 GHz. Detailed calculations are shown in \ref{supsec:Estimation_energy_area_latency_Macro}, \ref{supsec:Estimation_energy_area_latency_system} and Tab. \ref{Tab:tab3 Comparison}. Area efficiency of all works are normalized to 1 GHz clock and 16 nm CMOS process.
    \textbf{f} Energy efficiency comparison (this work, conventional ADC model, a chip for speech recognition using LSTM model\cite{Nature2023analogAIchipIBM}) at various levels: MAC array, NL-processing, full system. Full system includes MAC and NL-processing and other modules that assist MAC and NL-processing.     
    \label{fig:fig.4}
\end{figure}

On-chip inference shown in Fig. \ref{fig:fig.4}\textbf{a} is performed after training.  After feature extraction (MFCC extraction and normalization), a single input audio signal is divided into $49$ arrays, each with a feature length of $40$. These extracted features, along with the previous 32-dimensional output vector $h^t$ from the LSTM, are then sent to the memristor crossbar. The architecture of LSTM layer, along with the direction of data flow, is depicted in Fig. \ref{fig:fig.4}\textbf{b}. The equations of the LSTM layer are given by \Cref{eq:LSTM} and \Cref{eq:LSTM2} as below:
\begin{equation}
\left[\begin{array}{c} h_f^t \\ h_a^t \\ h_i^t \\ h_o^t\end{array}\right] = \left[\begin{array}{c} \sigma \\ \tanh \\ \sigma \\ \sigma \end{array}\right]\left[\begin{array}{lll} x^t & h^{t-1} \end{array}\right]  \left[\begin{array}{l} W \\ U \end{array}\right]
\label{eq:LSTM}
\end{equation}
\begin{equation}
h_c^t=h_f^t \odot h_c^{t-1}+h_i^t \odot h_a^t \\
h^t=h_o^t \odot \tanh \left(h_c^t\right)
\label{eq:LSTM2}
\end{equation}
where $x^t$ represents the input vector at the current step,  $h^t$ and $h^{t-1}$ denote hidden state vector also known as output vector of the LSTM cell at the current and previous time steps, and $h_c^t$ represents cell state vector at the current time step. The model parameters, including the weights $W$ and recurrent weights $U$, are stored for $h_f^t$, $h_i^t$, $h_o^t$ and $h_a^t$ (forget gate, input/update gate, output gate, cell input, respectively). $\odot$ denotes element-wise multiplication and $\sigma$ is the sigmoid function. 
The parameters ($W$ and $U$) and functions ($\sigma$ and tanh ) in \Cref{eq:LSTM} are programmed in the memristor crossbar (Methods) and the conductance difference map of the memristor crossbar is depicted in Fig. \ref{fig:fig.4}\textbf{b}. Therefore, all MAC and nonlinear operations specified in \Cref{eq:LSTM} are executed on chip removing the need to transfer weights back and forth. The four vectors outputted by the chip ($h_f^t$, $h_a^t$, $h_i^t$, $h_o^t$ ) are all digital, allowing them to be directly read by the off-chip processor without requiring an additional ADC. This approach provides notable benefits, such as a substantial reduction in the latency due to nonlinearity calculation in a digital processor and decreased energy consumption. \\
Fig. \ref{fig:fig.4}\textbf{d} depicts inference accuracy results for 12-class KWS. After adding the NL-ADC model to replace the nonlinear functions in the LSTM cell (\Cref{eq:LSTM}), 91.1\%, 90\% and 89.4\% inference accuracy were obtained in the 5-bit, 4-bit and 3-bit ADC models respectively which compare favorably with a floating-point baseline of 91.6\%.  To enhance the model's robustness against hardware non-ideal factors and minimize the decrease in inference accuracy from software to chip, we injected hardware noise (Methods) in the weight crossbar and NL-ADC crossbar conductances during the training process. The noise model data used in this process is obtained from the actual memristor crossbar and follows a normal distribution $\simeq$ N(0,\SI{5} {\micro\siemens}) (Supplementary Fig. \ref{supfig:nladc_prog2}\textbf{c}). As a result, we attain inference accuracies in software of 89.4\%, 88.2\%, and 87.1\% for the 5-bit, 4-bit, and 3-bit NL-ADC cases models respectively. Note that the drop in accuracy is much less for feedforward models as shown in \ref{supsec:accuracy_write_noise} Further, the robustness of the classification was verified by conducting $10$ runs; the small standard deviation shown in Fig. \ref{fig:fig.4}\textbf{d} confirming the robustness. Through noise-aware training, we obtain weights that are robust against write noise inherent in programming memristor conductances. These weights are then mapped to corresponding conductance values through a scaling factor by matching the maximum weight after training to the maximum achievable conductance (Methods). Both the conductance values associated with the weights and the NL-ADC are programmed on the memristor crossbar, facilitating on-chip inference.\\
Fig. \ref{fig:fig.4}\textbf{c} shows the performance of weight mapping after programming the memristor conductance using iterative methods. The error between the programmed conductance value and the theoretical value follows a normal distribution. The on-chip, experimentally measured inference accuracies achieved are 88.5\%, 86.6\%, and 85.2\% for the 5-bit, 4-bit, and 3-bit ADC models, respectively where the redundancy techniques in \ref{supsec:redundancy} were used for the 3-bit version. The experimental results indicate that 5-bit and 4-bit NL-ADC models can achieve higher inference accuracy than previous work\cite{Nature2023analogAIchipIBM,ISSCCKWS8603} (86.14\% and 86.03\%) based on same dataset and class number and are also within 2\% of the software estimates,  while that of the 3-bit version is marginally inferior. Nonetheless, it is important to highlight that the LSTM layer with the 3-bit ADC model significantly outperforms the 5-bit NL-ADC models in terms of area efficiency and energy efficiency as shown in Fig. \ref{fig:fig.4}\textbf{e} ( 31.33TOPS/W, 60.09  TOPS/W and 114.40  TOPS/W for 5-bit, 4-bit, and 3-bit NL-ADC models, respectively). The detailed calculation to assess the performance of our chip under different bit precision is done following recent work\cite{jiang2023efficient,cai_estimate} and shown in Supplementary \textnormal{Tab. }\ref{Tab:tab3 Comparison}. The earlier measurements were limited to the LSTM macro alone and did not consider the digital processor for pointwise multiplications (Eq. \ref{Seq:LSTM2}) and the small FC layer. The whole system level efficiencies are estimated in detail in \ref{supsec:Estimation_energy_area_latency_system}\textbf{a} for all three bit resolutions.

Table \ref{tab:LSTM work comp} compares the performance of our LSTM implementation with other published work on LSTM showing advantages in terms of energy efficiency, throughput and area efficiency. For a fair comparison of area efficiencies, throughput and area are normalized to a 1 GHz clock and 16 nm process respectively. Fig. \ref{fig:fig.4}\textbf{e} also graphically compares the energy efficiency and normalized area efficiency of the LSTM layer in our chip for KWS task with other published LSTM hardware. The results demonstrate that our chip with 5-bit NL-ADC exhibits significant advantages in terms of normalized area efficiency ($\approx 9.9X$) and energy efficiency ($\approx 4.5X$ system level) compared to the closest reported works. It should be noted that comparing the raw throughput is less useful since it can be increased by increasing the number of cores. In order to dig deeper into the reason for the superiority of our chip compared to conventional linear ADC chips\cite{liu202033,zhang2023edge}, detailed comparisons with a controlled baseline were also done using two models as shown in \textnormal{Fig. }\ref{figs:two models} where $k=1$ digital processor is assumed. While their inputs and outputs remain the same, the key distinction lies in the nonlinear operation component. Utilizing these two architectures, we conducted evaluations on the energy consumption and area of the respective chips (\textnormal{Tab. }\ref{Tab:our work full sysytem Energy, area and latency estimation} and \textnormal{Tab. }\ref{Tab:conventional full system model Energy, area and latency estimation}) to find the proposed one has $\approx 1.5X$ and $\approx 2.4X$ better metrics at the system level respectively.
\textnormal{Fig. }\ref{fig:fig.4}\textbf{f} presents a comprehensive comparison of energy efficiency for its individual subsystems among this work (5-bit NL-ADC), the conventional ADC model, and a reference chip\cite{Nature2023analogAIchipIBM} using IMC -- the MAC array, NL-processing, and the full system comprising the MAC array, NL-processing, and other auxiliary circuits (\textnormal{Tab. }\ref{Tab:tabs3 comparison  at various levels}). Our approach exhibits significant energy efficiency advantages, particularly in NL-processing, with a remarkable 3.6 TOPS/W compared to 0.3 TOPS/W and 0.9 TOPS/W for the other two chips. This substantial improvement in NL-processing energy efficiency is a crucial factor contributing to the superior energy efficiency of our chip, as depicted in \textnormal{Fig. }\ref{fig:fig.4}\textbf{e}.

The improvements in area efficiency also come about due to the improved throughput of the NL-processing. \textnormal{Fig. }\ref{figs: Energy and area breakdown} shows the energy and area breakdown of the main chip components of this work (5-bit NL-ADC) and the conventional model (5-bit ADC).  Our work demonstrates superior area efficiency, with a value of 130.82 TOPS/mm\textsuperscript{2}, compared to the conventional ADC model's 9.56 TOPS/mm\textsuperscript{2} in Tab. \ref{Tab:tab3 Comparison}. This $\approx 13X$ improvement is attributed mostly to throughput improvement of $4.6X$ over the digital processor in conventional systems. Table \ref{tab: ADC comp} compares our proposed NL-ADC with other ADC used in IMC systems. While some works have used Flash ADCs that require single cycle per conversion, they have a higher level of multiplexing (denoted by \# of columns per ADC) since these require exponentially more comparators than our proposed ramp ADC. Hence the effective AC latency in terms of number of clock cycles for our system is comparable with others. Moreover, since our proposed NL-ADC is the only one with integrated activation function (AF) computation, the latency in data conversion followed by AF computation (denoted by AF latency in the table) is significantly lower for our design. Here, we assume LUT based AF computation using $N_{cyc}=2$ clocks and 1 digital processor per 1024 neurons like other work\cite{natIBM64core}. Compared with other ramp converters, the area occupied by our 5-bit NL-ADC is merely 558.03 \si{\micro\metre}\textsuperscript{2} due to usage of only one row of memristors, while the traditional 5-bit Ramp-ADC\cite{liu202033,zhang2023edge} together with nonlinear processor occupy an area of 4665.47 \si{\micro\metre}\textsuperscript{2}. This substantial disparity in ADC area due to a capacitive DAC-based ramp generator\cite{liu202033,zhang2023edge} leads to a further difference in the area efficiency of the two chips. Using oscillator-based ADCs\cite{yi2018_elm,bonan2019} instead of ramp-based ones will reduce the area of the ADC in traditional systems, but the throughput, area and energy-efficiency advantages of our proposed method will still remain significant. Also, note that due to the usage of memristors for reference generation, our system is robust to perturbations (such as changes in temperature or read voltage) similar to other designs using replica biasing\cite{rram_replicabias}. Lastly, for monotonic AF, our design can directly generate pulse width modulated (PWM) output (like other ramp based designs\cite{Nature2023analogAIchipIBM}) which can be passed to the next stage avoiding the need for DAC circuits at the input of the next stage as well as counters in the ADC. This is known to further increase energy efficiencies\cite{symp_vlsi_jiang}, an aspect we have not explored yet.

\newpage
\begin{landscape}

\begin{table}[]
\caption{ \textbf{Comparison of LSTM performance with previous works}}
\begin{tabular}{|c|cc|c|c|c|c|c|c|}
\hline
\textbf{Metric}                                                                                    & \multicolumn{2}{c|}{\textbf{\begin{tabular}[c]{@{}c@{}}This work\\      (KWS/NLP task)\end{tabular}}} & \textbf{Nature’23\cite{Nature2023analogAIchipIBM}} & \textbf{\begin{tabular}[c]{@{}c@{}}Nat. \\      Electron.'23\cite{natIBM64core}\end{tabular}}  & \textbf{VLSI’17\cite{yin20171}} & \textbf{JSSC’20\cite{kadetotad20208}} & \textbf{ISSCC’17\cite{ shin201714}} & \textbf{CICC’18\cite{conti2018chipmunk}} \\ \hline
CMOS   technology                                                                                  & \multicolumn{2}{c|}{16 nm}                                                                            & 14 nm              & 14 nm                         & 65 nm            & 65 nm            & 65 nm             & 65 nm            \\ \hline
Memory   technology                                                                                & \multicolumn{2}{c|}{RRAM}                                                                             & PCM                & PCM                           & SRAM             & SRAM             & SRAM              & SRAM             \\ \hline
Operation   Frequency (MHz)                                                                        & \multicolumn{2}{c|}{1000}                                                                             & 1000               & 1000                          & 200              & 80               & 200               & 168              \\ \hline
IMC                                                                                                & \multicolumn{2}{c|}{Y}                                                                                & Y                  & Y                             & N                & N                & N                 & N                \\ \hline
Input/weight/output   precision                                                                    & \multicolumn{2}{c|}{5/Analog/5,8/Analog/8}                                                            & 8/Analog/8         & 8/Analog/8                    & 8/8/--           & 13/6/13          & 16/16/--          & 8/8/8            \\ \hline
Memory size   (kB)                                                                                 & \multicolumn{1}{c|}{1.125}                                  & 623                                     & 4250               & 272                           & 348              & 288              & 10                & 82               \\ \hline
KWS task on GSCD (Accuracy \%)                                                                     & \multicolumn{1}{c|}{88.5 (12 classes)}                    & --                                      & 86.1 (12 classes) & --                            & --               & --               & --                & --               \\ \hline
NLP task on PTB (BPC)                                                                                         & \multicolumn{1}{c|}{--}                                     & 1.349                                   & --                 & 1.439                         & --               & --               & --                & --               \\ \hline
Area (mm\textsuperscript{2})                                                                                         & \multicolumn{1}{c|}{0.003}                                  & 0.71                                    & 111.18             & 144                           & 19               & 7.74             & 2.6               & 0.93             \\ \hline
Power (mW)                                                                                          & \multicolumn{1}{c|}{3.7 (5b),4.6(8b)}                       & 406.5 (5b),766.8(8b)                    & 3450               & 3465                          & 296              & 65               & 2.3               & 29.03            \\ \hline
Peak   Throughput (TOPS)                                                                           & \multicolumn{1}{c|}{0.11(5b),0.02(8b)}                      & 19.5(5b),5.5(8b)                        & 23.94              & 4.9                           & 0.38             & 0.16/0.02             & 0.025             & 0.03             \\ \hline
Energy   Efficiency (TOPS/W)                                                                       & \multicolumn{1}{c|}{31.0(5b),4.0(8b)}                       & 47.9(5b),7.2(8b)                        & 6.94               & 1.96                          & 1.28             & 2.45/8.93             & 1.1               & 1.11             \\ \hline
Area   Efficiency (TOPS/mm\textsuperscript{2})                                                                       & \multicolumn{1}{c|}{39.48 (5b),6.1(8b)}                     & 27.6 (5b),7.8(8b)                       & 0.17               & 0.32                          & 0.02             & 0.02/0.0025             & 0.01              & 0.02             \\ \hline
\begin{tabular}[c]{@{}c@{}}Normalized Area Efficiency\\       (TOPS/mm\textsuperscript{2}, 1GHz, 16 nm)\end{tabular} & \multicolumn{1}{c|}{39.48 (5b),6.1(8b)}                     & 27.6 (5b),7.8(8b)                       & 0.22               & 0.32                          & 1.6              & 4/0.5                & 0.8               & 1.92             \\ \hline
\end{tabular}

\label{tab:LSTM work comp}
\end{table}

\begin{table}[hbtp]
\caption{ \textbf{Comparison of ADC performance with previous works}}
\begin{tabular}{|c|c|c|c|c|c|c|c|c|c|}
\hline
\textbf{}                                                                        & \textbf{This work} & \textbf{\begin{tabular}[c]{@{}c@{}}Trans. on   Electron \\      Devices'20\cite{yin2020high}\end{tabular}} & \textbf{SSCL'20\cite{he20202}} & \textbf{\begin{tabular}[c]{@{}c@{}}Nat. \\      Electron.'19\cite{cai2019fully}\end{tabular}} & \textbf{\begin{tabular}[c]{@{}c@{}}Nat. \\      Electron.'23\cite{natIBM64core}\end{tabular}} & \textbf{\begin{tabular}[c]{@{}c@{}}Nat. \\      Electron.'22\cite{huo2022computing}\end{tabular}} & \textbf{JSSC'22\cite{rram_replicabias}} & \textbf{Nature'20\cite{memristor_conv}} & \textbf{Science'23\cite{zhang2023edge}} \\ \hline
ADC type                                                                         & Ramp               & Flash                                                                                    & Flash            & SAR                                                                        & CCO-based                                                                  & SAR                                                                        & Flash            & SAR                & Ramp                \\ \hline
ADC resolution (bit)                                                             & 5                  & 3                                                                                        & 1                & 9                                                                          & 12                                                                         & 8                                                                          & 3                & 8                  & 8                   \\ \hline
ADC clk freq. (MHz)                                                               & 1000               & 150                                                                                      & 140              & 148                                                                        & 3300                                                                       & 8                                                                          & 100              & 20                 & 200                 \\ \hline
\# of column per ADC                                                             & 1                  & 8                                                                                        & 8                & 1                                                                          & 1                                                                          & 64                                                                         & 8                & 4                  & 1                   \\ \hline
Effective fs (MHz)                                                               & 31.25              & 18.75                                                                                      & 17.5              & 16.44                                                                      & 7.93                                                                       & 0.015                                                                          & 12.5              & 0.625                & 0.78                \\ \hline
\begin{tabular}[c]{@{}c@{}}Effective ADC \\      latency (\# clock)\end{tabular} & 32                 & 8                                                                                        & 8               & 9                                                                          & 128                                                                        & 512                                                                         & 8                & 32                 & 256                 \\ \hline
\begin{tabular}[c]{@{}c@{}}AF latency\\       (\# clock, KWS/NLP)\end{tabular}   & 32/32              & 257/1025                                                                                 & 257/1025         & 265/1033                                                                   & 384/1152                                                                   & 264/1032                                                                   & 257/1025         & 264/1032           & 512/1280            \\ \hline
Power (\textmu W)                                                                       & 9.3                & --                                                                                     & --             & --                                                                       & --                                                                       & 33.18                                                                      & --             & 51                 & 11.9                \\ \hline
FOM (pJ)                                                                         & 0.0186             & --                                                                                     & --             & --                                                                       & --                                                                      & 0.1296                                                                     & --             & 0.01             & 0.06                \\ \hline
Process (nm)                                                                     & 16                 & 90                                                                                       & 90               & 180                                                                        & 16                                                                         & 55                                                                         & 40               & 130                & 130                 \\ \hline
Replica  Bias                                                                    & Y                  & N                                                                                        & N                & N                                                                          & N                                                                          & N                                                                          & Y                & N                  & N                   \\ \hline
AF included                                                                      & Y                  & N                                                                                        & N                & N                                                                          & N                                                                          & N                                                                          & N                & N                  & N                   \\ \hline
PWM mode                                                                         & Y                  & N                                                                                        & N                & N                                                                          & Y                                                                          & N                                                                          & N                & N                  & Y                   \\ \hline
\end{tabular}

\label{tab: ADC comp}
\end{table}

\end{landscape}


\subsection*{Scaling to large RNNs for Natural Language Processing}
\begin{figure}[p]
    \centering 
    \includegraphics[width=0.85\textwidth]{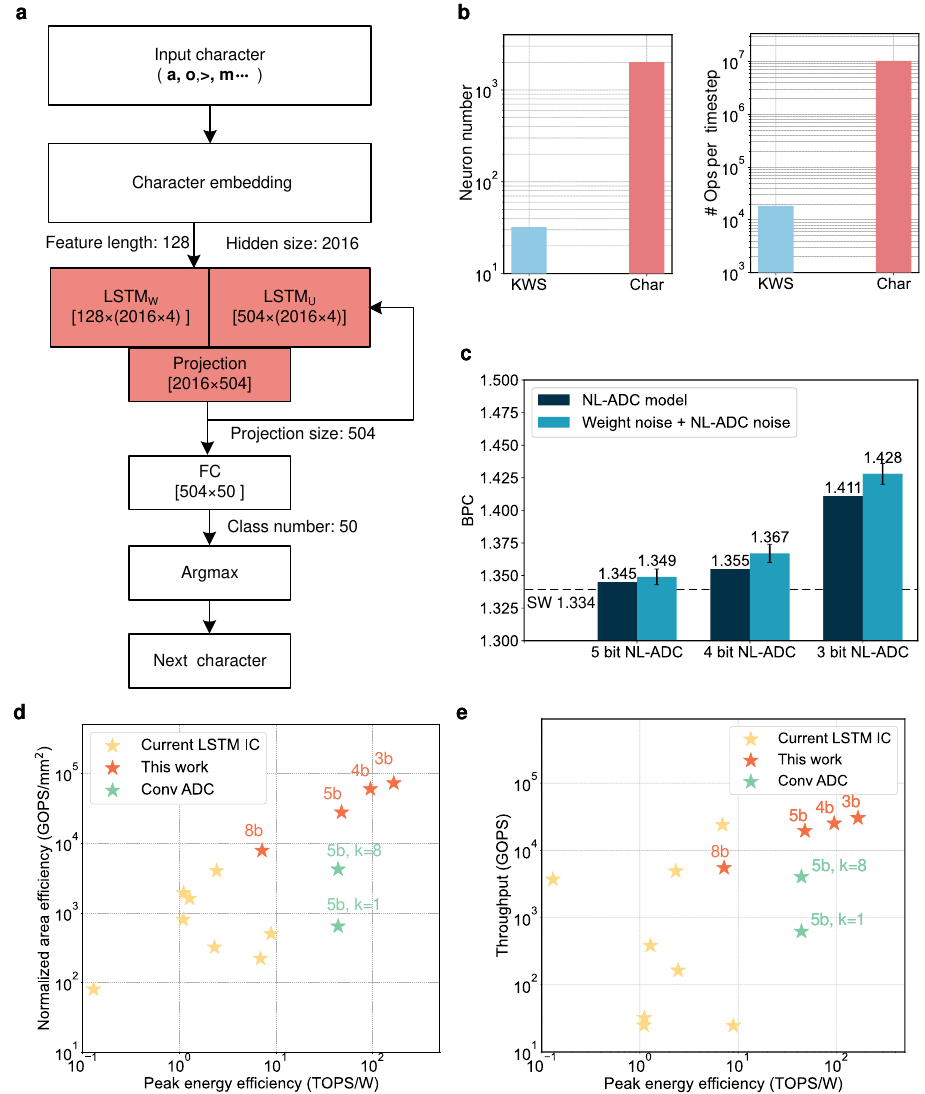}
    \caption{
    \textbf{| LSTM for NLP task.}
    \textbf{a} Architecture of LSTM network for on-chip inference in character prediction task.
    \textbf{b} Comparison in the LSTM layer between the number of neurons and operations per timestep in the NLP model for character prediction and the KWS model. 
    \textbf{c} Simulation results under different bit resolution of NL-ADC models and hardware-measured weight noise compared with software baseline using the ideal model. BPC results follow the "smaller is better" principle, meaning that lower values indicate better performance.
    \textbf{d} Energy efficiency and area efficiency comparison: our LSTM IC, conventional ADC model and recently published LSTM ICs from research papers\cite{yue20197,kadetotad20208,shin201714,yin20171,conti2018chipmunk,natIBM64core,Nature2023analogAIchipIBM,jouppi2017datacenter}. Detailed calculation of energy efficiency and throughput for both macro and system levels are shown in \ref{supsec:Estimation_energy_area_latency_Macro}, \ref{supsec:Estimation_energy_area_latency_system}and Tab. \ref{Tab:tab3 different ADC bit Comparison NLP}. Area efficiency of all works are normalized to 1 GHz clock and 16 nm CMOS process.
    \textbf{e} Energy efficiency and throughput comparison: our LSTM IC, conventional ADC model and recently published LSTM ICs from research papers\cite{yue20197,kadetotad20208,shin201714,yin20171,conti2018chipmunk,natIBM64core,Nature2023analogAIchipIBM,jouppi2017datacenter}.
    }
    \label{fig:fig5}
\end{figure}

Although KWS is an excellent benchmark for assessing the performance of small models\cite{Nature2023analogAIchipIBM,MFCC2020510}, our nonlinear function approximation method is also useful in handling significantly larger networks for applications such as character prediction in Natural Language Processing (NLP). To demonstrate the scalability of our method, we conduct simulations using a much larger LSTM model on the Penn Treebank Dataset\cite{PTB1993} (PTB) for character prediction. There are a total of 50 different characters in PTB. Each character is embedded into a unique random orthogonal vector of 128 dimensions, which are taken from a standard Gaussian distribution. Additionally, at each timestep, a loop of 128 cycles will be executed.  The training method is shown in Methods and \textnormal{Fig. }\ref{fig:fig5}\textbf{a} illustrates the inference network architecture.  The number of neurons and parameters in the PTB character prediction network ($6,112,512$ weight parameters and $8,568$ biases in LSTM and projection layers) are $\approx 66X$ and $\approx 600X$ more than  the corresponding numbers in the KWS network (\textnormal{Fig. }\ref{fig:fig5}\textbf{b}) leading to a much larger number of operations per timestep.

To map the problem onto memristive arrays, the LSTM layer alone needs a $633\times 8064$ prohibitively large crossbar. Instead, we partition the problem and map each section to a $633\times 512$ crossbar (similar to recent approaches\cite{Nature2023analogAIchipIBM}) with $16$ such crossbars for the entire problem. Within each crossbar, only $256$ input lines are enabled at one phase to prevent large voltage drops along the wires, with a total of $3$ phases to present the whole input, with the concomitant cost of $3X$ increase in input presentation time. An architecture for further reduced crossbars of size $256\times 256$ is shown in \ref{supsec: Capacitor-based accumulation method for large model} To assess the impact of on-chip buffers and interconnects in performing data transfer between tiles, Neurosim\cite{DNN_NeuroSim_V2.1} is used to perform system level simulations where the ADC in the tile is replaced by our model (details in \ref{supsec:Estimation_energy_area_latency_system}\textbf(b)).
First, the accuracy of the character prediction task is assessed using bits per character (BPC)\cite{natIBM64core}, which is a metric that measures the model's ability to predict samples from the true underlying probability distribution of the dataset, where lower values indicate better performance. Similar to the earlier KWS case, memristor write noise and NL-ADC quantization effects from the earlier hardware measurements are both included in the simulation. The results of inference are displayed in \textnormal{Fig. }\ref{fig:fig5}c, with a software baseline of 1.334. BPC results of 1.345, 1.355, and 1.411 are obtained with the 5-bit, 4-bit, and 3-bit ADC models, respectively when considering perfect weights, and exhibits a drop of only $0.011$ for the 5-bit NL-ADC model compared to the software baseline. Finally, the write noise of the memristors are included during both the training and testing phases using the same method as the KWS model resulting in BPC values of of 1.349, 1.367, and 1.428 are obtained with the 5-bit, 4-bit, and 3-bit NL-ADC models (with error bars showing standard deviations for $10$ runs). Compared to other recent work\cite{natIBM64core} on the same dataset that obtained a BPC of 1.358, our results are promising and show nonlinear function approximation by NL-ADC can be successfully applied to large-scale NLP models.\\
The throughput, area, and energy efficiencies are estimated next (details in \ref{supsec:Estimation_energy_area_latency_Macro}, Sup. Tab \ref{Tab:this work Energy, area and latency estimation NLP } at the macro level and in \ref{supsec:Estimation_energy_area_latency_system}\textbf(b) at the system level) and compared with a conventional architecture (Sup Tab. \ref{Tab:conventional model Energy, area and latency estimation NLP k=1},\ref{Tab:conventional model Energy, area and latency estimation NLP k=8}) for two different cases of $k=1$ and $k=8$ digital processors. These are compared along with current LSTM IC metrics in Fig. \ref{fig:fig5}\textbf{d} and Fig. \ref{fig:fig5}\textbf{e}. Considering the 5-bit NL-ADC, our estimated throughput and energy efficiency of $19.5$ TOPS and $47.9$ TOPS/W at the system level are $\approx 4.0X$  and $24.4X$ better for system level than earlier reported metrics\cite{natIBM64core} of $4.9$ TOPS and $2$ TOPS/W respectively. Lastly, the normalized area efficiency of our NL-ADC based LSTM layer is $\approx 86X$ better at system level than earlier work\cite{natIBM64core} (reporting results on same benchmark) due to the increased throughput and reduced area (we also estimated that an 8-bit version of our system will still be $24X$ more area efficient and $3.6X$ more energy efficient). Compared to the conventional IMC architecture baseline for this LSTM layer, we estimate an energy efficiency advantage of $1.1X$ at the system level similar to the KWS case, but the throughput and area efficiency advantages of $\approx 4.8X$ and $\approx 6.6X$ for system level respectively remain even for $k=8$ digital processors.

%% file: sections/03_discussion.tex

\section*{Discussion} 
In conclusion, we proposed and experimentally demonstrated a novel paradigm of nonlinear function approximation through a memristive in-memory ramp ADC. By predistorting the ramp waveform to follow the inverse of the desired nonlinear activation, our NL-ADC removes the need for any digital processor to implement nonlinear activations. The analog conductance states of the memristor enable the creation of different programmable voltage steps using a single device, resulting in great area savings over a similar SRAM-based implementation. Moreover, the in-memory ADC is shown to be more robust to voltage fluctuations compared to a conventional ADC with memristor crossbar based MAC. Using this approach, we implemented a LSTM network using 9216 weights programmed in the $72\times 128$ memristor chip to solve a 12-class keyword spotting problem using the Google speech commands dataset. The results for the 5-bit ADC show better accuracy of $88.5\%$ than previous hardware implementations\cite{Nature2023analogAIchipIBM,KWS202223} with significant advantages in terms of normalized area efficiency ($\approx 9.9X$) and energy efficiency ($\approx 4.5X$) compared to previous LSTM circuits. We further tested the scalability of our system by simulating a much larger network (6,112,512 weights) for natural language processing using the experimentally validated models. Our network with 5-bit NL-ADC again achieves better performance in terms of BPC than recent reports\cite{natIBM64core} of IMC based LSTM ICs while delivering $\approx 86X$ and  $\approx 24.4X$ better area and energy efficiencies at the system level. Our work paves the way for very energy efficient in-memory nonlinear operations that can be used in a wide variety of applications.

%% file: sections/04_methods.tex

\section*{Methods}
\label{sec:methods}
\subsection*{Memristor Integration}
The memristors are incorporated into a CMOS system manufactured in a commercial foundry using a 180 nm technology node. The integration process starts by eliminating the native oxide from the surface metal through reactive ion etching (RIE) and a buffered oxide etch (BOE) dip. Subsequently, chromium and platinum are sputtered and patterned with e-beam lithography (EBL) to serve as the bottom electrode. This is followed by the application of reactively sputtered 2 nm tantalum oxide as the switching layer and sputtered tantalum metal as the top electrode. The device stack is completed with sputtered platinum for passivation and enhanced electrical conduction.
\subsection*{Memristor Programming Methods}
In this work, we adopt the iterative-write-and-verify programming method to map the weights to the analog conductances of memristors. Before programming, a tolerance value (\SI{5}{\micro\siemens}) is added to the desired conductance value to allow certain programming errors. The programming will end if the measured device conductance is within the range of \SI{5}{\micro\siemens} above or below the target conductance. During programming, successive SET or RESET pulses with \SI{10}{\micro\second} pulse width are added to each single 1T1R structure in the array. Each SET or RESET pulse is followed by a \SI{20}{\nano\second} READ pulse. A RESET pulse is added to the device if its conductance is above the tolerated range while a SET pulse will be added if its conductance is below the range. We will gradually increase the amplitude of the SET/RESET voltage and the gate voltage of transistors between adjacent cycles. For SET pulse amplitude, we start from \SI{1}{\volt} to \SI{2.5}{\volt} with an increment of \SI{0.1}{\volt}. For RESET pulse amplitude, we start from \SI{0.5}{\volt} to \SI{3.5}{\volt} with an increment of \SI{0.05}{\volt}.For gate voltage of SET process, we start from \SI{1.0}{\volt} to \SI{2.0}{\volt} with an increment of \SI{0.1}{\volt}. For gate voltage of RESET process, we start from \SI{5.0}{\volt} to \SI{5.5}{\volt} with an increment of \SI{0.1}{\volt}. 
Detailed programming process is illustrated in Supplementary Fig. \ref{supfig:weight_mapping}.
\subsection*{Hardware Aware Training}
Directly mapping weights of the neural network to crossbars will heavily degrade the accuracy. This is mainly due to the programming error of the memristors. To make the network more error-tolerant when mapping on a real crossbar chip, we adopted the defect-aware training proposed in previous work\cite{hw_aware}. During training, we inject the random Gaussian noise into every weight value in the forward parts for gradient calculation. Then the back-propagation happens on the weights after noise injection. Weight updating will occur on the weight before the noise injection. We set the standard deviation to \SI{5}{\micro\siemens} which is relatively larger than experimentally measured programming error (~\SI{2.67}{\micro\siemens} shown in Supplementary Fig.\ref{supfig:nladc_prog2}\textbf{c}) to make the model adapt more errors when mapping to real devices.
Detailed defect-aware training used in this work is described in Algorithm \ref{algo:DAT}. 
In this work, we set $\sigma$ to \SI{5}{\micro\siemens} and $g_{\text{max}}$ to \SI{150}{\micro\siemens}.

\begin{algorithm}[H]
    \caption{Defect-aware training}
    \label{algo:DAT}
    \RestyleAlgo{ruled} 
    \KwData{Weight matrix at training iteration$t$: $\mathbf{W_{\mu}^t}$; input data $X$; learning rate: $\alpha$; weight-to-conductance ratio $g_{\text{ratio}}$; maximum conductance in crossbar $g_{\text{max}}$; injected noise $\sigma$; loss function $L$}
    \KwResult{Weight at time step $t+1$: $\mathbf{W_{\mu}^{t+1}}$}
    $\mathbf{W_{\mu}^t[g_{\text{ratio}} W_{\mu}^t} > g_{\text{max}}] = g_{\text{max}} / g_{\text{ratio}}$ \\
    $\mathbf{G_{\mu}^t \leftarrow |W_{\mu}^t|} \cdot g_{\text{ratio}}$ (differential mapping)\\
    Initialize $\mathbf{G_{\sigma}^t}$: Conductance standard deviation  \\
    $\mathbf{W_{\sigma}^t \leftarrow G_{\sigma}^t} / g_{\text{ratio}}$ \\
    $\mathbf{W^t} \leftarrow \mathbf{W_{\mu}^t + \epsilon \cdot W_{\sigma}^t \quad \epsilon \sim \mathcal{N}(0, I)}$ \\
    Compute loss $L(X, \mathbf{W^t})$ \\
    Update $\mathbf{W_{\mu}}$ through back propagation \\
    $\mathbf{W_{\mu}^{t+1} \leftarrow W_{\mu}^t} - \alpha \cdot \frac{\partial{L}}{\partial{\mathbf{W_{\mu}}}}$
\end{algorithm} 

\subsection*{Weight clipping and mapping}
\label{sec:weight_clipping}
We clip weights between -2 to 2 to avoid creating excessively large weights during training. Weights can be mapped to the conductance of the memristors when doing on-chip inference, nearly varying from \SI{0} {\micro\siemens} to \SI{150} {\micro\siemens}. The clipping method is defined according to \Cref{eq:w_clipping} and the mapping method is shown in \Cref{eq:g=yw}.
\begin{equation}
    w= \begin{cases}-2 & w<-2 \\ w & -2 \leq w \leq 2 \\ 2 & w>2\end{cases}
    \label{eq:w_clipping}
\end{equation}
\begin{equation}  
    g=\gamma w
     \:(\gamma=\frac{g_{max}}{|w|_{max}})
    \label{eq:g=yw}
\end{equation}
where $w$ is weights during training, $g$ is the conductance value of memristors and $\gamma$ is a scaling factor used to connect the weights to $g$. The maximum conductance of memristors ($g_{max}$) is \SI{150} {\micro\siemens} and the maximum absolute value of weights ($|w|_{max}$) is 2, therefore the scaling factor ($\gamma$) is equal to \SI{75} {\micro\siemens}.\\

\subsection*{Training for LSTM Keyword Spotting model}
The training comprises of two processes: \textbf{preprocessing} and \textbf{LSTM model training}.\\ 
\textbf{Preprocessing}: The GSCD\cite{GSD2018speech} is used to train model. It has 65,000 one-second-long utterances of 30 short words and several background noises, by thousands of different people, contributed by members of the public\cite{MFCC2020510}. We reclassify the original 31  classes into the following 12 classes \cite{KWS202223}: \textbf{yes}, \textbf{no}, \textbf{up}, \textbf{down}, \textbf{left}, \textbf{right}, \textbf{on}, \textbf{off}, \textbf{stop},\textbf{ }\textbf{go}, \textbf{background noise}, \textbf{unknown}. The \textbf{unknown} class contains the other 20 classes.  
For every one-second-long utterance, the number of sampling points is 16000. MFCC\cite{MFCC2020510} is applied to extract Mel-frequency cepstrum of voice signals. 49 windows are used to divide a one-second-long audio signal and extract 40 feature points per window. \\
\textbf{LSTM model training}: The custom LSTM layer is the core of this training model. The custom layer is necessary for modifying the parameters, including adding the NL-ADC algorithm to replace the activation function inside, adding weight noise training, quantizing weights, etc. Although the custom LSTM layer will increase the training time of the network, this is acceptable by weighing its advantages and disadvantages. 
The input length is 40 and the hidden size is 32. The sequence length is 49, which means the LSTM cell will iterate 49 times in one batch size of 256. \\
A FC layer is added after the LSTM layer to classify the features output by LSTM. The input size of the FC layer is 32 and the output size is 12 (class number). Cross-entropy loss is used to calculate loss. We train the ideal model (without NL-ADC and noise) for 128 epochs and update weights using the Adam optimizer with a learning rate (LR = 0.001). After finishing the ideal model training, NL-ADC-aware training and hardware noise-aware training are added. All models' performance is evaluated with top-1 accuracy.

\subsection*{Training for LSTM character prediction model}
\textbf{Preprocessing}: The PTB\cite{PTB1993}  is a widely used corpus in language model learning, which has 50 different characters. Both characters in the training dataset and the validation dataset of PTB are divided into many small sets. Each set consists of 128 characters and each character is embedded into a random vector (dimension D=128) obtained from the standard Gaussian distribution and then perform Gram–Schmidt orthogonalization on these vectors. \\
\textbf{LSTM model training}: We use a one-layer custom LSTM with  projection \cite{natIBM64core}. The input length is 128 and the hidden size is 2016. The length hidden state and the LSTM output are both 504. The sequence length is 128, which means the LSTM cell will loop 128 times in one batch (batch size = 8).\\
The FC layer after the LSTM layer will further extract the features output by LSTM and convert them into an output of size 50 (class number). Cross-entropy loss is used to calculate loss. We train the model for 30 epochs and update weights using the Adam\cite{adam} optimizer with a learning rate (LR = 0.001). The model's performance is evaluated through the BPC\cite{bpc} metric and the data of BPC is smaller the better. After finishing the ideal model training, we use the same training method to train the model after adding NL-ADC and hardware noise.

\subsection*{ Inference with the addition of write noise and read  noise}
 During the inference stage, we performed 10 separate simulations with different write noise following the measured distribution N(0, \SI{2.67}{\micro\siemens}) (Fig. \ref{supfig:nladc_prog2}\textbf{c}) in each case (simulating 10 separate chips). For each of the simulation, read noise following measured read noise distribution N(0,\SI{3.5}{\micro\siemens}) (Fig. \ref{figs: drift effect of the RRRAM conductances}\textbf{b}) is included. It is worth noting that in each chip simulation, a consistent write noise was introduced into the inference process applied to the entire test dataset. However, in relation to read noise, the normal distribution N(0,\SI{3.5}{\micro\siemens}) is employed for each mini-batch to generate distinct random noises. These noises are subsequently incorporated into the simulation.  Then we obtain the inference accuracy with the addition of write noise and read noise.

%% file: Supplementary.tex
\section*{Supplementary} 
\label{sec:Supplementary}

\renewcommand{\thesubsection}{Supplementary Note S\arabic{subsection}.}
\setcounter{subsection}{0}

\renewcommand{\thefigure}{S\arabic{figure}}
\setcounter{figure}{0}

\setcounter{table}{0}
\renewcommand{\thetable}{S\arabic{table}}

\setcounter{equation}{0}
\renewcommand{\theequation}{S\arabic{equation}}

\subsection{Nonlinear function approximation by ramp ADC }
\label{supsec:nlfunc_compare}
\begin{figure}[t]
    \centering 
   \includegraphics[width=1\textwidth]{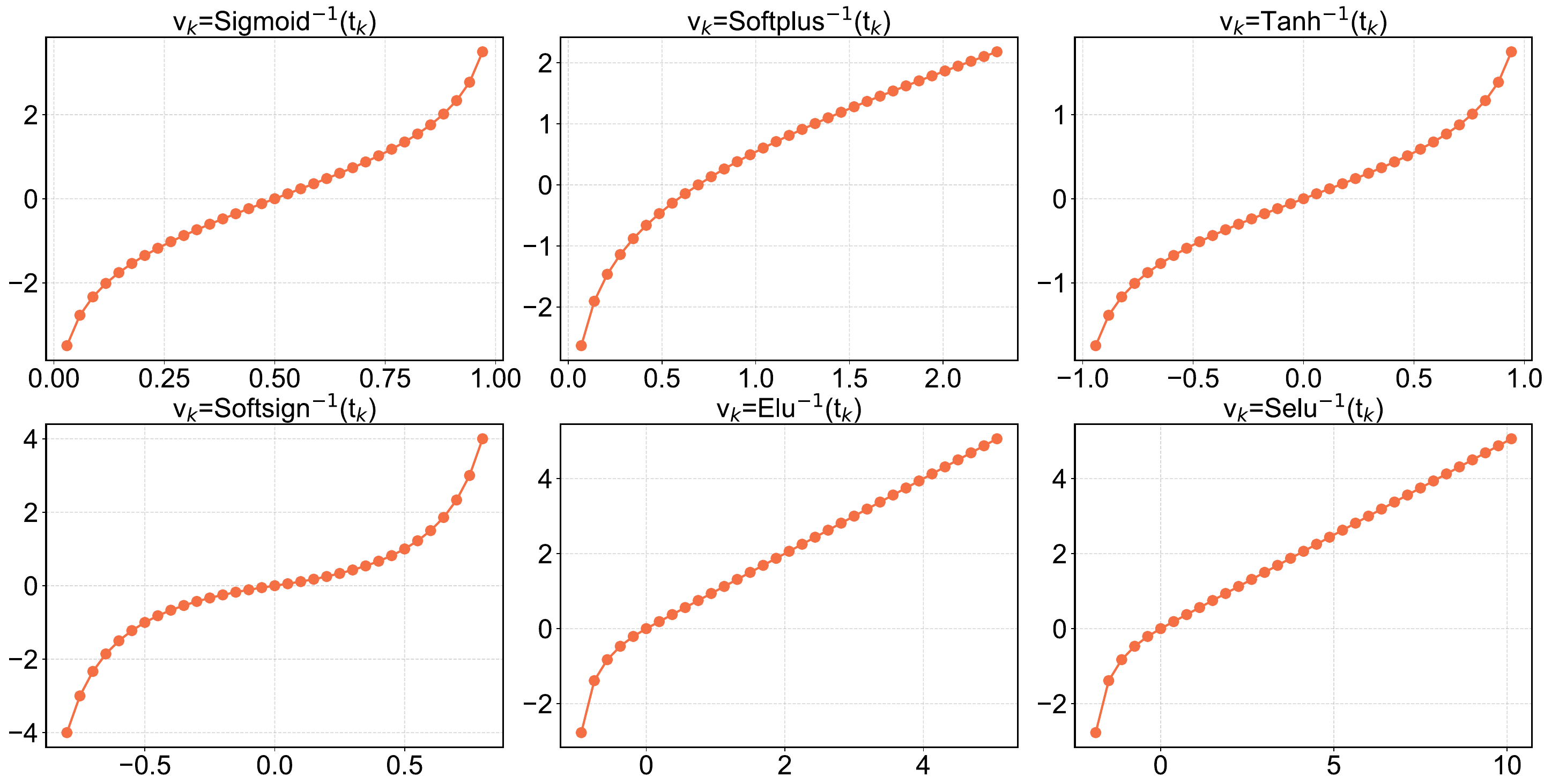}
    \caption{
    \textbf{ Plot of the six inverse functions.}
    }
    \label{supfig:nlfunc_compare}
\end{figure}

Six commonly used nonlinear functions in neural networks and their inverse functions are presented in \textnormal{Tab. }\ref{Tab: six NL}. To illustrate how the nonlinear functions are approximated, we consider the utilization of a 5-bit NL-ADC and sample 33 points in the inverse function of each nonlinear function. The curves of ($t_k$,$V_k$) tuples in \Cref{eq:ramp_discrete} Results for six inverse functions are presented in Fig. \ref{supfig:nlfunc_compare}. Following the acquisition of 33 sampling points (($t_k$,$V_k$)), the calculation of $\Delta V_k$ can be performed using: 
\begin{equation}
\Delta V_k=V_k-V_{k-1} \quad k \in[1,32]
    \label{eq:delt vk}
\end{equation}
The advantage associated with the tunability of memristor conductance allows for the encoding of each  $\Delta V_k$ using a single memristor. However, in the case of the SRAM model, each cell possesses the capacity to store only two levels. Consequently, representing a $\Delta V_k$ within the SRAM framework may require the utilization of multiple cells. The determination of the number of SRAM cells required per $\Delta V_k$ can be computed based on the provided formula ($\operatorname{round}\left(\frac{\Delta V_K}{\min \left(\Delta V_K\right)}\right)$). The values of $\Delta V_k$, along with the corresponding number of SRAM cells required to store $\Delta V_k$ for six functions, are presented in \textnormal{Tab. }\ref{Tab: delt v and cell number of 6 NL functions }. By comparing the number of SRAM cells and the number of memristor required for each $\Delta V_k$ (each $\Delta V_k$ only requires one memristor) and considering the inherent smaller bitcell size of memristors in comparison to SRAM, our utilization of memristor for nonlinear function approximation demonstrates a distinct advantage in terms of overhead. 

\begin{table}[ht]
\caption{Six commonly used nonlinear functions in neural networks and their inverse functions.}
\label{suptab:nlfunc_compare1}
\centering
\begin{tabular}{|c|c|c|}
\hline
Name     & Nonlinear function & Inverse function  \\ \hline
Sigmoid  & $f(x)=\frac{1}{1+e^{-x}}$                   & $f^{-1}(t)=\ln \frac{t}{1-t}$                 \\ \hline
Softplus & $f(x)=\ln \left(1+e^x\right)$                   & $f^{-1}(t)=\ln \left(e^t-1\right)$                 \\ \hline
Tanh     & $f(x)=\frac{e^x-e^{-x}}{e^x+e^{-x}}$                   & $f^{-1}(t)=0.5\ln \frac{1+t}{1-t}$                 \\ \hline
Softsign & $f(x)= \begin{cases}\frac{x}{1+x} & (x \geqslant 0) \\ \frac{x}{1-x} & (x<0)\end{cases}$                   & $f^{-1}(t)= \begin{cases}\frac{t}{1-t} & (t \geqslant 0) \\ \frac{t}{1+t} & (x<0)\end{cases}$                 \\ \hline
Elu      & $f(x)= \begin{cases}x & (x \geqslant 0) \\ e^x-1 & (x<0)\end{cases}$                   & $f^{-1}(t)= \begin{cases}t & (t \geqslant 0) \\ \ln (t+1) & (t<0)\end{cases}$                 \\ \hline
Selu     & $f(x)= \begin{cases}0.5x & (x \geqslant 0) \\ 2(e^x-1) & (x<0)\end{cases}$                   & $f^{-1}(t)= \begin{cases}2t & (t\geqslant0) \\ \ln (0.5t+1) & (t<0)\end{cases}$                 \\ \hline
\end{tabular}
\label{Tab: six NL}
\end{table}


\begin{table}[ht]
\caption{$\Delta V_k$ and SRAM cell numbers for six inverse functions.}
\label{suptab:nlfunc_compare2}
\centering
\scalebox{0.92}{

\begin{tabular}{|c|cc|lc|cc|cc|cc|cc|}
\hline
\multirow{2}{*}{\textbf{k}} & \multicolumn{2}{c|}{\textbf{Sigmoid}}                                           & \multicolumn{2}{c|}{\textbf{Softplus}}                                          & \multicolumn{2}{c|}{\textbf{Tanh}}                                              & \multicolumn{2}{c|}{\textbf{Softsign}}                                        & \multicolumn{2}{c|}{\textbf{Elu}}                                               & \multicolumn{2}{c|}{\textbf{Selu}}                                              \\ \cline{2-13} 
                            & \multicolumn{1}{c|}{\textbf{$\Delta V_k$}}   & \multicolumn{1}{l|}{\textbf{\# SR. cell}} & \multicolumn{1}{c|}{\textbf{$\Delta V_k$}}   & \multicolumn{1}{l|}{\textbf{\# SR. cell}} & \multicolumn{1}{c|}{\textbf{$\Delta V_k$}}   & \multicolumn{1}{l|}{\textbf{\# SR. cell}} & \multicolumn{1}{c|}{\textbf{$\Delta V_k$}} & \multicolumn{1}{l|}{\textbf{\# SR. cell}} & \multicolumn{1}{c|}{\textbf{$\Delta V_k$}}   & \multicolumn{1}{l|}{\textbf{\# SR. cell}} & \multicolumn{1}{c|}{\textbf{$\Delta V_k$}}   & \multicolumn{1}{l|}{\textbf{\# SR. cell}} \\ \hline
1                           & \multicolumn{1}{c|}{0.724}          & 6                                         & \multicolumn{1}{l|}{0.728}          & 9                                         & \multicolumn{1}{c|}{0.362}          & 6                                         & \multicolumn{1}{c|}{1}            & 19                                        & \multicolumn{1}{c|}{1.386}          & 7                                         & \multicolumn{1}{c|}{1.386}          & 7                                         \\ \hline
2                           & \multicolumn{1}{c|}{0.437}          & 4                                         & \multicolumn{1}{l|}{0.441}          & 6                                         & \multicolumn{1}{c|}{0.219}          & 4                                         & \multicolumn{1}{c|}{0.667}        & 13                                        & \multicolumn{1}{c|}{0.56}           & 3                                         & \multicolumn{1}{c|}{0.56}           & 3                                         \\ \hline
3                           & \multicolumn{1}{c|}{0.32}           & 3                                         & \multicolumn{1}{l|}{0.324}          & 4                                         & \multicolumn{1}{c|}{0.16}           & 3                                         & \multicolumn{1}{c|}{0.476}        & 9                                         & \multicolumn{1}{c|}{0.357}          & 2                                         & \multicolumn{1}{c|}{0.357}          & 2                                         \\ \hline
4                           & \multicolumn{1}{c|}{0.257}          & 2                                         & \multicolumn{1}{l|}{0.26}           & 3                                         & \multicolumn{1}{c|}{0.129}          & 2                                         & \multicolumn{1}{c|}{0.357}        & 7                                         & \multicolumn{1}{c|}{0.262}          & 1                                         & \multicolumn{1}{c|}{0.262}          & 1                                         \\ \hline
5                           & \multicolumn{1}{c|}{0.217}          & 2                                         & \multicolumn{1}{l|}{0.219}          & 3                                         & \multicolumn{1}{c|}{0.109}          & 2                                         & \multicolumn{1}{c|}{0.278}        & 5                                         & \multicolumn{1}{c|}{0.208}          & 1                                         & \multicolumn{1}{c|}{0.208}          & 1                                         \\ \hline
6                           & \multicolumn{1}{c|}{0.191}          & 2                                         & \multicolumn{1}{l|}{0.191}          & 2                                         & \multicolumn{1}{c|}{0.095}          & 2                                         & \multicolumn{1}{c|}{0.222}        & 4                                         & \multicolumn{1}{c|}{0.188}          & 1                                         & \multicolumn{1}{c|}{0.188}          & 1                                         \\ \hline
7                           & \multicolumn{1}{c|}{0.171}          & 1                                         & \multicolumn{1}{l|}{0.171}          & 2                                         & \multicolumn{1}{c|}{0.086}          & 1                                         & \multicolumn{1}{c|}{0.182}        & 3                                         & \multicolumn{1}{c|}{0.188}          & 1                                         & \multicolumn{1}{c|}{0.188}          & 1                                         \\ \hline
8                           & \multicolumn{1}{c|}{0.157}          & 1                                         & \multicolumn{1}{l|}{0.156}          & 2                                         & \multicolumn{1}{c|}{0.079}          & 1                                         & \multicolumn{1}{c|}{0.152}        & 3                                         & \multicolumn{1}{c|}{0.188}          & 1                                         & \multicolumn{1}{c|}{0.188}          & 1                                         \\ \hline
9                           & \multicolumn{1}{c|}{0.146}          & 1                                         & \multicolumn{1}{l|}{0.144}          & 2                                         & \multicolumn{1}{c|}{0.073}          & 1                                         & \multicolumn{1}{c|}{0.128}        & 2                                         & \multicolumn{1}{c|}{0.188}          & 1                                         & \multicolumn{1}{c|}{0.188}          & 1                                         \\ \hline
10                          & \multicolumn{1}{c|}{0.138}          & 1                                         & \multicolumn{1}{l|}{0.134}          & 2                                         & \multicolumn{1}{c|}{0.069}          & 1                                         & \multicolumn{1}{c|}{0.11}         & 2                                         & \multicolumn{1}{c|}{0.188}          & 1                                         & \multicolumn{1}{c|}{0.188}          & 1                                         \\ \hline
11                          & \multicolumn{1}{c|}{0.131}          & 1                                         & \multicolumn{1}{l|}{0.126}          & 2                                         & \multicolumn{1}{c|}{0.066}          & 1                                         & \multicolumn{1}{c|}{0.095}        & 2                                         & \multicolumn{1}{c|}{0.188}          & 1                                         & \multicolumn{1}{c|}{0.188}          & 1                                         \\ \hline
12                          & \multicolumn{1}{c|}{0.127}          & 1                                         & \multicolumn{1}{l|}{0.12}           & 2                                         & \multicolumn{1}{c|}{0.063}          & 1                                         & \multicolumn{1}{c|}{0.083}        & 2                                         & \multicolumn{1}{c|}{0.188}          & 1                                         & \multicolumn{1}{c|}{0.188}          & 1                                         \\ \hline
13                          & \multicolumn{1}{c|}{0.123}          & 1                                         & \multicolumn{1}{l|}{0.114}          & 1                                         & \multicolumn{1}{c|}{0.061}          & 1                                         & \multicolumn{1}{c|}{0.074}        & 1                                         & \multicolumn{1}{c|}{0.188}          & 1                                         & \multicolumn{1}{c|}{0.188}          & 1                                         \\ \hline
14                          & \multicolumn{1}{c|}{0.12}           & 1                                         & \multicolumn{1}{l|}{0.109}          & 1                                         & \multicolumn{1}{c|}{0.06}           & 1                                         & \multicolumn{1}{c|}{0.065}        & 1                                         & \multicolumn{1}{c|}{0.188}          & 1                                         & \multicolumn{1}{c|}{0.188}          & 1                                         \\ \hline
15                          & \multicolumn{1}{c|}{0.119}          & 1                                         & \multicolumn{1}{l|}{0.105}          & 1                                         & \multicolumn{1}{c|}{0.059}          & 1                                         & \multicolumn{1}{c|}{0.058}        & 1                                         & \multicolumn{1}{c|}{0.188}          & 1                                         & \multicolumn{1}{c|}{0.188}          & 1                                         \\ \hline
16                          & \multicolumn{1}{c|}{0.118}          & 1                                         & \multicolumn{1}{l|}{0.102}          & 1                                         & \multicolumn{1}{c|}{0.059}          & 1                                         & \multicolumn{1}{c|}{0.053}        & 1                                         & \multicolumn{1}{c|}{0.188}          & 1                                         & \multicolumn{1}{c|}{0.188}          & 1                                         \\ \hline
17                          & \multicolumn{1}{c|}{0.118}          & 1                                         & \multicolumn{1}{l|}{0.099}          & 1                                         & \multicolumn{1}{c|}{0.059}          & 1                                         & \multicolumn{1}{c|}{0.053}        & 1                                         & \multicolumn{1}{c|}{0.188}          & 1                                         & \multicolumn{1}{c|}{0.188}          & 1                                         \\ \hline
18                          & \multicolumn{1}{c|}{0.119}          & 1                                         & \multicolumn{1}{l|}{0.096}          & 1                                         & \multicolumn{1}{c|}{0.059}          & 1                                         & \multicolumn{1}{c|}{0.058}        & 1                                         & \multicolumn{1}{c|}{0.188}          & 1                                         & \multicolumn{1}{c|}{0.188}          & 1                                         \\ \hline
19                          & \multicolumn{1}{c|}{0.12}           & 1                                         & \multicolumn{1}{l|}{0.094}          & 1                                         & \multicolumn{1}{c|}{0.06}           & 1                                         & \multicolumn{1}{c|}{0.065}        & 1                                         & \multicolumn{1}{c|}{0.188}          & 1                                         & \multicolumn{1}{c|}{0.188}          & 1                                         \\ \hline
20                          & \multicolumn{1}{c|}{0.123}          & 1                                         & \multicolumn{1}{l|}{0.091}          & 1                                         & \multicolumn{1}{c|}{0.061}          & 1                                         & \multicolumn{1}{c|}{0.074}        & 1                                         & \multicolumn{1}{c|}{0.188}          & 1                                         & \multicolumn{1}{c|}{0.188}          & 1                                         \\ \hline
21                          & \multicolumn{1}{c|}{0.127}          & 1                                         & \multicolumn{1}{l|}{0.089}          & 1                                         & \multicolumn{1}{c|}{0.063}          & 1                                         & \multicolumn{1}{c|}{0.083}        & 2                                         & \multicolumn{1}{c|}{0.188}          & 1                                         & \multicolumn{1}{c|}{0.188}          & 1                                         \\ \hline
22                          & \multicolumn{1}{c|}{0.131}          & 1                                         & \multicolumn{1}{l|}{0.088}          & 1                                         & \multicolumn{1}{c|}{0.066}          & 1                                         & \multicolumn{1}{c|}{0.095}        & 2                                         & \multicolumn{1}{c|}{0.188}          & 1                                         & \multicolumn{1}{c|}{0.188}          & 1                                         \\ \hline
23                          & \multicolumn{1}{c|}{0.138}          & 1                                         & \multicolumn{1}{l|}{0.086}          & 1                                         & \multicolumn{1}{c|}{0.069}          & 1                                         & \multicolumn{1}{c|}{0.11}         & 2                                         & \multicolumn{1}{c|}{0.188}          & 1                                         & \multicolumn{1}{c|}{0.188}          & 1                                         \\ \hline
24                          & \multicolumn{1}{c|}{0.146}          & 1                                         & \multicolumn{1}{l|}{0.085}          & 1                                         & \multicolumn{1}{c|}{0.073}          & 1                                         & \multicolumn{1}{c|}{0.128}        & 2                                         & \multicolumn{1}{c|}{0.188}          & 1                                         & \multicolumn{1}{c|}{0.188}          & 1                                         \\ \hline
25                          & \multicolumn{1}{c|}{0.157}          & 1                                         & \multicolumn{1}{l|}{0.084}          & 1                                         & \multicolumn{1}{c|}{0.079}          & 1                                         & \multicolumn{1}{c|}{0.152}        & 3                                         & \multicolumn{1}{c|}{0.188}          & 1                                         & \multicolumn{1}{c|}{0.188}          & 1                                         \\ \hline
26                          & \multicolumn{1}{c|}{0.171}          & 1                                         & \multicolumn{1}{l|}{0.082}          & 1                                         & \multicolumn{1}{c|}{0.086}          & 1                                         & \multicolumn{1}{c|}{0.182}        & 3                                         & \multicolumn{1}{c|}{0.188}          & 1                                         & \multicolumn{1}{c|}{0.188}          & 1                                         \\ \hline
27                          & \multicolumn{1}{c|}{0.191}          & 2                                         & \multicolumn{1}{l|}{0.081}          & 1                                         & \multicolumn{1}{c|}{0.095}          & 2                                         & \multicolumn{1}{c|}{0.222}        & 4                                         & \multicolumn{1}{c|}{0.188}          & 1                                         & \multicolumn{1}{c|}{0.188}          & 1                                         \\ \hline
28                          & \multicolumn{1}{c|}{0.217}          & 2                                         & \multicolumn{1}{l|}{0.08}           & 1                                         & \multicolumn{1}{c|}{0.109}          & 2                                         & \multicolumn{1}{c|}{0.278}        & 5                                         & \multicolumn{1}{c|}{0.188}          & 1                                         & \multicolumn{1}{c|}{0.188}          & 1                                         \\ \hline
29                          & \multicolumn{1}{c|}{0.257}          & 2                                         & \multicolumn{1}{l|}{0.08}           & 1                                         & \multicolumn{1}{c|}{0.129}          & 2                                         & \multicolumn{1}{c|}{0.357}        & 7                                         & \multicolumn{1}{c|}{0.188}          & 1                                         & \multicolumn{1}{c|}{0.188}          & 1                                         \\ \hline
30                          & \multicolumn{1}{c|}{0.32}           & 3                                         & \multicolumn{1}{l|}{0.079}          & 1                                         & \multicolumn{1}{c|}{0.16}           & 3                                         & \multicolumn{1}{c|}{0.476}        & 9                                         & \multicolumn{1}{c|}{0.188}          & 1                                         & \multicolumn{1}{c|}{0.188}          & 1                                         \\ \hline
31                          & \multicolumn{1}{c|}{0.437}          & 4                                         & \multicolumn{1}{l|}{0.078}          & 1                                         & \multicolumn{1}{c|}{0.219}          & 4                                         & \multicolumn{1}{c|}{0.667}        & 13                                        & \multicolumn{1}{c|}{0.188}          & 1                                         & \multicolumn{1}{c|}{0.188}          & 1                                         \\ \hline
32                          & \multicolumn{1}{c|}{0.724}          & 6                                         & \multicolumn{1}{l|}{0.077}          & 1                                         & \multicolumn{1}{c|}{0.362}          & 6                                         & \multicolumn{1}{c|}{1}            & 19                                        & \multicolumn{1}{c|}{0.188}          & 1                                         & \multicolumn{1}{c|}{0.188}          & 1                                         \\ \hline
\textbf{Sum}                & \multicolumn{1}{c|}{\textbf{6.992}} & \textbf{58}                               & \multicolumn{1}{c|}{\textbf{4.813}} & \textbf{59}                               & \multicolumn{1}{c|}{\textbf{3.498}} & \textbf{58}                               & \multicolumn{1}{c|}{\textbf{8.0}} & \textbf{150}                              & \multicolumn{1}{c|}{\textbf{7.849}} & \textbf{41}                               & \multicolumn{1}{c|}{\textbf{7.849}} & \textbf{41}                               \\ \hline
\end{tabular}
}
\label{Tab: delt v and cell number of 6 NL functions }
\end{table}

\newpage
\subsection{Spice simulation}
\label{supsec:Spice}
\begin{figure}[ht]
    \centering 
   \includegraphics[width=0.8\textwidth]{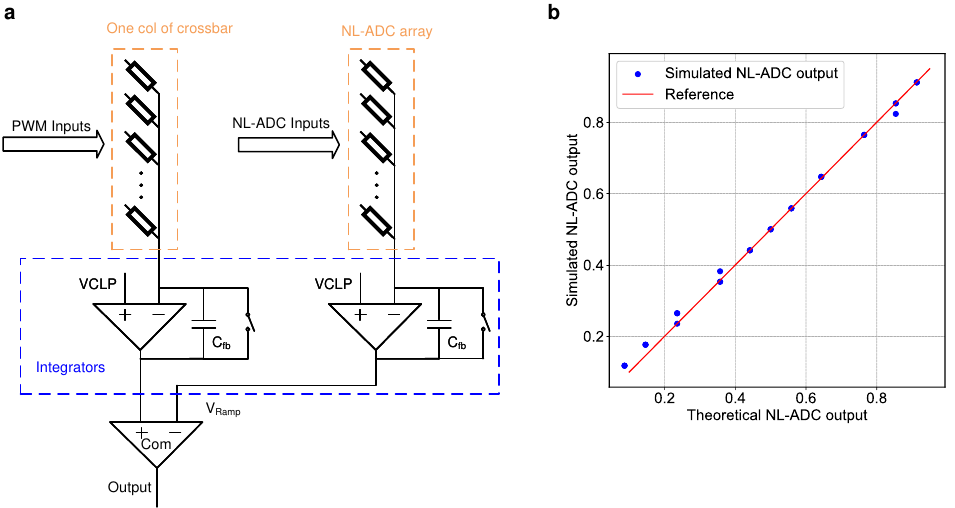}
    \caption{
    \textbf{Spice simulation.}
    \textbf{a} Architecture of circuit simulation. 
    \textbf{b} simulation result.
     In our simulation setup, we utilize a column of Resistive Random Access Memory (RRAM) to simulate the Multiply-Accumulate (MAC) operation. Additionally, we employ 33 RRAMs to simulate the Nonlinear Analog-to-Digital Converter (NL-ADC), with one RRAM dedicated to NL-ADC compensation. The integrator in our system has a Gain-Bandwidth Product (GBW) of 200MHz and a direct current (DC) gain of 1000. Throughout our simulations, we analyze a total of 30 cases, where the theoretical output is $sigmoid^{-1}(V_{MAC})$.
     }  
    \label{figs:SPICE sim result}
\end{figure}

In order to evaluate the effect of circuit non-idealities in the charge integrator, we performed spice simulations (Fig. \ref{figs:SPICE sim result}) using a TSMC $65$ nm CMOS process models. Finite gain and bandwidth of the integrator stage were varied to find acceptable values where the error in integration is $<1\%$. A finite DC gain of $\approx 1000$ and gain bandwidth product of $\approx 200$ MHz was found to be sufficient in this case. A sigmoidal nonlinear activation was considered here.
We also performed more SPICE simulations with more realistic memristor models to make sure our energy estimation based on first order model of a memristor as a resistor are sufficiently accurate. First, we estimate the capacitance of the Ta/TaOx/Pt vertically stacked memristor device based on the parallel plate capacitor model. The lateral dimension of the device is 50 nm $\times$ 50 nm. The thickness of the switching layer between two metal electrodes is 8 nm. The dielectric constant of the TaOx is assumed to be $30$ and the overall estimated capacitance of the devices is 0.083 fF. The SPICE simulation was redone using a more conservative estimate of $C_p$ = 1 fF, resulting in $\approx$ 0.18\% increase in energy.

Further, we also chose the popularly used detailed HSPICE model of memristors\cite{spice_memristor} and redid the simulation. The HSPICE model simulation was redone using the more conservative estimate of $C_p$ = 1 fF, resulting in ~0.19\% increase in energy.

\newpage
\subsection{Energy, Area and Latency estimation at Macro level}
\label{supsec:Estimation_energy_area_latency_Macro}
We estimate our work from two levels, Macro level and full system level. The Macro level includes most of the LSTM layer operations and full system level consists of all LSTM layer operations and FC layer operations.

\begin{figure}[t]
    \centering 
   \includegraphics[width=0.7\textwidth]{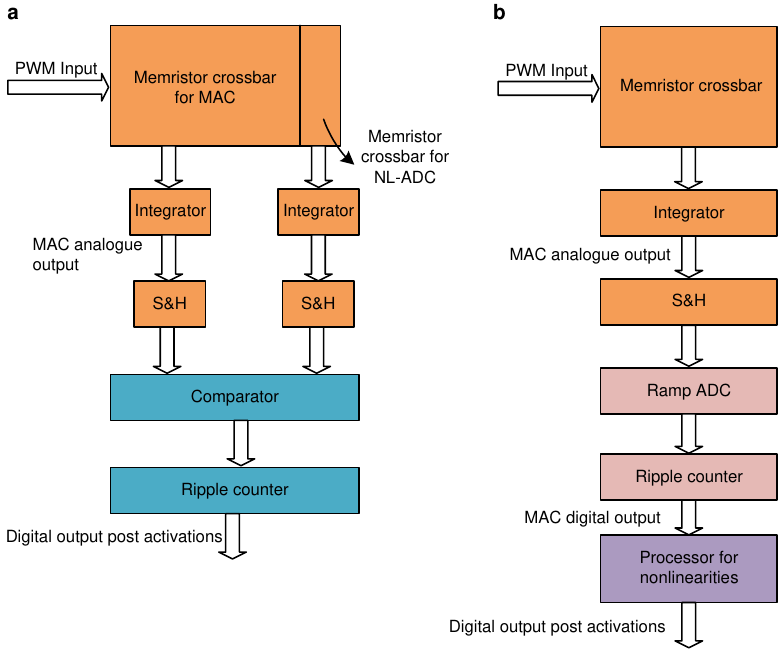}
    \caption{
    \textbf{Architectures of this work and conventional ADC model.}
    \textbf{a} Architecture of this work. 
    \textbf{b} Architecture of conventional ADC model.
    The massive matrix multiplications are distributed in 72×128 crossbar arrays for KWS model.  For our work, crossbar arrays comprises a 72 bit line (BL) drivers and 9216 1T1R devices. The peripheral circuits of crossbar include 128 integrators, 128 S$\&$H blocks and 128 ripple counters. Additionally, an extra column of crossbar arrays, one integrator, one S$\&$H block and 128 comparators are utilized for NL-ADC. The 5-bit ripple counter efficiently converts the thermometer code output from Macro into binary code. In order to process the computations, each real-valued input is divided into 5 binary pulse-width modulation (PWM) inputs. Compared to our work, the conventional model does not include NL-ADC circuits, but instead incorporates 128 ramp analog-to-digital converters (ADCs) and a processor for nonlinear operations.}
    
    \label{figs:two models}
\end{figure}

In order to have a fair comparison between our proposed method combining ADC and non-linear function computation in IMC with the traditional method of IMC for the MAC operation followed by a conventional ramp ADC and digital processor, we created two models as shown in Fig. \ref{figs:two models} where other parameters such as input and output bit-width, clock frequency etc. were kept the same. These two models consist of crossbar arrays, integrators, sample and hold (S$\&$H) blocks, comparators, conventional ramp ADCs, ripple counters and processors. We acquired data pertaining to  crossbar arrays, drivers and S$\&$H from reference \cite{jiang2023efficient}. The data regarding integrators and comparators were obtained from reference \cite{cai_estimate} and  reference \cite{yu202265} respectively.  In the case of the conventional ramp ADC, we get the data from reference \cite{liu202033}.  Ripple counter efficiently converts the thermometer code output from comparators into binary code, thus simplifying subsequent processing by the digital domain processor. The energy consumption and area information of the ripple counter are determined through Spice simulation and reference paper\cite{cordic} respectively. As for processors, we get the data from reference \cite{cordic}. We consider a programming 8-bit ADC for every sub-array of size 512x512 for RRAM writing. This ADC is only used to program RRAM and it is inactive during the inference process. Therefore, we only need to pay attention to its area information and do not need to consider its energy when we calculate efficiency for inference. The area of ADC is 280 µm\textsuperscript{2} from reference\cite{natIBM64core}. We followed the methods in recent papers\cite{jiang2023efficient,cai_estimate} to estimate the energy, latency and area of different sub-circuits in the whole system. For fair comparison, all of the numbers are scaled to the same process node of $16$ nm and frequency of 1 GHz.

We train two models, and evaluate the two models separately. In Keyword Spotting (KWS) model, the  LSTM parameter size (72 × 128) is small enough such that it can fit in one crossbar of size 128 × 128. But for Natural language processing (NLP) model, the LSTM parameter size (633 × 8064) is too big to fit in one crossbar. We assume a crossbar size of $633\times 512$ similar to recent work\cite{Nature2023analogAIchipIBM} and use $16$ such crossbars. The number of rows is kept at $633$ to ensure that it can fit the input dimension of the LSTM layer. However, to avoid large IR drop penalties, we assume only $256$ inputs are enabled at one phase. Hence, it requires $\lceil\frac{633}{256}\rceil=3$ phases to provide the inputs where each phase requires $32$ ns for a 5-bit input with $1$ GHz clock. As an alternative, the number of rows could be limited to $256$ and $3$ separate columns could be used to store these weights with a 3-phase multiplexing of these columns to the same integrator being implemented at the periphery. In that case, the same input lines have to be reused to provide different inputs for the three phases. This would increase the number of required crossbars by $3X$ to $48$ assuming the maximum number of columns is kept fixed at $512$. If the crossbar size is reduced to $256\times 256$ similar to other works\cite{natIBM64core}, the number of cores would increase to $96$ which can be split across multiple chips if needed. For both cases, the advantages of our proposed IMC architecture is retained over the conventional IMC one. We provide detailed estimates of our work (Tab \ref{Tab:this work Energy, area and latency estimation NLP }) and the conventional model for the two cases of $k=1$ (Tab \ref{Tab:conventional model Energy, area and latency estimation NLP k=1}) and $k=8$ (Tab \ref{Tab:conventional model Energy, area and latency estimation NLP k=8}) digital processors. Table \ref{Tab:tab3 different ADC bit Comparison NLP} presents a summary comparing the two methods showing $16X$, $42X$ and $1.1X$ improvement in throughput, area efficiency and energy efficiency respectively for the 5-bit ADC case with $k=8$ digital processors in the conventional model.

\textit{\textbf{A. KWS model (Macro level)}}
\newpage
\begin{table}[ht]
\caption{Energy, area and latency estimation for \textbf{this work (5-bit NL-ADC) at Macro level for KWS task.}}
\centering
\begin{tabular}{|c|c|c|c|c|c|}
\hline
Module       & Number  & On time (ns) & Area (µm\textsuperscript{2}) & Energy (pJ) & Delay (ns)     \\ \hline
MAC array    & 72×128 & 32  & 126.45    & 188.74     & \textless{}0.3 \\ \hline
NL-ADC array & 32     & 32  & 0.44       & 0.12        & \textless{}0.3 \\ \hline
Drivers      & 72     &32   & 198.40    & 3.92       & 0.1            \\ \hline
Integrator   & 129    &32   & 1253.88    & 324.42      & 0.2            \\ \hline
S\&H         & 129    &32   & 4.08      & 0.41        & --            \\ \hline
Comparator   & 128    &32   & 547.84     & 33.10       & 0.1             \\ \hline
Ripple counter  & 128    &32  & 36.48     & 7.09       & --            \\ \hline
ADC (for writing)  & 1    &--  & 280     & --        & --            \\ \hline
Sum          & 9835   &--   & 2447.57    & 557.79     & \textless{}1   \\ \hline
\end{tabular}
\label{Tab:this work Energy, area and latency estimation }
\end{table}
Tab \ref{Tab:this work Energy, area and latency estimation } shows energy, area and latency estimation for this work (5-bit NL-ADC)  based on process node of $16$ nm and frequency of 1 GHz. For MAC array, we use formula ($E_{\text {MAC }}=N_{\text {row }} N_{\text {col }}\left({G}_{\text {on }}+{G}_{\text {off }}\right) V_{\text {read }}^2 \bar{T}_{\text {on }}$) to calculate energy consumption. $N_{\text {row }}$ and $N_{\text {col }}$  is 72 and 128 respectively, we assume that $G_{\text {off }}$ is \SI{5}{\micro\siemens} and $G_{\text {on }}$ is based on the value corresponding to the actual weight.  $V_{\text {read }}$ is 0.2 V.
The input is  5-bit PWM waves, which means that for MAC operation, the maximum on-time of MAC array is 32 ns, and the minimum on-time is 0 ns. To calculate the energy, we take the average of these two values as  $\bar{T}_{\text {on }}$. For NL-ADC, we also use aforementioned equation but $G_{\text {on }}$ is from Fig. \ref{fig:nladc_overview}d and $\bar{T}_{\text {on }}$ is 1 ns. We acquire data pertaining to  drivers and S$\&$H from reference \cite{jiang2023efficient}. The energy consumption and area information of the ripple counter are determined through Spice simulation and reference paper\cite{cordic} respectively. The data regarding integrators and comparators are obtained from reference \cite{cai_estimate} and  reference \cite{yu202265} and scaled respectively.  According to timing of in-memory computing circuits in Fig. \ref{fig:nladc_overview}b, MAC and NL-ADC operation requires a total of 64 clock cycles. Taking into account the delay of the circuit, we take 65 ns as an evaluation period to evaluate the energy consumption. The power dissipation of the system is estimated by dividing the energy consumption for total latency ( 557.79 pJ) by the time needed for total latency (65 ns). The result of this calculation is 8.58 mW.

\begin{table}[ht]
\caption{Energy, area and latency estimation for \textbf{conventional ADC model (5-bit ADC) at Macro level for KWS task.}}
\centering
\begin{tabular}{|c|c|c|c|c|c|}
\hline
Module         & Number & On time (ns) & Area (µm\textsuperscript{2}) & Energy (pJ) & Delay (ns)      \\ \hline
MAC array      & 72×128 & 32        & 126.45    & 188.74     & \textless{}0.3 \\ \hline
Drivers        & 72     & 32        & 198.40    & 3.92       & 0.1             \\ \hline
Integrator     & 128    & 32         & 1244.16    & 321.90      & 0.2             \\ \hline
S\&H           & 128    & 32         & 4.04      & 0.41        & --              \\ \hline
5-bit Ramp-ADC & 128    & 32         & 4546.30   & 256      & 0.2              \\ \hline
Ripple counter  & 128    &32  & 36.48     & 7.09       & --             \\ \hline
Processor      & 1      & 256       & 119.17     & 256     & 0.2          \\ \hline
Sum            & 9801   & --            & 6275.01   & 829.26     & \textless{}1            \\ \hline
\end{tabular}
\label{Tab:conventional model Energy, area and latency estimation}
\end{table}
Tab \ref{Tab:conventional model Energy, area and latency estimation} shows energy, area and latency estimation for  conventional ADC model(5-bit ADC)  based on process node of $16$ nm and frequency of 1 GHz.  In the case of the conventional ramp ADC and processor, we get the data from reference \cite{liu202033} and reference \cite{cordic}  and scaled respectively. The evaluation method for other modules is the same as the previous one.  The total latency of one period is 321 ns, which includes four components: 1 ns for circuit delay time, 32 ns for PWM input time, 32 ns for ADC conversion time, and 256 ns for the processor to calculate the nonlinear functions (One nonlinear function needs 2 clock cycles).  The power dissipation of the system is estimated by dividing the energy consumption for total latency ( 829.26 pJ) by the time needed for total latency (321 ns). The result of this calculation is 2.58 mW.

\newpage

\newpage
\begin{figure}[t]
    \centering 
   \includegraphics[width=0.7\textwidth]{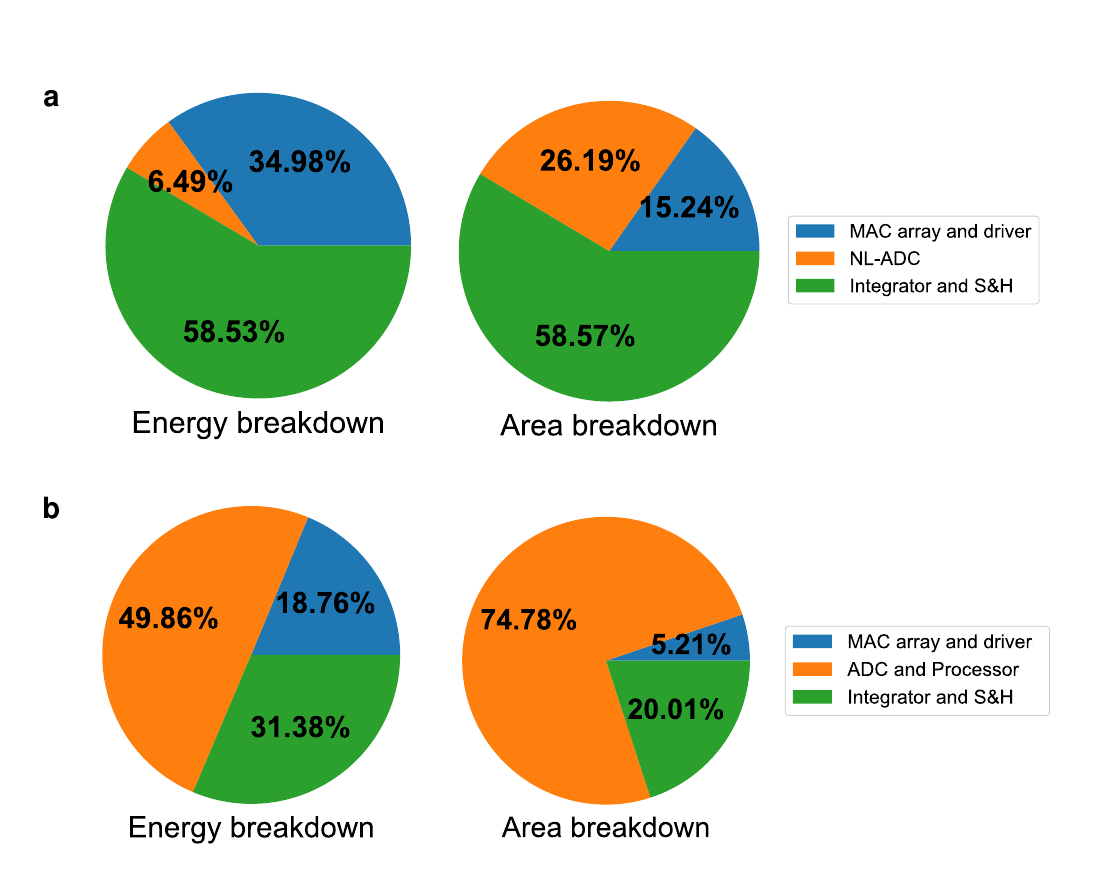}
    \caption{
    \textbf{Energy and area breakdown based on Tab. \ref{Tab:this work Energy, area and latency estimation } and Tab. \ref{Tab:conventional model Energy, area and latency estimation} at Macro level for KWS task.}
    \textbf{a} Energy and area breakdown of this work. 
    \textbf{b} Energy and area breakdown of conventional ADC model.
    The NL-ADC part in the figure includes the NL-ADC array, an integrator, a S$\&$H, and 128 comparators. Our ADC has two functions, nonlinear calculation and conversion of analog signals to digital signals. Therefore, this is a reasonable comparison to a conventional ADC coupled with a processor that primarily computes nonlinearities.
    }
    \label{figs: Energy and area breakdown}
\end{figure}

\begin{table}[ht]
\caption{Comparison of the performance of our work for different NL-ADC resolution and the performance of conventional ADC work \textbf{at Macro level for KWS task.}}
\centering
\begin{tabular}{|c|c|c|c|c|}
\hline
Benchmark metric           & This work(5-bit) & This work(4-bit) & This work(3-bit) & Conventional ADC model(5-bit) \\ \hline
Throughput (TOPS)          & 0.28                    & 0.56                    & 1.08                    & 0.06                   \\ \hline
Power (mW)                 & 8.58                   & 8.43                   & 8.12                   & 2.58                   \\ \hline
Energy-efficiency (TOPS/W) & 33.04                   & 66.24                  & 133.77                  & 23.26                \\ \hline
Area-efficiency (TOPS/mm\textsuperscript{2}) & 115.86                  & 228.87                 & 445.64                 &  9.56                  \\ \hline
\end{tabular}
\label{Tab:tab3 Comparison}
\end{table}


\newpage
\textit{\textbf{B. NLP model (Macro level)}}

\begin{table}[ht]
\caption{Energy, area and latency estimation for \textbf{this work (5-bit NL-ADC) at Macro level for NLP task}. }
\centering
\begin{tabular}{|c|c|c|c|c|c|}
\hline
Module       & Number  & On time (ns) & Area (µm\textsuperscript{2}) & Energy (pJ) & Delay (ns)     \\ \hline
MAC array    & 633×8064 & 32  & 70039.01    & 104540.41     & \textless{}0.3 \\ \hline
NL-ADC array & 512     & 32  & 7.02       & 1.86        & \textless{}0.3 \\ \hline
Drivers      & 10128     &32   &27908.7   & 551       & 0.1            \\ \hline
Integrator   & 8065    &96   & 78391.80    & 60847.52      & 0.2            \\ \hline
S\&H         & 8065    &32   & 254.85      & 25.81        & --            \\ \hline
Comparator   & 8064    &32   & 34513.92     & 2085.03       & 0.1             \\ \hline
Ripple counter  & 8064    &32  & 2298.24     & 446.42       & --            \\ \hline
ADC (for writing)  & 16    &--  & 4480     & --        & --            \\ \hline
Sum          & 5147426   &--   & 217893.57    &168498.01& \textless{}1   \\ \hline
\end{tabular}
\label{Tab:this work Energy, area and latency estimation NLP }
\end{table}
Due to the relatively large input data dimension in this model, we split the input data into three parts, each with a duration of 32 ns. Consequently, the operation time for the MAC array, integrator, and driver is 96 ns, while the comparator time remains unchanged at 32 ns. Additionally, considering a circuit delay of 1ns, the total latency is 129 ns. The energy evaluation method is the same as in Tab. \ref{Tab:this work Energy, area and latency estimation }.

\begin{table}[ht]
\caption{Energy, area and latency estimation for \textbf{conventional ADC model (5-bit ADC) at Macro level for NLP task (k=1)}. }
\centering
\begin{tabular}{|c|c|c|c|c|c|}
\hline
Module         & Number & On time (ns) & Area (µm\textsuperscript{2}) & Energy (pJ) & Delay (ns)      \\ \hline
MAC array      & 633×8064 & 32        & 70039.01    & 104540.41     & \textless{}0.3 \\ \hline
Drivers      & 10128     &32   &27908.7   & 551       & 0.1            \\ \hline
Integrator     & 8064    & 96         & 78382.08    & 60839.98      & 0.1             \\ \hline
S\&H           & 8064    & 32         & 254.82      & 25.80       & --              \\ \hline
5-bit Ramp-ADC & 8064    & 32         & 286417.15   & 16128      & 0.2              \\ \hline
Ripple counter  & 8064    &32  & 2298.24     & 446.42       & --   \\ \hline
Processor (k)     & 1      & 16128       & 119.17     & 16128     & 0.2          \\ \hline
Sum            & 5146897   & --            & 465419.19   & 185757.17     & \textless{}1            \\ \hline
\end{tabular}
\label{Tab:conventional model Energy, area and latency estimation NLP k=1}
\end{table}
The variable "k" represents the number of processors in the system, which signifies the degree of parallelism in processing nonlinear functions. A higher value of "k" indicates a greater degree of parallelism, meaning that more processors are employed simultaneously for processing the nonlinear functions and the nonlinear processing time will be reduced. For this case, k=1. The input method is the same as the previous one. The total latency is 16257 ns. The energy evaluation method is the same as in Tab. \ref{Tab:conventional model Energy, area and latency estimation}.

\begin{table}[ht]
\caption{Energy, area and latency estimation for \textbf{conventional ADC model (5-bit ADC) at Macro level for NLP task(k=8).} }
\centering
\begin{tabular}{|c|c|c|c|c|c|}
\hline
Module         & Number & On time (ns) & Area (µm\textsuperscript{2}) & Energy (pJ) & Delay (ns)      \\ \hline
MAC array      & 633×8064 & 32        & 70039.01    & 104540.41     & \textless{}0.3 \\ \hline
Drivers      & 10128     &32   &27908.7   & 551       & 0.1            \\ \hline
Integrator     & 8064    & 96         & 78382.08    & 60839.98      & 0.2             \\ \hline
S\&H           & 8064    & 32         & 254.82      & 25.80       & --              \\ \hline
5-bit Ramp-ADC & 8064    & 32         & 286417.15   & 16128      & 0.2              \\ \hline
Ripple counter  & 8064    &32  & 2298.24     & 446.42       & --             \\ \hline
Processor (k)     & 8      & 2016       & 953.36     & 16128     & 0.2          \\ \hline
Sum            &5146904   & --            & 466253.38   & 185757.17     & \textless{}1            \\ \hline
\end{tabular}
\label{Tab:conventional model Energy, area and latency estimation NLP k=8}
\end{table}
The evaluation conditions for the current scenario are identical to the previous one, with the exception that the value of "k" is now set to 8. The total latency of one period is 2145 ns. The energy evaluation method is the same as in Tab. \ref{Tab:conventional model Energy, area and latency estimation}.

\begin{table}[ht]
\caption{Comparison of the performance of our work for different NL-ADC resolution and the performance of conventional ADC work \textbf{at Macro level for NLP task}.}
\centering
\begin{tabular}{|c|c|c|c|c|c|}
\hline
Benchmark metric    & \begin{tabular}[c]{@{}c@{}}This   work\\ (5-bit )\end{tabular} & \begin{tabular}[c]{@{}c@{}}This   work\\ (4-bit)\end{tabular} & \begin{tabular}[c]{@{}c@{}}This   work\\ (3-bit)\end{tabular} & \begin{tabular}[c]{@{}c@{}}Conv ADC  model\\ (5-bit, k=1)\end{tabular} & \begin{tabular}[c]{@{}c@{}}Conv ADC model\\ (5-bit, k=8)\end{tabular} \\ \hline
Throughput (TOPS)          & 79.14                    & 157.06                    & 309.36         & 0.62                   & 4.8                 \\ \hline
Power (mW)                 & 1306.2                   & 1295.5                   & 1275.2     & 11.4                  & 86.35 \\ \hline
Energy-efficiency (TOPS/W) & 60.77                   & 121.62                  & 243.36                  & 55.11            & 55.11      \\ \hline
Area-efficiency (TOPS/mm\textsuperscript{2}) & 363.2                  &722.34                 & 1425.81               &1.35   & 10.21                \\ \hline
\end{tabular}
\label{Tab:tab3 different ADC bit Comparison NLP}
\end{table}

\newpage

\subsection{Energy, Area and Latency estimation at system level}
\label{supsec:Estimation_energy_area_latency_system}

\begin{figure}[ht]
    \centering 
   \includegraphics[width=0.8\textwidth]{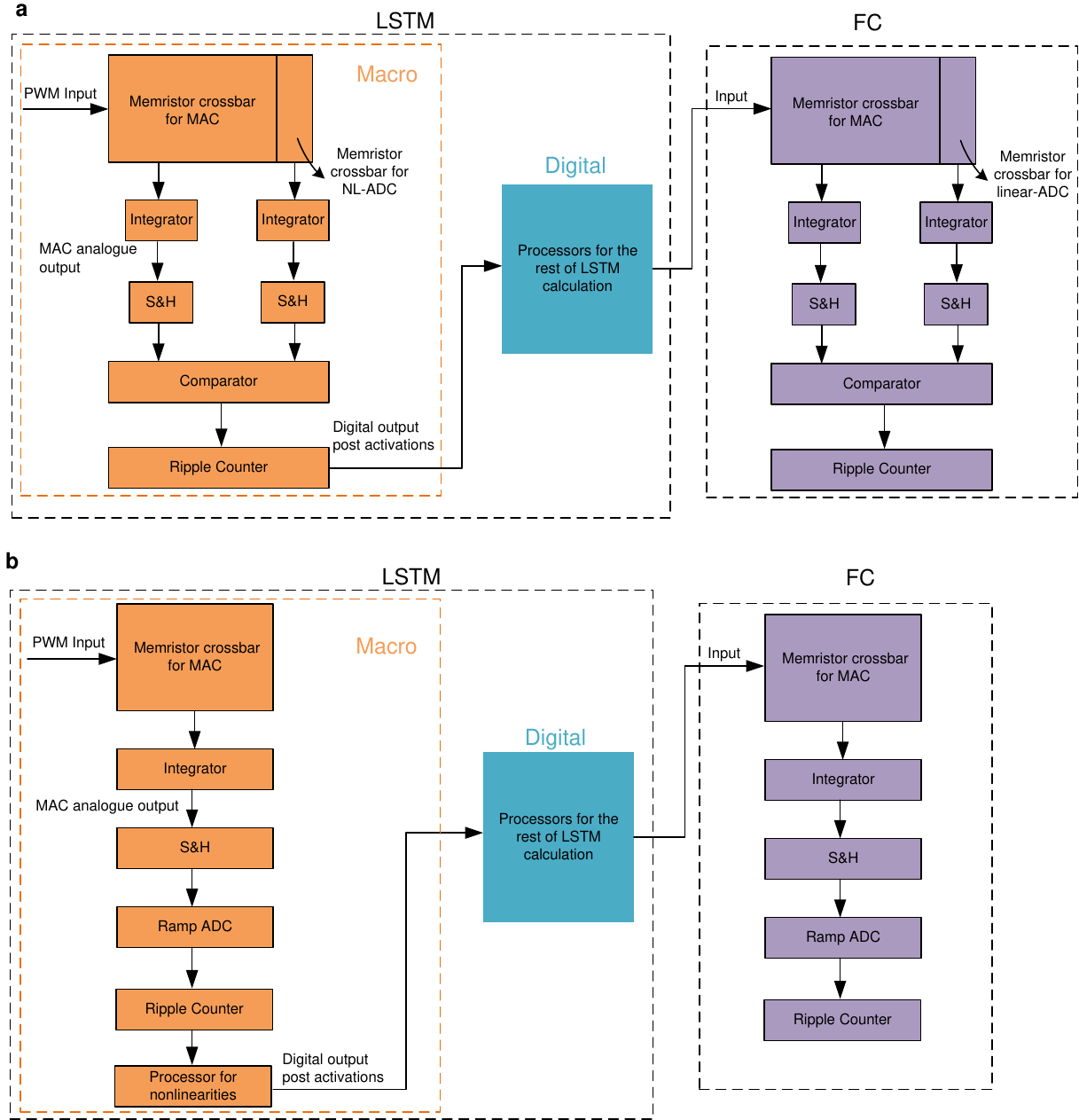}
    \caption{
    \textbf{Full system architectures of this work and conventional ADC model.}
    \textbf{a} Full system architecture of this work. 
    \textbf{b} Full system architecture of conventional ADC model.   
    }
    \label{figs:two full system models arch}
\end{figure}

We also consider the performance of full system comprising of two layers, LSTM layer and FC layer in Fig. \ref{figs:two full system models arch}a . The LSTM layer in Fig. \ref{figs:two full system models arch}a includes two parts, Macro part and digital part for calculating \Cref{Seq:LSTM} and \Cref{Seq:LSTM2} respectively. 

For the KWS model, the output from Macro consists of 128 5-bit binary numbers. This output data size is relatively small, eliminating the need for additional buffers or caches. The ripple counter module incorporates registers that serve as temporary storage for the output data. This allows for direct access and reading of the data by the subsequent processor for further processing and analysis. But for the NLP model with a much bigger network, a simulator (NeuroSim\cite{chen2018neurosim,DNN_NeuroSim_V2.1}) is used to estimate the latency and energy requirements of the buffer and peripheral circuits.

\begin{figure}[ht]
    \centering 
   \includegraphics[width=0.8\textwidth]{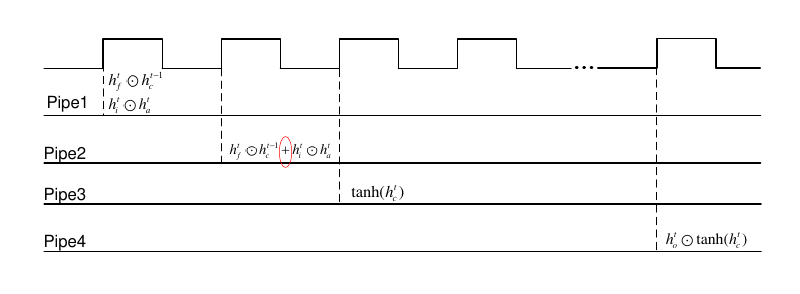}
    \caption{
    \textbf{Pipeline for \Cref{Seq:LSTM2} calculation of LSTM layer.}
    We use digital processor \cite{cordic} to calculate \Cref{Seq:LSTM2}.   
    }
    \label{figs:pipeline for LSTM2}
\end{figure}

The digital part of LSTM layer in Fig. \ref{figs:two full system models arch} is computed in digital domain. Pipeline method is used to estimate the energy and latency of computing \Cref{Seq:LSTM2} in digital part in LSTM layer as depicted in Fig. \ref{figs:pipeline for LSTM2}. Pipeline 1 implements two elementwise multiplication simultaneously in \Cref{Seq:LSTM2} (left equation), which needs one clock cycle. Pipeline 2 finishes adding using one clock cycle and then $h_c^t$ in \Cref{Seq:LSTM2} (left equation) is obtained . For one tanh($h_c^t$) in \Cref{Seq:LSTM2} (right equation), two clocks are needed at least and this can be done in  Pipeline 3. Then Pipeline 4 can calculate the last elementwise multiplication in  \Cref{Seq:LSTM2} (right equation) and then get the final LSTM output ($h^t$). Therefore, total latency of four Pipelines is $(2*N_{tanh}+3)*T_{clk}$, where $N_{tanh}$ is the number of Tanh function in \Cref{Seq:LSTM2} in one time step of LSTM layer and $T_{clk}$ is clock period (1 ns) of digital processor\cite{cordic}.

\Cref{Seq:LSTM} and \Cref{Seq:LSTM2} are computed within Macro and digital domain, respectively. To minimize system latency and energy consumption, the FC layer is also computed within the RRAM array and its implementation method is the same as Macro in LSTM layer. The following evaluates the system performance of the KWS model and the NLP model respectively.
\begin{equation}
\left[\begin{array}{c} h_f^t \\ h_a^t \\ h_i^t \\ h_o^t\end{array}\right] = \left[\begin{array}{c} \sigma \\ \tanh \\ \sigma \\ \sigma \end{array}\right]\left[\begin{array}{lll} x^t & h^{t-1} \end{array}\right]  \left[\begin{array}{l} W \\ U \end{array}\right]
\label{Seq:LSTM}
\end{equation}
\begin{equation}
h_c^t=h_f^t \odot h_c^{t-1}+h_i^t \odot h_a^t \\
h^t=h_o^t \odot \tanh \left(h_c^t\right)
\label{Seq:LSTM2}
\end{equation}

\textit{\textbf{A. KWS model (full system level)}}

For KWS model, the dimensions of $h_f^t,h_a^t, h_i^t, h_o^t$ in \Cref{Seq:LSTM2} are all 32. We leverage two digital processors to estimate the performance of digital part in LSTM layer. One processor calculates 16 Tanh operations. So according to this equation $(2*N_{tanh}+3)*T_{clk}$, the total latency for digital part in LSTM layer is (2*16+3)*1 ns=35 ns($N_{tanh}$=16, $T_{clk}$=1 ns).  We get power and area information of processor from reference paper\cite{cordic}. Therefore energy consumption of digital part in LSTM layer can be obtained (power times latency).

The energy consumption and area information of the ripple counter are determined through Spice simulation and reference paper\cite{natIBM64core} respectively. The evaluation method of other modules in the FC layer is the same as that in the Macro level in the section \textit{\textbf{A. KWS model (Macro level)}}.

 The performance of each module in digital part in LSTM layer and FC layer is shown in  Tab. \ref{Tab:our work full sysytem Energy, area and latency estimation} below.  The total area, latency, and energy consumption of digital part in LSTM layer and FC layer is 513.78µm\textsuperscript{2},98.3 ns and 46.24 pJ.

Combining Tab. \ref{Tab:this work Energy, area and latency estimation } and  estimate of digital part, we can get full system data as shown in Tab. \ref{Tab:our work full sysytem Energy, area and latency estimation}. The power dissipation of the full system is estimated by dividing the energy consumption (618.01 pJ) for total latency by the time needed for total latency (165.6 ns) and the result of this calculation is 3.73 mW.

\begin{table}[ht]
\caption{Energy, area and latency estimation for  \textbf{this work (5-bit NL-ADC) at system level for KWS task}.}
\centering

\begin{tabular}{|c|c|c|c|c|c|c|}
\hline
Layer                   & Module   & Number & On time (ns)           & Area (µm\textsuperscript{2}) & Energy (pJ) & \multicolumn{1}{l|}{Latency (ns)}    \\ \hline
                        & MAC array                                               & 9216   & 32                             & 126.45     & 188.74      &                                    \\ \cline{2-6}
                        & NLADC array                                             & 32     & 32                              & 0.4391     & 0.1165      &                                    \\ \cline{2-6}
                        & Drivers                                                 & 72     & 32                             & 198.4      & 3.9168      &                                    \\ \cline{2-6}
                        & Integrator                                              & 129    & 32                             & 1253.9     & 324.42      &                                    \\ \cline{2-6}
                        & Comparator                                              & 128    & 32                             & 547.84     & 33.096      &                                    \\ \cline{2-6}
                        & S\&H                                                    & 128    & 32                             & 4.0448     & 0.4096      &                                    \\ \cline{2-6}
\multirow{-7}{*}{LSTM} & Ripple counter                  & 128    & 32                             & 36.48      & 7.0861      & \multirow{-7}{*}{65.3}     \\ \hline
LSTM                   & Processors for the rest of LSTM & 2      & 35                             & 238.34     & 14          & 35                                \\ \hline
                        & MAC array                                               & 384    & 32                             & 5.2689     & 7.8643      &                                    \\ \cline{2-6}
                        & ADC array                                               & 32     & 32                             & 0.4391     & 0.1165      &                                    \\ \cline{2-6}
                        & Drivers                                                 & 32     & 32                             & 88.179     & 1.7408      &                                    \\ \cline{2-6}
                        & Integrator                                              & 13     & 32                             & 126.36     & 32.693      &                                    \\ \cline{2-6}
                        & Comparator                                              & 12     & 32                             & 51.36      & 3.1027      &                                    \\ \cline{2-6}
                        & S\&H                                                    & 13     & 32                             & 0.4108     & 0.0416      &                                    \\ \cline{2-6}
\multirow{-7}{*}{FC}    & Ripple counter                  & 12     & 32                             & 3.42       & 0.6643      & \multirow{-7}{*}{65.3} \\ \hline
                       &ADC (for writing)  & 1    &--  & 280    & --        & --                                 \\ \hline
                       & Sum                             & 10334  & \multicolumn{1}{l|}{\textbf{}} & 2961.32     & 618.01      & 165.6      \\ \hline
\end{tabular}
\label{Tab:our work full sysytem Energy, area and latency estimation}
\end{table}

\newpage
For conventional ADC model, we utilize conventional ADC to estimate both digital part in LSTM layer and FC layer as depicted in Tab. \ref{Tab:conventional full system model Energy, area and latency estimation}  and the method is same as  Tab. \ref{Tab:conventional model Energy, area and latency estimation}. The power dissipation of the full system is estimated by dividing the energy consumption (910.27 pJ) for total latency by the time needed for total latency (421.6 ns) and the result of this calculation is 2.16 mW.
\begin{table}[ht]
\caption{Energy, area and latency estimation for \textbf{conventional ADC model (5-bit ADC) at system level for KWS task}.}
\centering
\begin{tabular}{|c|c|c|c|c|c|c|}
\hline
Layer                  & Module                          & Number & On time (ns)  & Area (µm\textsuperscript{2}) & Energy (pJ) & \multicolumn{1}{l|}{Latency (ns)} \\ \hline
\multirow{8}{*}{LSTM} & MAC array                       & 9216   & 32                    & 126.452736 & 188.74368   & \multirow{8}{*}{321.3}   \\ \cline{2-6}
                       & 5 bit RA- ADC                   & 128    & 32                    & 4546.304   & 256         &                                 \\ \cline{2-6}
                       & Drivers                         & 72     & 32                    & 198.4032   & 3.9168      &                                 \\ \cline{2-6}
                       & Integrator                      & 128    & 32                    & 1244.16    & 321.90464   &                                 \\ \cline{2-6}  
                       & S\&H                            & 128    & 32                    & 4.0448     & 0.4096      &                                 \\ \cline{2-6}
                       & Ripple counter                  & 128    & 32                    & 36.48      & 7.08608     &                                 \\ \cline{2-6}
                       & Processor(NL)                   & 1      & 256                   & 119.17     & 51.2        &                                 \\ \hline
LSTM                  & Processors for the rest of LSTM & 2      & 35                    & 238.34     & 14          & 35                \\ \hline
\multirow{7}{*}{FC}    & MAC array                       & 384    & 32                    & 5.268864   & 7.86432     & \multirow{7}{*}{65.3}       \\ \cline{2-6}
                       & 5 bit RA- ADC                   & 12     & 32                    & 426.216    & 24          &                                 \\ \cline{2-6}
                       & Drivers                         & 32     & 32                    & 88.1792    & 1.7408      &                                 \\ \cline{2-6}
                       & Integrator                      & 13     & 32                    & 126.36     & 32.69344    &                                 \\ \cline{2-6} 
                       & S\&H                            & 13     & 32                    & 0.4108     & 0.0416      &                                 \\ \cline{2-6}
                       & Ripple counter                  & 12     & 32                    & 3.42       & 0.66432     &                                 \\ \hline
\textbf{}              & Sum                             & 10409  & \multicolumn{1}{l|}{} & 7163.21  & 910.27   & 421.6            \\ \hline
\end{tabular}
\label{Tab:conventional full system model Energy, area and latency estimation}
\end{table}

\newpage
\begin{table}[ht]
\caption{Comparison of the performance of full system for different NL-ADC resolution and the performance of conventional 5-bit ADC work\textbf{ at system level for KWS task}. In terms of throughput, energy efficiency and area efficiency, this work is 2 times, 1.5 times and 6.8 times that of traditional  conventional architectures at the system level.}
\centering
\begin{tabular}{|c|c|c|c|c|}
\hline
Benchmark metric           & This work(5-bit) & This work(4-bit) & This work(3-bit) & Conventional ADC model(5-bit) \\ \hline
Throughput (TOPS)          & 0.12                   & 0.19                    & 0.28                    & 0.06                   \\ \hline
Power (mW)                 & 3.73                   & 3.15                   & 2.41                   & 2.16                   \\ \hline
Energy-efficiency (TOPS/W) & 31.33                   & 60.09                  & 114.40                  & 21.27                  \\ \hline
Area-efficiency (TOPS/mm\textsuperscript{2}) & 39.48                  & 63.36                 & 92.78                 & 6.41                   \\ \hline
\end{tabular}
\label{Tab:full system Comparison under different bit KWS}
\end{table}

\begin{table}[ht]
\caption{Energy efficiency comparison \textbf{ at various levels for KWS task}: MAC array, NL-processing, full system. For this work, the energy-efficiency calculation of NL-processing module takes into account the NL-ADC array, an integrator, a S\&H, and 128 comparators.}
\centering
\begin{tabular}{|c|c|c|c|}
\hline
Energy-efficiency (TOPS/W) & This work(5-bit NL-ADC) & Conventional ADC model(5-bit ADC) & Nature’23\cite{Nature2023analogAIchipIBM} \\ \hline
MAC array                  & 97.6                    & 97.6                   & 20.0      \\ \hline
NL-processing              & 3.6                     & 0.3                    & 0.9       \\ \hline
Macro level                 & 33.0                    & 23.3                   & 7.09       \\ \hline
Full system                & 31.33                    & 21.27                   & 6.9       \\ \hline
\end{tabular}
\label{Tab:tabs3 comparison  at various levels}
\end{table}

\newpage
\textit{\textbf{B. NLP model (full system level)}\\}

The buffer and interconnect results are simulated in NeuroSim. The trained model is mapped onto the RRAM-based IMC architecture, where each tile consists of $2\times2$ PEs and each PE is composed of $4\times4$ synaptic arrays. The array size is set to $128\times 128$ 1T1R array at 16 nm technology node. Thus, the PE and tile sizes are $512\times 512$ and $1024\times 1024$, respectively. When mapping LSTM layer weight onto arrays, the weight matrices ($LSTM_w$ size: 128 $\times$ (2016 $\times$ 4) and $LSTM_u$ size: 504 $\times$ (2016 $\times$ 4)) are partitioned and assigned to different synaptic arrays due to the limited array size and then sum up the partial results from arrays to get final results. In the column dimension, we need 63 synaptic arrays (i.e., $2016 \times 4 / 128$), which is equivalent to 16 PEs, or 8 tiles. In the row dimension, 5 synaptic arrays are needed (i.e., (128+504)/128 ), which is equal to 2 PEs or 1 tile. So, $8\times 1$ tiles are used to implement LSTM layer. Similarly, 1 tile is enough to implement a FC layer weight with small size ($504\times 50$). In terms of the device characteristics, the transistor has a 1.1 V gate voltage and 15 k$\Omega$ access resistance. The memristor has 128 conductance states, on/off ratio of 30, on-state resistance of 7 k$\Omega$, and 0.2 V read voltage. Then the evaluated buffer and interconnect results are given as shown in Tab. \ref{Tab:our work full sysytem Energy, area and latency estimation NLP}.

Combining Tab. \ref{Tab:this work Energy, area and latency estimation NLP } with estimate of the digital processor, FC layer, interconnect and buffer, we can get full system data as shown in Tab. \ref{Tab:our work full sysytem Energy, area and latency estimation NLP}. The power dissipation of the full system is estimated by dividing the energy consumption (196,367 pJ) for total latency by the time needed for total latency (526.9 ns) and the result of this calculation is 372.683 mW.  

\begin{table}[ht]
\caption{Energy, area and latency estimation for \textbf{ this work (5-bit NL-ADC) at system level for NLP task}.}
\centering

\begin{tabular}{|c|c|c|c|c|c|c|}
\hline
Layer                  & Module                          & Number  &  On time (ns) & Area (µm\textsuperscript{2}) & Energy (pJ) & Latency (ns)          \\ \hline
\multirow{7}{*}{LSTM} & MAC array                       & 5104512 & 32                   & 70039.0092 & 104540.4058 & \multirow{7}{*}{129} \\ \cline{2-6}
                       & NLADC array                     & 512      & 32                    & 7.03   & 1.86     &                       \\ \cline{2-6}
                       & Drivers                         & 10128     & 32                   & 27908.7  & 550.96     &                       \\ \cline{2-6}
                       & Integrator                      & 8065    & 96                   & 78391.8    & 60847  &                       \\ \cline{2-6}
                       & Comparator                      & 8064    & 32                   & 34513.92   & 2085.02784  &                       \\ \cline{2-6}
                       & S\&H                        & 8065    & 32                   & 254.854    & 25.808      &                       \\ \cline{2-6}
                       & Ripple counter                  & 8064    & 32                   & 2298.24    & 446.42304   &                       \\ \hline
LSTM                  & Processors for the rest of LSTM & 30      & 137.4                & 3575.1     & 824.4       & 137.4                  \\ \hline
\multirow{7}{*}{FC}    & MAC array                       & 25200   & 32                   & 345.7692   & 516.096     & \multirow{7}{*}{65.3} \\ \cline{2-6}
                       & ADC array                       & 32     & 32                    & 0.44   & 0.12     &                       \\ \cline{2-6}
                       & Drivers                         & 504     & 32                   & 1388.8224  & 27.4176     &                       \\ \cline{2-6}
                       & Integrator                      & 51      & 32                   & 495.72     & 128.25888   &                       \\ \cline{2-6}
                       & Comparator                      & 50      & 32                   & 214        & 12.928      &                       \\ \cline{2-6}
                       & S\&H                        & 51      & 32                   & 1.6116     & 0.1632      &                       \\ \cline{2-6}
                       & Ripple counter                  & 50      & 32                   & 14.25      & 2.768       &                       \\ \hline
\multirow{2}{*}{All}   & Buffer                          & –       & 71.7                 & 50916      & 36751.8     & 71.7                  \\ \cline{2-7} 
                       & Interconnect                    & –       & 123.5                & 433261     & 7890.42     & 123.5                 \\ \hline
                       &ADC (for writing)  & 16    &--  & 4480     & --        & --                                 \\ \hline
                       & Sum                             & 5173394 &                      & 705793.78  & 214203.19    & 526.9                   \\ \hline
\end{tabular}
\label{Tab:our work full sysytem Energy, area and latency estimation NLP}
\end{table}

\begin{table}[ht]
\caption{Energy, area and latency estimation for \textbf{conventional ADC model (5-bit ADC) at system level for NLP task}. The variable "k" represents the
number of processors in the system, which signifies the degree of parallelism in processing nonlinear functions. A higher value
of "k" indicates a greater degree of parallelism, meaning that more processors are employed simultaneously for processing the
nonlinear functions and the nonlinear processing time will be reduced. \textbf{For this case, k=1.} }
\centering
\begin{tabular}{|c|c|c|c|c|c|c|}
\hline
Layer                  & Module                          & Number  &  On time (ns) & Area (µm\textsuperscript{2}) & Energy (pJ) & Latency (ns)          \\ \hline
\multirow{8}{*}{LSTM} & MAC array                       & 5104512 & 32                   & 70039.0092 & 104540.4058 & \multirow{7}{*}{129} \\ \cline{2-6}
                       & 5 bit RA- ADC                   & 8064    & 32                   & 286417.152 & 16128       &                       \\ \cline{2-6}
                       & Drivers                         & 10128     & 32                   & 27908.7  & 550.96     &                        \\ \cline{2-6}
                       & Integrator                      & 8064    & 96                   & 78382.08   & 60839 &                       \\ \cline{2-6} 
                       & S\&H                            & 8064    & 32                   & 254.8224   & 25.8048     &                       \\ \cline{2-6}
                       & Ripple Counter                  & 8064    & 32                   & 2298.24    & 446.42304   &                       \\ \cline{2-7} 
                       & Processor(\textbf{k})                   & \textbf{1 }      & 16128                & 119.17     & 3225.6      & 16128                 \\ \hline
LSTM                  & Processors for the rest of LSTM & 30      & 137.4                & 3575.1     & 824.4       & 137.4                   \\ \hline
\multirow{7}{*}{FC}    & MAC array                       & 100800  & 32                   & 1383.0768  & 2064.384    & \multirow{7}{*}{65.3} \\ \cline{2-6}
                       & 5 bit RA- ADC                   & 50      & 32                   & 1775.9     & 100         &                       \\ \cline{2-6}
                       & Drivers                         & 2016    & 32                   & 5555.2896  & 109.6704    &                       \\ \cline{2-6}
                       & Integrator                      & 51      & 32                   & 495.72     & 128.25888   &                       \\ \cline{2-6}
                       & S\&H                            & 51      & 32                   & 1.6116     & 0.1632      &                       \\ \cline{2-6}
                       & Ripple counter                  & 50      & 32                   & 14.25      & 2.768       &                       \\ \hline
\multirow{2}{*}{All}   & Buffer                          & –       & 71.7                  & 50916      & 36751.8     & 71.7                  \\ \cline{2-7} 
                       & Interconnect                    & –       & 123.5               & 433261     & 7890.42     & 123.5                 \\ \hline
                       & Sum                             & 5174352 & –                     & 961359.83  & 232080.75   & 16655.2              \\ \hline
\end{tabular}
\label{Tab:conventional full system model Energy, area and latency estimation NLP k=1}
\end{table}

\newpage
\begin{table}[ht]
\caption{Energy, area and latency estimation for \textbf{conventional ADC model (5-bit ADC) at system level for NLP task}. The variable "k" represents the
number of processors in the system, which signifies the degree of parallelism in processing nonlinear functions. A higher value
of "k" indicates a greater degree of parallelism, meaning that more processors are employed simultaneously for processing the
nonlinear functions and the nonlinear processing time will be reduced. \textbf{For this case, k=8.} }
\centering
\begin{tabular}{|c|c|c|c|c|c|c|}
\hline
Layer                  & Module                          & Number     &  On time (ns) & Area (µm\textsuperscript{2}) & Energy (pJ) & Latency (ns)          \\ \hline
\multirow{8}{*}{LSTM} & MAC array                       & 5104512    & 32                   & 70039.0092 & 104540.4058 & \multirow{7}{*}{129} \\ \cline{2-6}
                       & 5 bit RA- ADC                   & 8064       & 32                   & 286417.152 & 16128       &                       \\ \cline{2-6}
                       & Drivers                         & 10128     & 32                   & 27908.7  & 550.96     &                      \\ \cline{2-6}
                       & Integrator                      & 8064       & 96                   & 78382.08   & 60839       &                       \\ \cline{2-6} 
                       & S\&H                            & 8064       & 32                   & 254.8224   & 25.8048     &                       \\ \cline{2-6}
                       & Ripple Counter                  & 8064       & 32                   & 2298.24    & 446.42304   &                       \\ \cline{2-7} 
                       & Processor(\textbf{k})                   & \textbf{8} & 2016                & 119.17     & 3225.6      & 2016                  \\ \hline
LSTM                  & Processors for the rest of LSTM & 30         & 137.4                & 3575.1     & 824.4       & 137.4                   \\ \hline
\multirow{7}{*}{FC}    & MAC array                       & 100800     & 32                   & 1383.0768  & 2064.384    & \multirow{7}{*}{65.3} \\ \cline{2-6}
                       & 5 bit RA- ADC                   & 50         & 32                   & 1775.9     & 100         &                       \\ \cline{2-6}
                       & Drivers                         & 2016       & 32                   & 5555.2896  & 109.6704    &                       \\ \cline{2-6}
                       & Integrator                      & 51         & 32                   & 495.72     & 128.25888   &                       \\ \cline{2-6}
                       & S\&H                            & 51         & 32                   & 1.6116     & 0.1632      &                       \\ \cline{2-6}
                       & Ripple counter                  & 50         & 32                   & 14.25      & 2.768       &                       \\ \hline
\multirow{2}{*}{All}   & Buffer                          & –          & 71.7                  & 50916      & 36751.8     & 71.7                  \\ \cline{2-7} 
                       & Interconnect                    & –          & 123.5                 & 433261     & 7890.42      & 123.5                 \\ \hline
                       & Sum                             & 5174352    &–                      & 962194.02   & 232080.75   & 2543.2                \\ \hline
\end{tabular}
\label{Tab:conventional full system model Energy, area and latency estimation NLP k=8}
\end{table}

\newpage
\begin{table}[ht]
\caption{Comparison of the performance for different NL-ADC resolution and the performance of conventional ADC work \textbf{at system level for NLP task}. In terms of throughput, energy efficiency and area efficiency, this work is 4.9 times, 1.1 times and 7.9 times that of traditional  conventional architectures at the system level.}
\centering
\begin{tabular}{|c|c|c|c|c|c|}
\hline
Benchmark metric    & \begin{tabular}[c]{@{}c@{}}This   work\\ (5-bit )\end{tabular} & \begin{tabular}[c]{@{}c@{}}This   work\\ (4-bit)\end{tabular} & \begin{tabular}[c]{@{}c@{}}This   work\\ (3-bit)\end{tabular} & \begin{tabular}[c]{@{}c@{}}Conv ADC  model\\ (5-bit, k=1)\end{tabular} & \begin{tabular}[c]{@{}c@{}}Conv ADC model\\ (5-bit, k=8)\end{tabular} \\ \hline
Throughput (TOPS)          & 19.49                    & 25.14                    & 30.23         & 0.62                   & 4.03                 \\ \hline
Power (mW)                 &406.5                   & 263.87                   & 181.9      & 13.9                  & 91.2 \\ \hline
Energy-efficiency (TOPS/W) & 47.9                   & 95.3                  & 166.5                  & 44.2           & 44.2     \\ \hline
Area-efficiency (TOPS/mm\textsuperscript{2}) & 27.6                  & 59.51                 & 72.35                 & 0.64   & 4.2                 \\ \hline
\end{tabular}
\label{Tab:full system different ADC bit Comparison NLP}
\end{table}

\newpage
\subsection{Programming of NL-ADC on crossbar arrays}
\label{supsec:nladc_prog}

\begin{figure}[t]
    \centering 
   \includegraphics[width=0.9\textwidth]{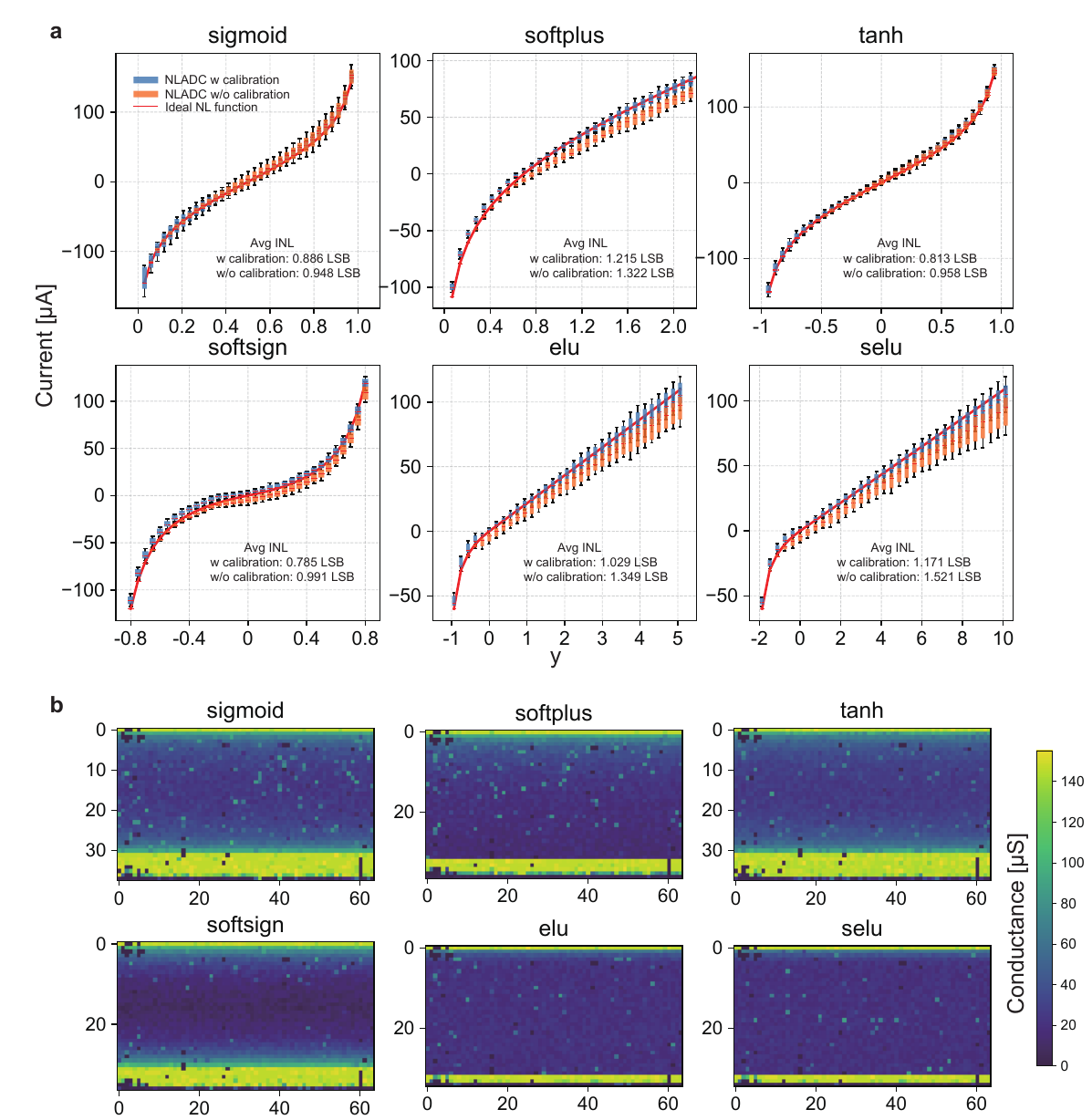}
    \caption{
    \textbf{Programming of NL-ADC on crossbar arrays.}
    \textbf{a} Transfer function of different NL-functions after mapping on real crossbar array with bias term and without bias term. 
    \textbf{b} Actual conductance map of 6 different NL-ADC weights and bias mapped to 64 crossbar columns.
    }
    \label{supfig:nladc_prog}
\end{figure}

\begin{figure}[t]
    \centering 
   \includegraphics[width=0.9\textwidth]{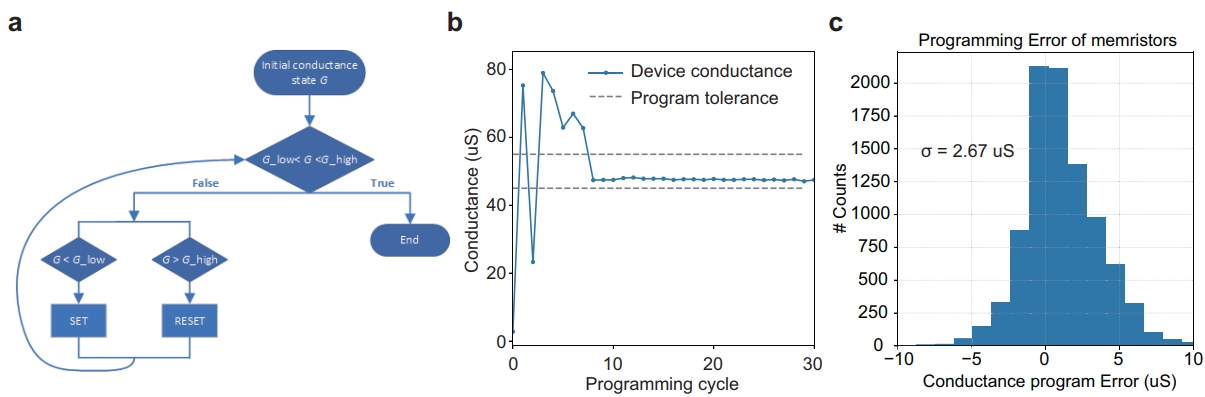}
    \caption{
    \textbf{Iterative programming on memristor crossbar arrays} 
    \textbf{a} Flow char showing the process of programming the entire array.
    \textbf{b} Conductance updating plot of a single device during iterative programming. It shows that under several programming cycles, the conductance finally lies in the tolerated range.
    \textbf{c} Programming error distribution shows that with our iterative programming method, we can achieve programming standard deviation about \SI{2.67}{\micro\siemens}.
    }
    \label{supfig:nladc_prog2}
\end{figure}

Typically 30-40 iterations are needed for accurate programming, even including repeated programming to counter drift \cite{rram_stable_programming}. As a conservative estimate, we consider 100 iterations to program each device. Using a 8-bit version of the current controlled oscillator (CCO) based ADC from \cite{natIBM64core}, we can estimate $256$ ns for each conversion cycle for a read operation and a few ns for any control operations. Do note that this is also a conservative estimate and ADCs with conversion time of ~5 ns are easily available\cite{sar_stable} Hence, the ADC operation time required for programming a $256\times256$ crossbar of memristors, sufficient for KWS, is $\approx1.68$ sec. For the large NLP task with $\approx~6$ millions parameters, the estimated ADC operation time is $157$ sec or less than 3 minutes. Hence, we can see that this is not a bottleneck in terms of programming time.

\newpage
\subsection{Latency estimation of MAC operation in In-memory computing and Digital Non-linear function implementations}
\label{supsec:latency_comparison_NL_Mac}
The latency for a MAC operation ($T_{MAC}$) using In-memory computing (IMC) comprises two parts--input generation ($T_{in}$) and output generation ($T_{out}$) by analog to digital conversion. Assuming PWM inputs with bit-width $b_{in}$ and ramp/CCO ADC for output generation with bit-width $b_{out}$, the latency is given by the following equation:
\begin{equation}
    \label{supeq:latency_mac}
    T_{MAC}=T_{in}+T_{out}=2^{b_{in}}T_{clk}+(2^{b_{out}}-1)T_{clk}
\end{equation}
where $T_{clk}$ denotes the clock period. We choose ramp/CCO ADC since these have areas small enough to be pitch-matched with memory and have been used in recent IMC research\cite{natIBM64core,Nature2023analogAIchipIBM,bonan2019,yi2018_elm}. The same equation can be generalized to other ADCs by modifying the equation of $T_{out}$ (e.g. $T_{out}=b_{out}$ for a Successive Approximation ADC).

The latency of nonlinear function approximation ($T_{NL}$) using digital processors depends on the method used and desired accuracy. For commonly used methods of cordic\cite{cordic}, look-up table (LUT) and piece wise linear functions\cite{giraldo2018laika,ime_pwl}, we estimate$N_{cyc}=2-5$ cycles being needed with a minimum of $2$ cycles needed in the case of LUT to read the input and fetch the corresponding data. A second factor affecting the latency is the number of hidden neurons $N_h$ for which this computation has to be done as well as the number $k$ of parallel digital processors available. The latency for nonlinear function approximation can then be expressed as:
\begin{equation}
    \label{supeq:latency_nl}
    T_{NL}=\frac{4\times N_h\times N_{cyc}}{k}T_{clk}
\end{equation}
where the multiple of $4$ is due to the four gates per neuron in an LSTM cell. The ratio $\frac{T_{NL}}{T_{MAC}}$ is plotted in Fig. \ref{fig:nladc_motivation}\textbf{c} for various values of $k$ with $b_{in}=b_{out}=5$, $N_{cyc}=2$ or $5$ and $N_h=512$ which is representative of common applications. Here, each digital processor handles $\frac{N_h}{k}$ neurons. As an example, a recent work\cite{natIBM64core} shares one digital processor among $8$ MAC cores with each core having $128$ neurons. Thus each digital processor handles $128\times8=1024$ neurons which is worse than all cases considered in Fig. \ref{fig:nladc_motivation}\textbf{c}. 

\newpage
\subsection{Circuit implementation}
\label{supsec:nladc_imp}
\begin{figure}[p]
    \centering 
   \includegraphics[width=0.9\textwidth]{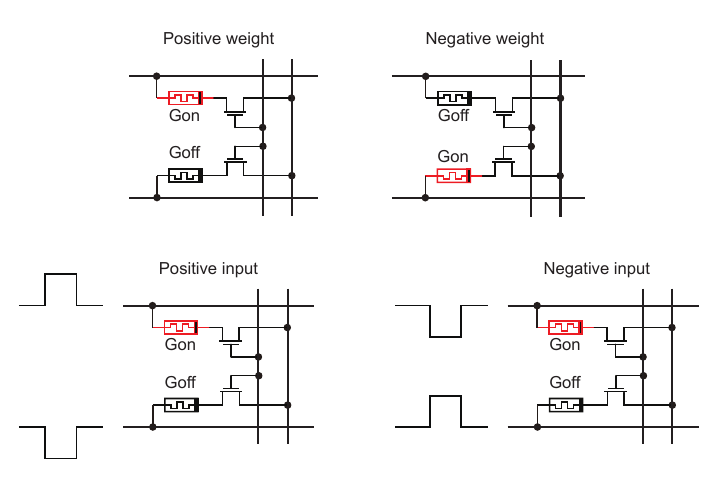}
    \caption{
    This figure illustrates how we map both the positive and negative weight/inputs to the crossbar arrays. We use differential encoding by using two memristors in a single column to represent a single weight and 2 input lines to represent a single input signal.
    }
    \label{supfig:weight_mapping}
\end{figure}

Fig. \ref{supfig:weight_mapping} illustrates how we map both the positive and negative weight/inputs to the crossbar arrays. 

\newpage
\subsection{Analysis of Accuracy Drop with Write Noise}
\label{supsec:accuracy_write_noise}

If we analyze the 5-bit NL-ADC case for the KWS task, there is a drop of 1.7\% in accuracy in software simulation including hardware non-idealities (weight noise and NL-ADC noise in Fig. 4d) from the baseline. Chip testing resulted in only 0.9\% drop from this earlier software simulation. This large drop is attributed to the fact that LSTMs are more sensitive to errors than conventionally reported fully connected (FC) or convolutional networks (CNN). To verify this, we have now simulated a 2 layer FC network with ReLU activations (mirroring the one in \cite{Nature2023analogAIchipIBM}) on the same KWS task. The software test accuracy, (exclusive of hardware non-ideal factors) was determined to be 86.1\%. Following the introduction of hardware noise (derived from measured write noise) into this network, the accuracy decreased marginally to 85.3\%. This experiment demonstrated a modest accuracy reduction of 0.8\%, much less than the 1.7\% drop for the case of LSTM with 5-bit NL-ADC, proving that the large drop is specific to recurrent architectures like LSTM. Further, we also tested the VGG-8 and VGG-16 networks on CIFAR-10 dataset with 5-bit ADC quantization and memristor write noise. For the VGG-16 networks, the accuracy reduced from 93.9\% to 92.9\%; in comparison, the accuracy for VGG-16 on CIFAR-10 in \cite{dac_charan} is 92.57\% when tested with hardware non-idealities. For the VGG-8 networks, the accuracy reduced from 93\% to 92.6\%; in comparison, the accuracy for VGG-8 on CIFAR-10 in \cite{rram_replicabias} and \cite{symp_vlsi_jiang} are 90.7\% and 89\% respectively when tested with hardware non-idealities. These results show that the hardware non-idealities and training methods used in our work are comparable to those used by other research groups.

Second, the drop in accuracy is largely due to the error in programming and weights and not due to the NL-ADC, which is the focus of this paper. In the current Fig. 4d, the effect of both are present and hence, the contribution of each is not obvious. To tease apart the two effects, we did simulations where the write noise was added only in weights or the NL-ADC. For the 5-bit case, the accuracies were 90.7\% (NL-ADC only) and 89.9\% (weight only) compared to a baseline of 91.1\%. For the 3-bit case, the accuracies were 88.9\% (NL-ADC only) and 87.8\% (weight only) compared to a baseline of 89.4\%. As can be seen, the effect of write noise on NL-ADC is less critical. Further, this can be reduced even more by using the redundancy technique described in Section \ref{supsec:redundancy}.

Lastly, improving the quality of devices\cite{rao_nature} as well as programming techniques\cite{song_science} for better accuracies in writing to memristors are subjects of ongoing research in the community. We hope with future improvements in both these areas, the drop in accuracy from software to hardware can be further minimized in future.

\newpage

\subsection{One-point calibration process}
\label{supsec:one-point calibration scheme}

\begin{figure}[t]
    \centering 
   \includegraphics[width=0.9\textwidth]{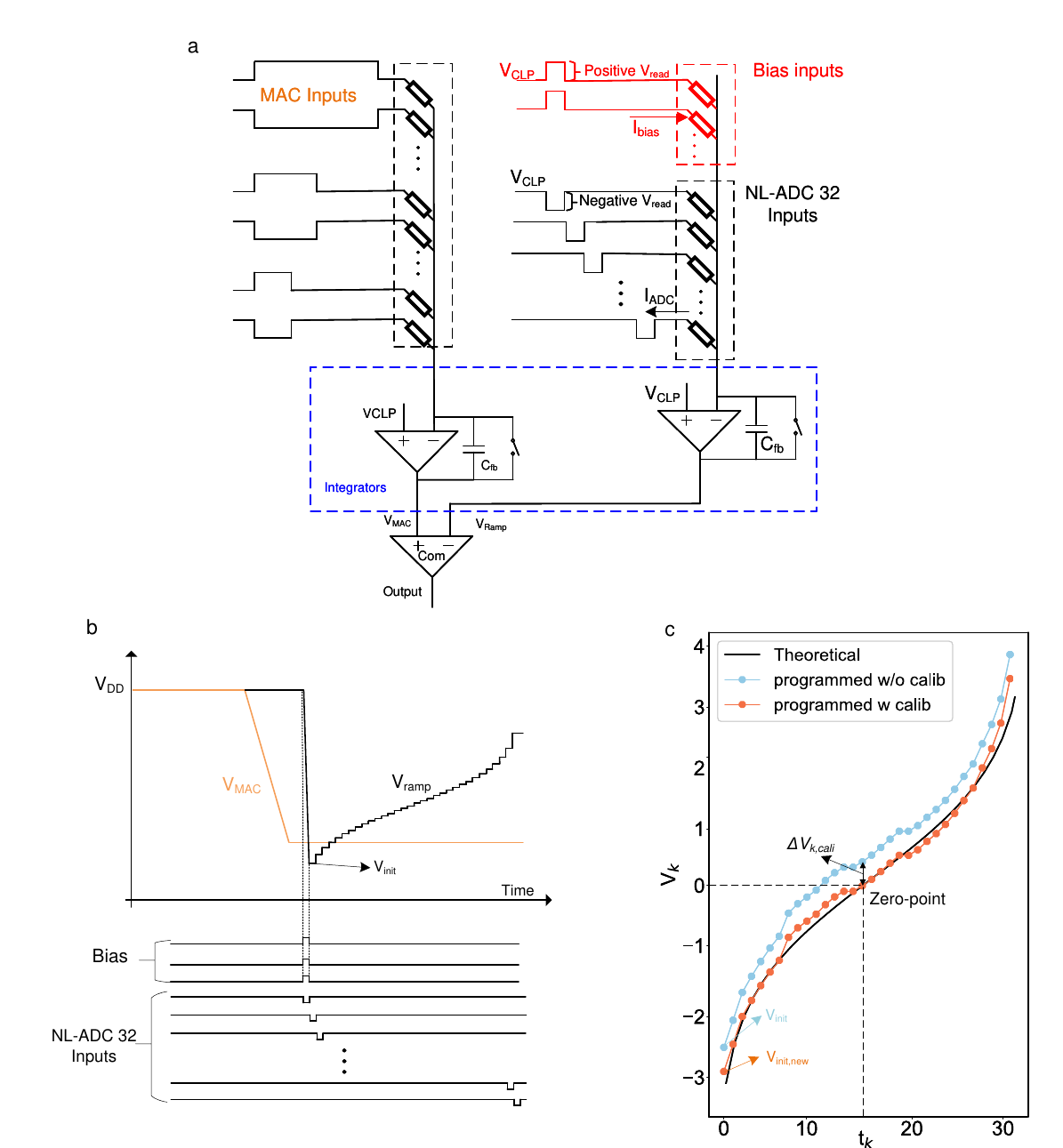}
    \caption{
    \textbf{One-point calibration.}
    \textbf{a} Hardware block diagram of V$_{MAC}$ and V$_{ramp}$. 
    \textbf{b} Timing diagram of V$_{MAC}$ and V$_{ramp}$.
    \textbf{c} Comparison of actual programmed V${_k}$ (same as V$_{ramp}$) with calibration and without calibration.
    }
   \label{figs:one-point calibration scheme}
\end{figure}

In the experimental setup, where we are using a 5-bit ADC as an example, the initial step involves programming the conductance values of 32 ADCs and bias conductance value into a column of RRAM. Subsequently, 32 pulses are sequentially sent to the RRAM corresponding to each ADC, as depicted in Fig. \ref{figs:one-point calibration scheme}\textbf{a}. The desired ramp voltage initially (V$_{init}$ in Fig. \ref{figs:one-point calibration scheme}\textbf{b}) starts from the smallest value in the first clock cycle and then increases progressively in every clock cycle after that to cover the full range of MAC values (Fig. \ref{figs:one-point calibration scheme}\textbf{b}). The initial negative voltage drop is created using the bias memristors. 

It is important to note that the pulses for bias are configured as positive input, while the pulses for the 32 NL-ADC cell inputs are configured as negative input (as illustrated in Fig. \ref{figs:one-point calibration scheme}\textbf{a}). This polarity distinction arises due to the negative $V_{read}$ of the NL-ADC pulses, resulting in the current flowing from the BL to the input of the RRAM. Consequently, these currents generate a positive voltage at the output after passing through the integrator, thereby causing all $V_{ramp}$ to be positive. However, in actuality, $V_{ramp}$ typically spans both positive and negative values. Thus, the pulse for bias necessitates a positive $V_{read}$ value to generate the desired negative voltage at the initial point (V$_{init}$ in Fig. \ref{figs:one-point calibration scheme}b). The current direction for bias RRAM cells is from the input of the RRAM to BL. Upon passing through the integrator, a negative voltage is generated, serving as the starting point (V$_{init}$) for the NL-ADC.

The blue curve in Fig. \ref{figs:one-point calibration scheme}\textbf{c} is the actual programmed value of $V_{ramp}$ without calibration. We can see that it deviates a lot from the theoretical value. To reduce the average INL (or average deviation of programmed ramp from the desired ramp), we use the one-point calibration method where the original starting point V$_{init}$ is modified to become V$_{init,new}$, such that the zero crossing point of the actual and desired ramps overlap. Based on the formula $\Delta V_{k,cali}=\frac{1}{C_{fb}}{V_{read}G_{cali}T_{adc}}$ ($\Delta V_{k,cali}$ is shown in Fig. \ref{figs:one-point calibration scheme}\textbf{c}, which is the vertical distance between two middle points of blue curve and red curve), the required RRAM conductance ($G_{cali}$) for calibration is determined. The new bias conductance value is obtained by adding the original bias conductance value to $G_{cali}$. Then we can get practical programmed $V_{k}$ with calibration in Fig. \ref{figs:one-point calibration scheme}\textbf{c}, where we can see $V_{k}$ with calibration is much closer to the theoretical value than $V_{k}$ without calibration.

\newpage

\subsection{Capacitor-based accumulation method for high-dimensional inputs}
\label{supsec: Capacitor-based accumulation method for large model}
\begin{figure}[t]
    \centering 
   \includegraphics[width=0.9\textwidth]{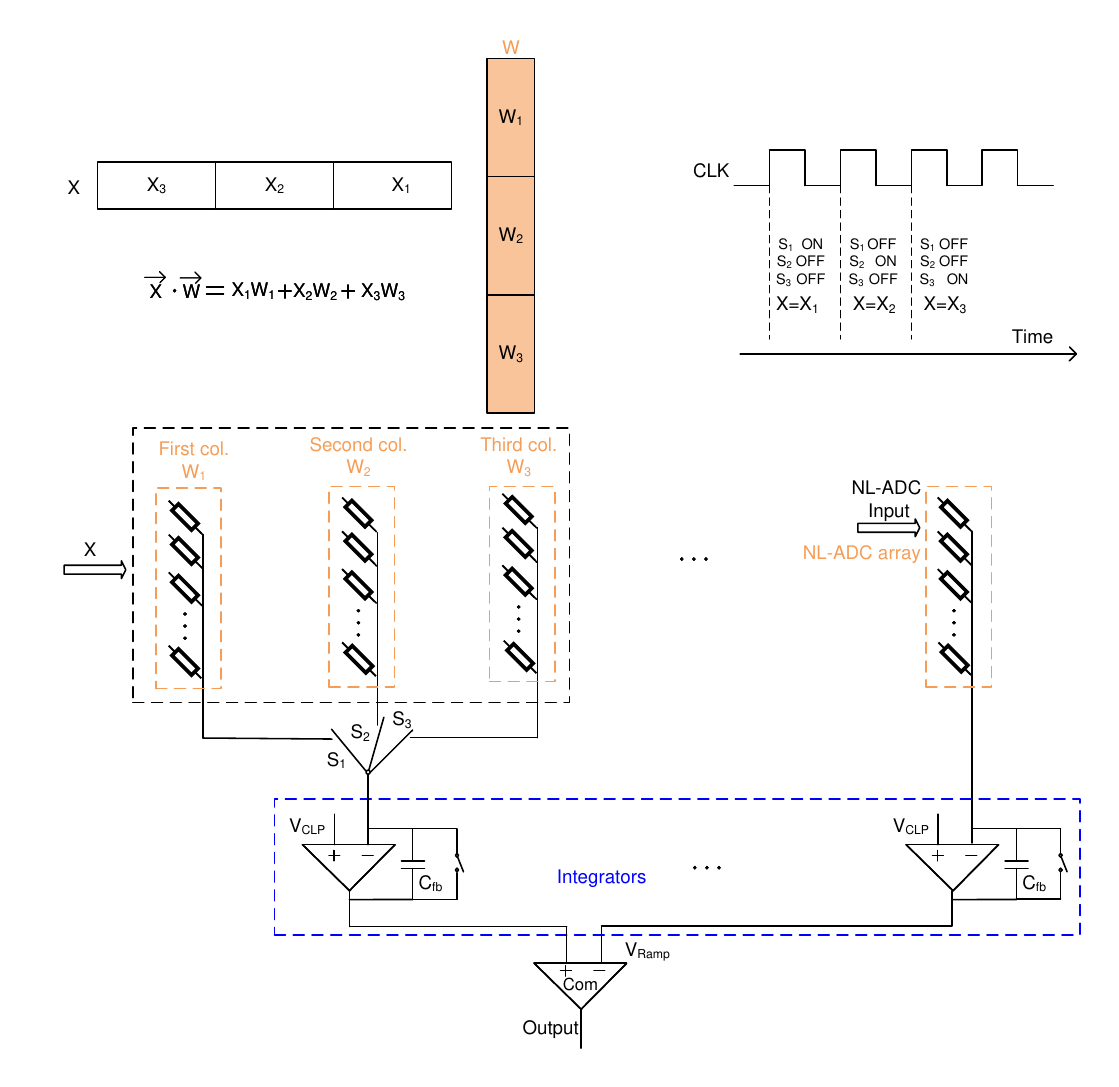}
    \caption{
    \textbf{Capacitor-based accumulation method for large model.} Input vectors with dimension larger than the number of rows can be applied by splitting it into parts and applying them sequentially over time. The corresponding weights are programmed on different neighbouring columns and are switched to an integrator following the same sequence. The capacitor accumulates the final MAC value over multiple cycles.
    }
   \label{figs: Capacitor-based accumulation method for large model}
\end{figure}

The method shown earlier can only handle input vectors with dimension less than the number of rows in the RRAM Macro. We show here how this method can be extended to calculate partial dot products and combine them later. In this case, the partial dot products can be stored as charge on the integrating capacitor. As shown in Fig. \ref{figs: Capacitor-based accumulation method for large model}, if the input vector dimension is more than the number of rows of the memory, multiple columns (in this case, 3 columns are used) may be used to store the weights. Then the input vector is also split into the same number of parts and is applied to the memory array in different clock cycles. A switch is used to connect the integrator to different columns in each of these cycles, where each of these columns store the weights for the respective part of the inputs. In the example shown, the input vector X is split into 3  parts–$X_1$, $X_2$ and $X_3$. So in the first clock cycle, the dot product $X_1\cdot W_1$ is calculated and stored on the capacitor. In the 2nd cycle, $X_2\cdot W_2$ is computed and added to the same capacitor and so on. Thus by using partial products in the analog domain, our method can be extended to handle input vectors beyond the row-depth of the memory. 

\newpage
\subsection{Redundancy for improved NL-ADC programming}
\label{supsec:redundancy}

\begin{figure}[t]
    \centering 
   \includegraphics[width=0.9\textwidth]{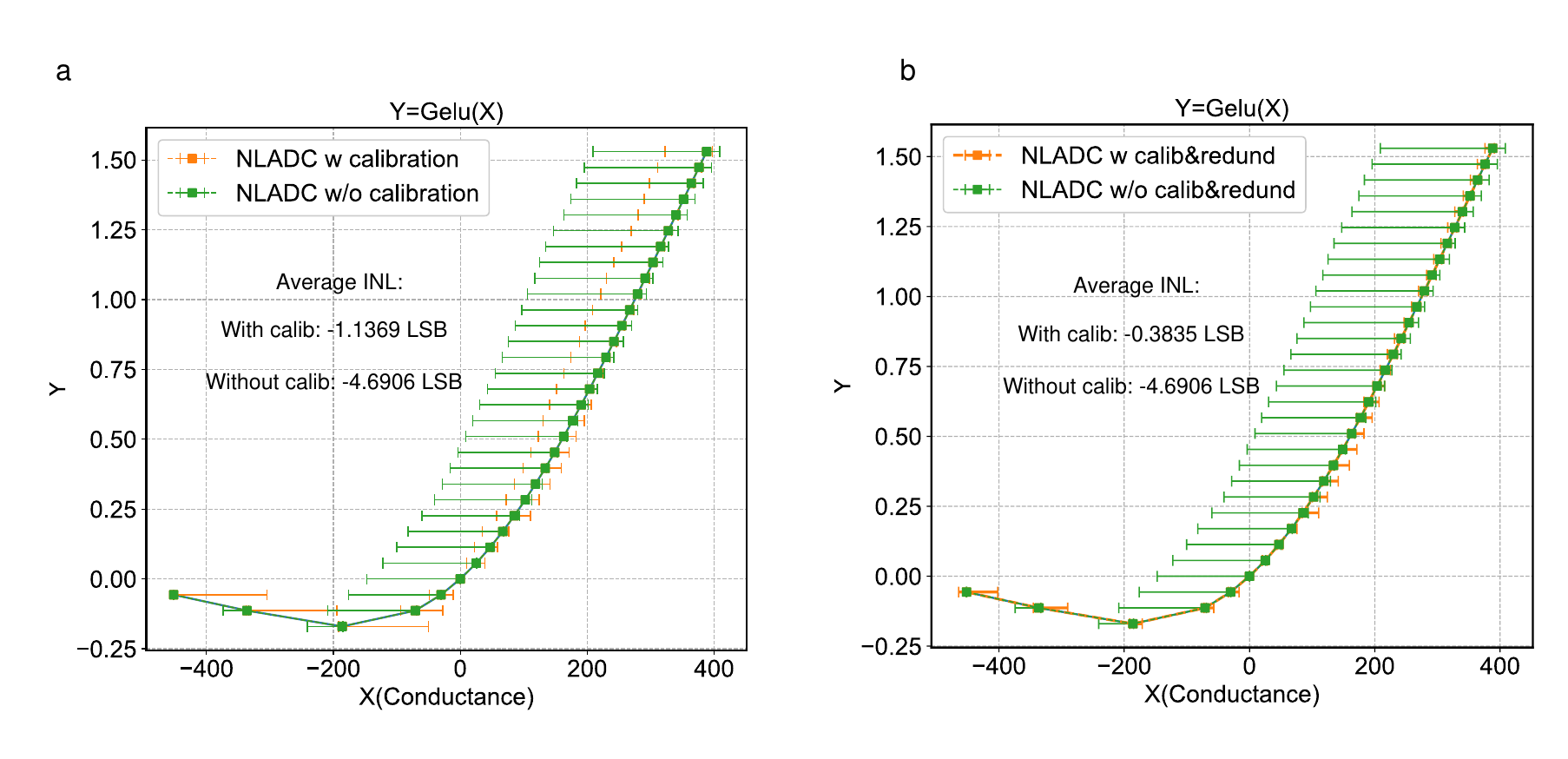}
    \caption{
    \textbf{GELU function approximation by ramp ADC with/without redundancy method. } 
    \textbf{a} GELU function approximation by ramp ADC without redundancy method.
    \textbf{a} GELU function approximation by ramp ADC with redundancy method.
    The average INL reduces to -0.38 LSB from -1.14 LSB.
    }
   \label{figs: Gelu with redundency}
\end{figure}

We reduce variability and improves accuracy a lot with a redundancy based method. Briefly, the column used to generate the ramp for the ADC consists of the same number of memristors as those used for MAC, i.e. 64. However, when we use only 32 out of these for a 5-bit NL ADC, the remaining memristors in the same column are unused. The number of unused memristors are even more for 4-bit and 3-bit versions of the NL-ADC. We propose to use them and program redundant copies of the ADC reference in the case of 3-bit and 4-bit ADCs. For the 5-bit ADC, the starting location (row address in that column) of the ramp can be varied as well as multiple programming attempts can be made to get the best NL-ADC characteristic. This requires minimal overhead of only an extra register of 6-bits to store the base or starting address of the ramp for every crossbar. Alternatively, additional columns may also be used for programming redundant copies of the NL-ADC–this is the approach we have taken to get the new measured results below. For larger crossbars like 128x128, many such redundant copies can be fit into the same column with no extra overhead. We show an example where a 5-bit NL ADC for the GELU function is programmed by choosing the best out of 4 possible NL-ADC characteristics(Fig. \ref{figs: Gelu with redundency}). The average INL reduces to -0.38 LSB from -1.14 LSB proving the efficacy of this method. More details of implementing non-monotonic functions such as GELU and Swish are provided in \ref{supsec:Non-monotonic nonlinear function approximation by ramp ADC}.

\subsection{Non-monotonic nonlinear function approximation by ramp ADC}
\label{supsec:Non-monotonic nonlinear function approximation by ramp ADC}
\begin{figure}[t]
    \centering 
   \includegraphics[width=1.1\textwidth]{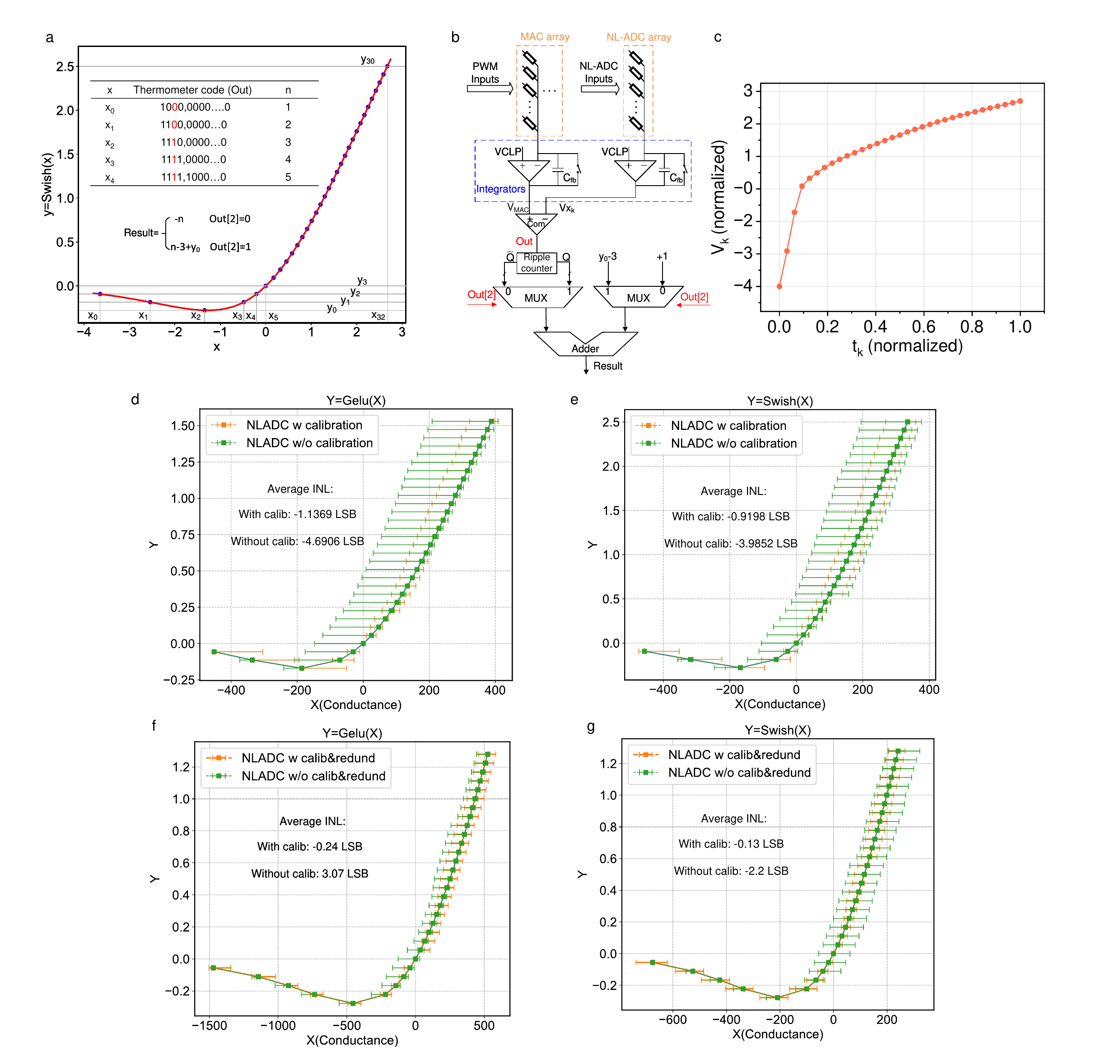}
    \caption{
    \textbf{Non-monotonic nonlinear function approximation by ramp ADC}
    \textbf{a} Method of non-monotonic nonlinear function approximation by ramp ADC requires splitting the curve into monotonic sections and using different linear equations for each section. Which section is to be chosen depends on the bit corresponding to the minima (Out[2] in this case) and two separate equations are used for the final ``result" depending on which section is selected.
    \textbf{b} Circuit diagram to implement the above method only requires the addition of one flip-flop to store the second output bit, one full-adder and two multiplexors.
    \textbf{c} The ramp voltage waveform for the Swish function obtained using the proposed method.
    \textbf{d} Swish function approximation by programmed ramp ADC. Average INL reduces to -1.1 LSB.
    \textbf{e} Gelu function approximation by programmed ramp ADC. Average INL reduces  to 0.9 LSB.
    \textbf{f} Gelu function approximation by programmed ramp ADC with more points on negative side and redundancy method. Average INL reduces to -0.24 LSB. 
    \textbf{g} Swish function approximation by programmed ramp ADC with more points on negative side and redundancy method. Average INL reduces to 0.13 LSB.
    }
   \label{figs: Swish and Gelu}
\end{figure}

For non-monotonic functions, the inverse function does not exist and hence the method of using the ramp ADC to approximate the nonlinear function cannot be directly used. \emph{However, the concept can still be applied by splitting the function into sub-parts where it is monotonic and using a logic to decide which sub-part has to be used}. Therefore, we directly divide the original function by selecting the extrema (maxima or minima) as the key points and then obtain the conductance values for each sub-part following the earlier technique. Taking the Swish function\cite{SWISH_ACTIVATION} (Fig. \ref{figs: Swish and Gelu}\textbf{a}) and 5-bit NL-ADC as an example, first the minima is identified as $(x_2,y_0)$ and this is used to split the function into two parts--the left and right of the minima. As before, starting from the minimum point $y_0$ of the function, the range of the function is divided into 30 equal intervals  ($y_0$-$y_{30}$). The spacing between two consecutive `y' values is the resolution of the NL-ADC. For $y_1$ and $y_2$, each value corresponds to two x values ($x_1$ and $x_3$, $x_0$ and $x_4$), so 33 x values can be obtained. Using the formula  $\Delta X_k=X_{k+1}-X_k \quad k \in[0,31]$, 32 $\Delta X_k$ can be obtained. Then, according to the formula $G_k=X_k 150 / \max \left(\Delta X_k\right) \quad k \in[0,31]$, 32 $G_k$ values can be obtained as conductance to be programmed into 32 RRAM cells.

After obtaining 32 conductance values, the other operations are the same as the previous monotonic function method. The control timing of the pulse is exactly the same as in Fig.\ref{figs:one-point calibration scheme}. In this way, a ramp waveform starting from $x_0$ and passing through  $x_1$, $x_2$, ..., $x_{31}$ can be generated as shown in Fig. \ref{figs: Swish and Gelu}\textbf{c}. The output result is still the thermometer code that generates positive and negative values by comparing the MAC value and the step wave of $x_0$-$x_{32}$. However, the output has to be obtained by two different equations depending on which sub-part has to be chosen corresponding to the input x. This can be done easily based on the output bits of the thermometer code. In the example shown, the minima is at $(x_2,y_0)$; hence, we have to use the left sub-part of the function if $Out[2]=0$ and the right sub-part otherwise. Here, $Out[k]$ denotes the k-th bit of the thermometer code produced by the sense amplifier (SA). In the case of monotonic functions, these bits of  thermometer code can be converted to a binary code (representing the decimal number $n$) using a ripple counter (with bits denoted by $Q[i]$) that counts the number of 1's in the thermometer code. However, in this case of non-monotonic functions, two different equations have to be used for the two sub-parts to produce the final ``result". For the Swish and Gelu functions, it is given by: 
\begin{equation}
\label{supeq:non-monotonic}
result= \begin{cases}-n, & Out[2]=0 \\ n+y_0-3, &  Out[2]=1\end{cases} 
\end{equation}
as also shown in Fig. \ref{figs: Swish and Gelu}\textbf{a}. In this case, $y_0=-3$ as seen in \ref{figs: Swish and Gelu}\textbf{a}.

The hardware circuit implementation for this is shown in Fig. \ref{figs: Swish and Gelu}\textbf{b}. The inputs to the comparator are the MAC values and the ramp waveform corresponding to $x_0$-$x_{32}$ as earlier. Now, in addition we need a flip-flop to store $Out[2]$ that determines which sub-part has to be selected as explained earlier. The output of this flip-flop controls two multiplexors (MUX) to create the final result according to equation \ref{supeq:non-monotonic}. If $Out[2] < 0$, the result is given by $result=-n$ represented by $\bar{Q}+1$ in two's complement. On the other hand, if $Out[2]=1$, the result is given by $result=n-3+y_0$. We use two MUXs and one adder to implement this simple comparison and addition as shown in Fig. \ref{figs: Swish and Gelu}\textbf{b}. The mathematical relationship can be summarized as follows:
\begin{equation}
result= \begin{cases}\bar{Q}+1, & Out[2]=0 \\ Q+y_0-3, &  Out[2]=1\end{cases} 
\end{equation}
For example, when MAC value is $x_1$, $n$ is equal to 2 as shown in the inserted table in Fig.\ref{figs: Swish and Gelu}\textbf{a}. In that case, $Out[2]= 0$ and we can get the two results from left MUX and right MUX: $11101$(i.e., $\bar{Q}$) and $1$ respectively, obtaining the sum result of $11110$ equivalent to $-2$ in two's complement. Similarly, when MAC value is $x_4$, $n$ is 5, $Out[2]=1$ and we can get the two results from left MUX and right MUX: $00101$(i.e., $Q$) and $y_0-3=-6$ respectively. Then the sum result is $5+y_0-3=-1$ as desired.

Two different commonly used non-monotonic activation functions of Gelu\cite{hendrycks2016gaussian} and Swish\cite{SWISH_ACTIVATION} were programmed on the memristor array and the results are shown in \ref{figs: Swish and Gelu}\textbf{d} and \ref{figs: Swish and Gelu}\textbf{e} respectively. As done earlier in Fig. \ref{fig:nladc_performance}, $64$ copies of the same function were programmed to see the variability and the one point calibration was used to reduce average INL to $-1.1$ LSB for Gelu and $-0.91$ LSB for Swish respectively.

 We reprogrammed the memristors for both Gelu and Swish taking more number of sample points corresponding to negative outputs ( Fig. \ref{figs: Swish and Gelu}\textbf{f} and  Fig. \ref{figs: Swish and Gelu}\textbf{g}). This programmability is an advantage of our memristive ADC over other ones with fixed reference. Combined with the earlier described redundancy methods, the achieved average INL are -0.24 LSB and -0.13 LSB respectively. 

Finally, to assess whether our method of approximating the non-monotonic AF with uniform distribution of Y is effective in practical situations, two experiments are conducted using a Vision Transformer \cite{vision_transformer} (26 layers with 86M parameters) with the Gelu function on the CIFAR-100 dataset and a mixed-depth convolutional network \cite{mixed_depth} (124 layers with 2.6M parameters) with the Swish function on the CIFAR-10 dataset. Firstly, we train these two networks to obtain the software-level baseline accuracy: 92.2\%  and 93.7\%. When implementing these models on hardware, the reference voltage range of ADC is limited, which leads to clipping in the MAC (Multiply-Accumulate) results. So the accuracy of Vision Transformer and mixed-depth convolutional network degrade to 91.4\% and 93.2\%. In this case, we modified the training method to be: quantized activation functions are employed for forward propagation, while unquantized activation functions are used for backward propagation. As a reference, ReLU functions are also used to check if the quantized non-monotonic AF have any advantage over simpler but high precision AF. The networks utilizing 5-bit Gelu and Swish achieve 91.3\% and 93.2\% accuracy, respectively. This represents a reduction of only 0.9\% and 0.5\% compared to the SW baseline, while also outperforming the use of ReLU in place of the non-monotonic functions (accuracy when using ReLU was 91\% and 91.65\% respectively for these datasets). These results prove that 5-bit approximations of the AF incur very low loss in accuracy even for complicated networks such as vision transformers.

\newpage
\subsection{Effect of long-term drift of the RRAM conductances}
\label{supsec: Test results after adding long-term drift effect}

\begin{figure}[t]
    \centering 
   \includegraphics[width=0.95\textwidth]{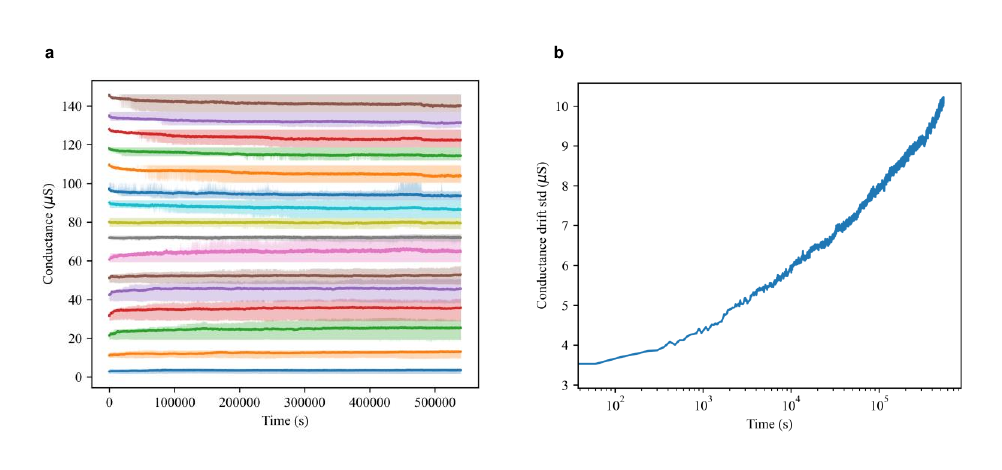}
    \caption{
    \textbf{Long-term drift effect of the RRAM conductances.}
    \textbf{a} RRAM conductance change over time for 16 different initial values. These are taken as reference values for the later simulations on classification accuracy change with time.  
    \textbf{b} Standard deviation of RRAM conductance change over time.
    }
   \label{figs: drift effect of the RRRAM conductances}
\end{figure}
The initial step involves acquiring the drift data of RRAM conductance. The conductance range of our RRAM device spans from 0 to \SI{150} {\micro\siemens}. This range is divided into 16 equidistant intervals, each with a resolution of \SI{9.375} {\micro\siemens}. Consequently, 16 distinct RRAM conductance values can be obtained, such as \SI{0} {\micro\siemens}, \SI{9.375} {\micro\siemens}, \SI{18.75} {\micro\siemens}, \SI{28.125} {\micro\siemens}, and so on. Subsequently, a 64x64 RRAM array is partitioned into 16 sub-arrays, each measuring 16x16. Within each sub-array, the 256 RRAM cells are programmed with the same conductance value, selected from the aforementioned set of 16 values. Following the programming phase, the conductance value of each sub-array is measured every 60 seconds. The average and standard deviation of these conductance values are calculated and visualized through graphical representation, as depicted in Fig. \ref{figs: drift effect of the RRRAM conductances}. The total duration of the measurement spans 500,000 seconds.

\begin{figure}[t]
    \centering 
   \includegraphics[width=0.95\textwidth]{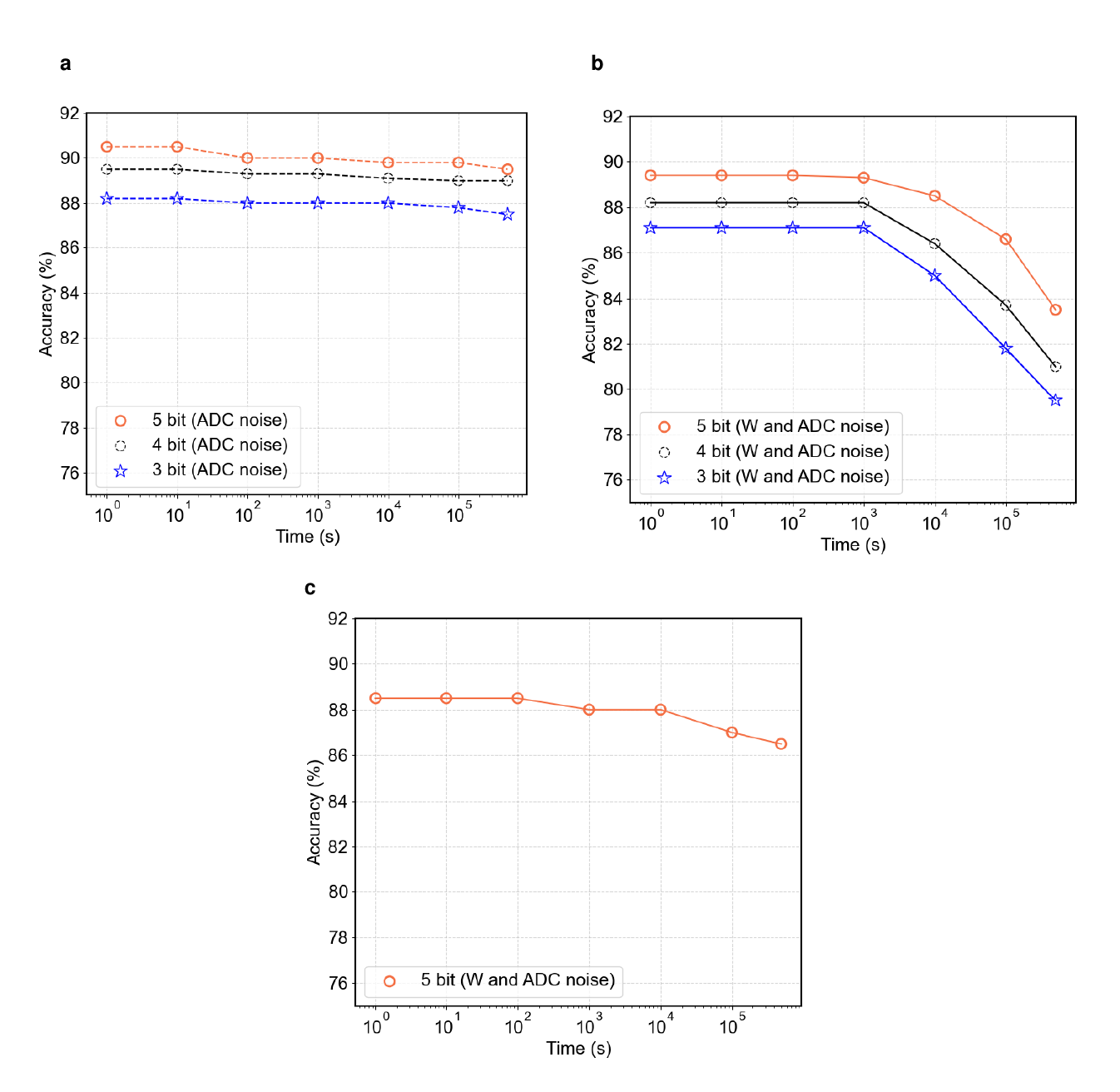}
    \caption{
    \textbf{Test accuracy with drift effect of the RRAM conductances (KWS model)}
    \textbf{a} Drift noise is added only in NL-ADC module for different resolutions. This shows minimal drop in accuracy over the entire simulation duration. 
    \textbf{b} Drift noise is added in both  NL-ADC module (for different resolutions) and weight module. Accuracy starts degrading after $\approx 1000$ seconds with a maximum degradation of $\approx 6\%$ at $5\times10^5$ seconds.
    \textbf{c} Modifying the training by adding a larger amount of noise ($N(0,8 \mu S)$) during training, and then testing with added drift noise in both NL-ADC and weights show reduced drop in accuracy and stable performance over time.
    }
   \label{figs: Test accuracy with drift effect of the RRRAM conductances (KWS model)}
\end{figure}

During the inference stage, in addition to incorporating write noise N(0, 2.67/75) (Fig. \ref{supfig:nladc_prog2}\textbf{c}, where $\gamma=75$ represents the scaling factor from RRAM conductance to weight as described in \Cref{eq:g=yw}) into the weights, we account for the impact of conductance drift using the weighted weight method. Initially, the 16 average RRAM values from Fig. \ref{figs: drift effect of the RRRAM conductances}\textbf{a} are divided by the scaling factor (75) and mapped to the weights, obtaining a set of 16 average weight curves depicting their temporal evolution (w${_i}$, where $i\in [0, 15]$). Subsequently, the drift effect of conductance is introduced to the weights using the following formula.

 Using this data in Fig. \ref{figs: drift effect of the RRRAM conductances}, we simulate the neural network for the KWS task to check the degradation of performance over time. To do this simulation, all the conductances in the KWS task are written as weighted average of the two nearest conductance values among the initial 16 reference conductances in this plot. For example, we can write for the k-th conductance,
\begin{equation}
G_k=a * G_{\text {ref, } p}(0)+b * G_{\text {ref, } p+1}(0)   \\ \\
\end{equation}

Where $\mathrm{G}_{\text {ref,p }}(0)<\mathrm{G}_{\mathrm{k}}<\mathrm{G}_{\text {ref,p+1 }}(0)$, and $\mathrm{G}_{\text {ref,p}}(0)$  indicates the values of the p-th reference conductance at time t = 0.

$\mathrm{a}=\left(G_{ref,p+1}(0)-G_k\right) /\left(G_{re,p+1}(0)-G_{ref,p}(0)\right)$ and $b=\left(\mathrm{G}_{\mathrm{k}}-\mathrm{G}_{\text {ref,p}}(0)\right) /\left(\mathrm{G}_{\text {ref,p} +1}(0)-\mathrm{G}_{\text {ref,p }}(0)\right)$   are weighting coefficients.  Then, the value of $G_k$ at time $t_k$ is obtained by the same weighted average of the drifted values of these reference conductances at time $t_k$ as follows: $G_k(t k)=a * G_{\text {ref,p }}(t k)+b^* G_{\text {ref,p+1 }}(t k)$.

Using this method, we show that if the RRAM drift affects the NL-ADC alone, the drop in accuracy is negligible (<1\%) for the 5-bit ADC (Fig. \ref{figs: Test accuracy with drift effect of the RRRAM conductances (KWS model)}\textbf{a}). However, if the drift affects both the weights and the NL-ADC, then the drop in accuracy starts increasing to ~6\% for the 5-bit ADC after 500,000 seconds (Fig. \ref{figs: Test accuracy with drift effect of the RRRAM conductances (KWS model)}\textbf{b}). In our work, we show that modifying the training by adding a larger amount of noise during training, the drop in accuracy can be restricted to <2\% even at 500,000 seconds (Fig. \ref{figs: Test accuracy with drift effect of the RRRAM conductances (KWS model)}\textbf{c}). This is much larger than the time needed for ADC operation during programming ($\approx3$ minutes) even when one single ADC is used for read operations during programming, as shown in \ref{supsec:nladc_prog}. Hence, the programming can be finished much before conductance drift starts affecting results.

\newpage
\subsection{Memristor programming circuits and overhead}
\label{supsec:memristor_prog_overhead}

In our prototype system, the write operation of memristors is done by serially accessing one device at each time. Multiplexers at each row and column of the crossbar arrays are used to select one memristor device at each time. For the read process, a constant 0.2V voltage drop is applied on the memristor device. The current flowing through the device is collected by a transimpedance amplifier and converted to voltage signal, which is then digitized by a conventional ADC. For the writing process, a positive/negative voltage drop is applied on the memristor cell to SET/RESET the device. The pulse width of SET/RESET voltage is set to 20 ns. The amplitude of SET/RESET voltage and the voltage on the gate terminal of the transistor of the 1T1M cell (which is used for current compliance) are adaptively changed in the write-and-verify process. It is worth noting that the ADC/DAC needed in the read/write process can be shared across all the rows and columns of the array, causing very limited overhead. To accurately tune the conductance of the memristor to an arbitrary value, multiple iterations of write-and-verify might be needed, which consumes relatively long time (100 iterations is a conservative estimate \cite{rram_stable_programming}). However, thanks to the non-volatile property of memristor devices, the conductance tuning is a one-time overhead and does not influence the inference latency and throughput.  Once the weights of the neural network and references of the ADCs are programmed to the memristor array, they can be retained for any later usage without the need of programming the memristors again. Nevertheless, we assume usage of one ADC per crossbar array for better scalability in programming and have included its overhead (area of $\approx 280\mu m^2$ \cite{natIBM64core}) in area calculations at system level.
